\documentclass[cernpreprint,texlive=2011,texmf,txfonts,USenglish]{atlasdoc} 
\pdfoutput=1
\usepackage[subfigure=true,lineno=false,biblatex=false]{atlaspackage} 
\usepackage{atlasphysics}

\graphicspath{{logos/}{plots/}}

\usepackage{subfigure}
\usepackage{mathrsfs}
\usepackage{units}
\usepackage{graphics}
\usepackage{graphicx}
\usepackage{amsmath}
\usepackage{multirow}
\usepackage{url}
\usepackage{xspace}
\newcommand{\ETM}{\MET}

\newcommand{\dfour}{\ensuremath{B}}

\newcommand{\MT}{\ensuremath{M_{\mathrm T}}}
\newcommand{\Njets}{\ensuremath{N_{\mathrm jets}}}
\newcommand{\Nbjets}{\ensuremath{N_{\mathrm bjets}}}
\newcommand{\NV}{\ensuremath{N_{\mathrm V}}}

\newcommand{\obslimitB}{810}
\newcommand{\explimitB}{760}
\newcommand{\obslimitBsinglet}{640}
\newcommand{\explimitBsinglet}{505}
\newcommand{\obslimitT}{840}
\newcommand{\explimitT}{780}

 \PreprintIdNumber{CERN-PH-EP-2015-036} 

\AtlasTitle{Search for vector-like $B$ quarks in events with one isolated lepton, missing transverse momentum and jets at $\sqrt{s}=8$~TeV with the ATLAS detector}

\author{The ATLAS Collaboration}

\AtlasAbstract{
A search has been performed for pair production of heavy vector-like down-type ($B$) quarks.
The analysis explores the lepton-plus-jets final state, characterized by events 
with one isolated charged lepton (electron or muon), significant 
missing transverse momentum and multiple jets.
One or more jets are required to be tagged as arising from $b$-quarks, and at 
least one pair of jets must be tagged as arising from the hadronic decay of an electroweak boson.
The analysis uses the full data sample of 
$pp$ collisions recorded in 2012 by the ATLAS detector at the LHC, operating at a center-of-mass energy of  8~TeV, corresponding to an integrated luminosity of 20.3~fb$^{-1}$. 
No significant excess of events is observed above the expected background. Limits are set on vector-like
$B$ production, as a function of the $B$ branching ratios, assuming the allowable
decay modes are $B \rightarrow Wt/Zb/Hb$. 
In the chiral limit with a branching ratio of 100\% for
the decay $B \rightarrow Wt$, the observed (expected) 
95\% CL lower limit on the vector-like $B$ mass is 810~GeV  (760~GeV). 
In the case where the vector-like $B$ quark has
branching ratio values corresponding to those of an $SU(2)$ singlet state, the observed (expected)
95\% CL lower limit on the vector-like $B$ mass is 640~GeV (505~GeV). 
The same analysis, when used to investigate
pair production  of a colored, charge 5/3 exotic fermion $T_{5/3}$, with subsequent decay $T_{5/3} \rightarrow Wt$,
sets an observed (expected) 95\% CL lower limit on the $T_{5/3}$ mass of 840~GeV  (780~GeV).
}

\begin{document}

\maketitle

\section{Introduction}
\label{sec:intro}

A natural extension of the Standard Model (SM) is the introduction of new
fermions besides the usual three generations of chiral leptons and quarks.
Extensions of this type have been used to introduce new scales that assist in the
unification of gauge couplings in nonsupersymmetric models as well as certain
supersymmetric
models~\cite{Dermisek:2012ke,Dermisek:2012as,Martin:2012dg,Fajfer:2013wca} and
to provide new sources of $CP$ violation~\cite{Holdom:2009rf,Alok:2010zj}.
Additional quarks also arise naturally when considering Little Higgs
models~\cite{ArkaniHamed:2002qy,Schmaltz:2005ky},
models with a composite Higgs~\cite{Kaplan:1983sm,Agashe:2004rs,Contino:2006qr,Gillioz:2013pba},
nonminimal supersymmetric models~\cite{Kang:2007ib},
and nonsupersymmetric ``natural'' models~\cite{Berger:2012ec}.

The recent discovery~\cite{2012gk,Chatrchyan:2012ufa} of a Higgs boson with a mass near 125~GeV
 has modified rather than eliminated expectations for additional quarks.  SM-like ``sequential'' 
 fourth-generation quarks are disfavored, though not completely
excluded~\cite{Buchkremer:2012yy,Djouadi:2012ae,Chanowitz:2012nk,Banerjee:2013hxa}.
On the other hand, so-called ``vector-like'' quarks (VLQs), for which both the
left- and right-handed fields transform identically under the $SU(2)\times U(1)$ 
gauge transformations~\cite{AguilarSaavedra:2009es}, remain viable.
Indeed, VLQs could play a role in stabilizing the electroweak vacuum in light
of the observed Higgs mass~\cite{Xiao:2014kba}.  In contrast to sequential
chiral quarks, VLQs neither acquire mass through electroweak symmetry
breaking nor modify precision electroweak observables significantly.
Another distinguishing feature is that VLQs can have large
flavor-changing neutral current (FCNC) decay rates.  For instance, whereas a heavy
new chiral down-type quark ($d_4$) would decay predominantly via $d_4 \rightarrow Wt$, a
vector-like down-type quark of charge $-1/3$ (denoted hereafter by $B$) could decay  via $B \rightarrow Zb$
and $B\rightarrow Hb$, in addition to
via $B \rightarrow Wt$.  Likewise, a vector-like up-type quark of charge $2/3$ (denoted by $T$) could decay not
only via $T \rightarrow Wb$ but also via $T \rightarrow Zt$ and $T \rightarrow Ht$.  The values
of the VLQ $B$ and VLQ $T$  branching ratios are
determined by the VLQ multiplet structure and
effects such as the VLQ mixing with SM
quarks~\cite{AguilarSaavedra:2009es,Aguilar-Saavedra:2013pxa}.  It is usually assumed that new heavy
quarks of either type would couple primarily to the third generation of SM
quarks, for instance in order to suppress FCNC interactions among the SM 
quarks~\cite{Aguilar-Saavedra:2013wba}, but couplings to lighter generations
are not excluded~\cite{Atre:2008iu,Atre:2011ae}.

The ATLAS collaboration has published searches for sequential down-type
fourth-generation quarks~\cite{ATLAS:2012aw} and for heavy quarks decaying via
a neutral current~\cite{ATLAS:2012aka}.  More recently, limits on
VLQ $B$ masses have been quoted from an ATLAS 
analysis of  the $Zb$ final state in terms of VLQ multiplet structure,
with a 95\% CL lower limit of 685~GeV for an $SU(2)$ singlet and 755~GeV for
an $SU(2)$ doublet~\cite{Aad:2014efa}.  
The CMS collaboration has reported exclusion limits on $T_{5/3}$ at 800~GeV~\cite{CMS_T53}
and on $B$ assuming 100\% branching ratio to $Wt$ at 675~GeV~\cite{CMS_Wt}.

At a hadron collider, strong interactions among the initial state partons
can lead to the production of quark-antiquark pairs.
This article  presents a search for VLQ $B$ pair production, 
using  the full dataset of proton--proton ($pp$) collision events at a center-of-mass
energy of $\sqrt{s} = 8$~TeV recorded  in 2012 with the ATLAS detector at the
CERN LHC.  
The analysis explores
 the lepton-plus-jets final state, characterized by events 
with one isolated charged lepton (electron or muon), significant 
missing transverse momentum (the magnitude of which is referred to as \met), and multiple jets. 
The main SM backgrounds to this signature are events with production of a top quark and
antitop quark ($t\overline{t}$) and events with a $W$ boson produced in association
with  jets ($W$ + jets).  The
analysis requires that one or more jets are tagged as arising from $b$ quarks and that
at least one pair of jets is tagged as arising from the hadronic decay of an electroweak boson.
These requirements are designed primarily to address the signature
$B\bar{B} \rightarrow W^+W^-t\overline{t}\rightarrow W^+W^-W^+W^-b\overline{b}$ but also to retain
sensitivity to other VLQ $B$ (and $T$) decay signatures. 
In addition, since the analysis does
not distinguish the charges of the hadronically decaying $W$ bosons, it also has
sensitivity to pair production of a colored charge 5/3 exotic fermion, denoted by $T_{5/3}$, that
decays via $T_{5/3} \rightarrow Wt$ to a $W$ boson and top quark of the same charge sign.
The $T_{5/3}$ is predicted, for example, in some composite Higgs models~\cite{Contino:2006qr}.

After a brief description of the ATLAS detector in Sec.~\ref{sec:detector},
Sec.~\ref{sec:mc:simulation} describes the samples of data and Monte Carlo (MC)
simulation events used.  Section~\ref{sec:objects} describes how the various
reconstructed objects in the final state are reconstructed and identified.   The event selection is described in
Sec.~\ref{sec:selection}, followed in Sec.~\ref{sec:optimize} by a discussion of
the analysis strategy.
Section~\ref{sec:bg} describes how the
background is characterized, and Sec.~\ref{sec:syst} describes the systematic
uncertainties.  Section~\ref{sec:result} presents the results and,
since no signal is observed, provides the limits that are set on VLQ $B$ and $T_{5/3}$
production.  A summary and conclusions are given in
Sec.~\ref{sec:conclusions}.

\section{The ATLAS Detector}
\label{sec:detector}

 The ATLAS detector~\cite{Aad:2008zzm} covers nearly the entire solid 
 angle around the collision point and consists of an inner tracking detector surrounded 
 by a solenoid, electromagnetic (EM) and hadronic calorimeters, 
 and a muon spectrometer incorporating three large toroidal magnet systems,
 each with eight coils.

The inner-detector system (ID) is immersed in a 2~T axial magnetic field, provided
by a thin superconducting solenoid located before the calorimeters and 
provides charged-particle tracking in the pseudorapidity range $|\eta| < 2.5$.~\footnote{ATLAS uses a 
right-handed coordinate system with its origin at the nominal interaction point (IP) 
in the center of the detector and the $z$ axis along the beam pipe. The $x$ axis 
points from the IP to the center of the LHC ring, and the $y$ axis points upward. 
Cylindrical coordinates $(r,\phi)$ are used in the transverse plane, $\phi$ being 
the azimuthal angle around the beam pipe. The pseudorapidity is defined in terms 
of the polar angle $\theta$ as $\eta=-\ln\tan(\theta/2)$, and
angular distance is measured in terms of $\Delta R
\equiv \sqrt{(\Delta\eta)^{2} + (\Delta\phi)^{2}}$.
The transverse energy is defined as $\ET = E \sin\theta$.}  
The ID consists of three detector subsystems, beginning closest to the beamline
with a high-granularity silicon pixel detector, 
followed at larger radii by a silicon microstrip tracker and then a straw-tube-based
transition radiation tracker.  The ID makes possible an accurate reconstruction of tracks from the primary
collision and precise determination of the location of the primary vertex, as well as
reconstruction of secondary vertices due to decays of long-lived particles, such as those
including $b$-quarks.

The ATLAS calorimeter system covers the pseudorapidity range $|\eta|< 4.9$. 
Finely segmented EM sampling calorimeters, using lead
as the absorber material and liquid argon (LAr) as the active medium,
cover the  barrel ($|\eta| < 1.475$) and endcap ($1.375 < |\eta| < 3.2$) regions. An additional thin LAr 
presampler covering $|\eta| < 1.8$ allows corrections for energy losses in material upstream of the EM calorimeters. 
Hadronic calorimetry is provided by a steel/scintillator-tile calorimeter, segmented into three barrel 
structures within $|\eta| < 1.7$, and two copper/LAr hadronic endcap calorimeters
that cover the region $1.5<|\eta|<3.2$. The solid 
angle coverage is completed with forward copper/LAr and tungsten/LAr calorimeter modules, 
optimized for EM and hadronic measurements, respectively, and covering the region $3.1<|\eta|<4.9$.  

Outside the calorimeters lies the
muon spectrometer, which identifies muons and measures their deflection 
 in a magnetic field 
 generated by superconducting air-core toroidal magnet systems.  The spectrometer is made up of separate
trigger and high-precision tracking chambers. The precision chambers cover the region
$|\eta| < 2.7$ with three stations of monitored drift-tube chambers, complemented by cathode-strip
chambers in the forward region. The trigger system covers the range $|\eta|<2.4$, using
resistive plate chambers in the barrel and thin-gap chambers in the endcap regions

ATLAS uses a three-level trigger and data acquisition system.
The first-level trigger system is implemented in custom electronics, using a subset of the detector information 
to reduce the maximum event rate to a design value of 75~kHz. The second and third levels use software 
algorithms running on computer farms to yield a recorded event rate of approximately 400~Hz.

\section{Data and Monte Carlo Simulation Samples}
\label{sec:mc:simulation}

This analysis uses  the full dataset of 8~TeV $pp$ collision events recorded  in 2012 
with the ATLAS detector at the LHC. 
The data sample, after applying quality criteria that require all ATLAS subdetector
systems to be functioning normally, corresponds to a total integrated luminosity of $20.3~\ifb$.
Events were required to pass either a single-electron or single-muon
trigger designed to result in  roughly constant efficiency for electrons and muons identified
with the criteria described in Sec.~\ref{sec:objects}.

Simulated MC samples of events with VLQ $B\bar{B}$ production were generated using
{\sc Protos}~\cite{AguilarSaavedra:2008gt}
(version 2.2) and {\sc Pythia}~\cite{Sjostrand:2006za} (version 6.421) with
the ATLAS AUET2B MC parameter set (tune)~\cite{ATL-PHYS-PUB-2011-014} for parton showering and the underlying event
and MSTW2008 parton distribution functions (PDFs)~\cite{Martin:2009iq}.  VLQ $B$ masses were
set to values from 350 to 850~GeV in 50~GeV steps.  Production cross
sections were normalized using predictions from {\sc
Top}++~\cite{Czakon:2013goa,Czakon:2011xx} (version 2.0) at
next-to-next-to-leading order (NNLO) in QCD
with resummed next-to-next-to-leading-logarithm (NNLL) soft gluon terms
and MSTW2008 NNLO PDFs~\cite{Martin:2009iq,Martin:2009bu}.  Branching ratios to $Wt$,
$Zb$, and $Hb$ were set to 1/3 each; alternative branching ratio values were investigated
by reweighting MC events according to generator-level information.
For example, the SU(2) singlet point corresponds to a branching ratio to $Wt$  between 32\% and 47\%
(increasing with the mass of the $B$) and to $Hb$ between 26\% and 29\%
(decreasing with the mass of the $B$), over the mass range considered in this analysis.
Events for a few mass points (400, 600, and 800~GeV) were passed through
a {\tt GEANT4}-based detector simulation~\cite{geant4,Aad:2010wqa}.
Further mass points were simulated using a
faster simulation~\cite{ATLAS:1300517}, but validated at the above
mass points with the {\tt GEANT4}-based simulation.
Kinematically similar chiral $d_4$ samples, at mass values from 400~GeV
to 1~TeV, in 50~GeV steps, were generated with {\sc Pythia}~\cite{Sjostrand:2006za}  and
also passed through fast detector simulation, and  were used to bolster samples of simulated decays of VLQ $B$ to $Wt$.
All MC samples were reconstructed using the same algorithms used for the data.

The signal process of $T_{5/3}$ pair production was simulated at $T_{5/3}$ mass points ranging
from 600 to 1100~GeV, in 50~GeV steps, using
{\sc Madgraph}~\cite{Maltoni:2002qb}
and {\sc Pythia}
(version 8.175), with the ATLAS AU2 tune~\cite{ATL-PHYS-PUB-2012-003}
and CTEQ6L1 PDFs~\cite{Pumplin:2002vw}, and
passed through the fast detector simulation. As for the VLQ $B$ MC samples,
the cross section for $T_{5/3}$ pair production was normalized using
the NNLO+NNLL prediction from {\sc Top}++~\cite{Czakon:2013goa,Czakon:2011xx}.

The dominant background in this analysis is due to production of
$t\overline{t}$ pairs with additional jets, followed by $W$ bosons produced in
association with high-energy jets (``$W$+jets'') and other, smaller
background contributions.  The $t\overline{t}$ background was modeled using the
{\sc Powheg-box} (version 1, r2330)
next-to-leading-order (NLO) generator~\cite{Alioli:2010xd,Frixione:2007nw}
interfaced to {\sc Pythia} (version 6.427) with CT10 PDFs~\cite{Lai:2010vv}
and the Perugia P2011C tune~\cite{Skands:2010ak}
for parton shower and underlying event modeling
and then normalized to the theoretical cross section calculated
at NNLO with resummation of NNLL soft gluon
terms~\cite{Cacciari:2011hy,Baernreuther:2012ws,Czakon:2012zr,Czakon:2012pz,Czakon:2013goa,Czakon:2011xx}.
The $W$+jets background, along with $Z$+jets,
was modeled with {\sc Alpgen}~\cite{alpgen} (version 2.14) with up to five additional partons,
and {\sc Pythia} (version 6.426) using CTEQ6L1 PDFs.  Both were normalized to
inclusive NNLO cross sections~\cite{Anastasiou:2003ds,Gavin:2010az}.

Among the smaller backgrounds, $t\overline{t}$ in association with a $W$ or $Z$ boson (``$t\overline{t} + V$'') 
was modeled with {\sc Madgraph}~\cite{Maltoni:2002qb} (version 5)
and {\sc Pythia} (version 6.425) with the CTEQ6L1 PDFs
and normalized to the NLO cross section prediction~\cite{Campbell:2012dh,Garzelli:2012bn}.
Electroweak single top production was simulated using
{\sc Powheg-box} (version 1, r2330) and {\sc Pythia} (version 6.425) with the
P2011C tune and CT10 PDFs for $s$-channel
and $Wt$ processes and
{\sc AcerMC}~\cite{Kersevan:2004yg,Kersevan:2006fq} (version 3.8)
and {\sc Pythia} (version 6.426)
for the $t$-channel process.
The combined single top sample, with overlaps between the $Wt$ and \ttbar{} samples 
removed~\cite{Frixione:2005vw}, was normalized to approximate
NNLO cross sections~\cite{Kidonakis:2011wy,Kidonakis:2010tc,Kidonakis:2010ux}
using the MSTW2008 NNLO PDF set.
Diboson ($WW$, $WZ$, $ZZ$) production was modeled using
{\sc Alpgen} (version 2.14) and
{\sc Jimmy}~\cite{Butterworth:1996zw}
for all processes except for the $WZ$ channel 
where the $Z$ boson decayed hadronically, in which case
{\sc Sherpa}~\cite{Gleisberg:2008ta} (version 01-04-01)
was used.
All diboson samples used CT10 PDFs and were
normalized to the NLO cross-section calculation~\cite{Campbell:2011bn}.

The normalizations and shapes of the background contributions
are validated using data control regions (see Sec.~\ref{sec:bg}).  The
multijet background contribution with misidentified lepton candidates is determined entirely
with data-driven techniques.

The effect of multiple $pp$ interactions in the same or nearby bunch
crossings (``pileup'') is taken into account in all simulations,
and the distribution of the number of interactions
per bunch crossing in the simulation is reweighted
to that observed in the data. During the  2012 data-taking period, the 
average number of $pp$ collisions per bunch crossing varied between 6 and 40, with a mean value of 20.7.

\section{Object Reconstruction and Identification}
\label{sec:objects}

The reconstruction and identification algorithms for electrons are described in 
Refs.~\cite{Aad:2011mk, Aad:2014fxa}. Electrons are identified as isolated EM calorimeter 
energy deposits, matched to reconstructed tracks in the inner detector, with transverse
energy $\ET>25$~GeV and pseudorapidity $|\eta|<2.47$, excluding the transition
region, $1.37<|\eta|<1.52$, between the barrel and endcap calorimeters.
The track must originate less than $2\;{\mathrm mm}$ along the
beamline from the primary vertex, which is defined as the
reconstructed vertex with the largest sum of associated track $\pT^2$.
In addition, nonprompt electrons are
suppressed by imposing isolation requirements:  the calorimeter $\ET$
within a surrounding cone of $\Delta R=0.2$ and track \pT \  within $\Delta R=0.3$,
excluding the electron candidate itself, are each required to be smaller than
$\ET$- (or $\pT$-) and $\eta$-dependent thresholds. The \ET \ and \pT \ thresholds are determined separately to accept
$98\%$ of electrons from $Z\rightarrow ee$ decays.  Since electrons are also 
reconstructed as jets, the closest jet within $\Delta R=0.2$ of an electron is removed,
in order to avoid using one reconstructed object multiple times.  Finally, electron
candidates within $\Delta R=0.4$ of jets, described below, are discarded.

Muon candidates are found by matching tracks in the muon spectrometer
and the inner detector with $|\eta|<2.5$, $\pT>25$~GeV and
originating within $2\;{\mathrm mm}$ of the primary vertex~\cite{Aad:2014zya}.
Muon isolation is enforced by calculating the ratio $I$ of the sum of the $\pT$ values of
tracks in
a cone of size $\Delta R=10$~GeV $/\pT^\mu$ to the transverse momentum
$\pT^\mu$ of the muon candidate itself.  A requirement of $I<0.05$ is applied, which has
an efficiency of $97\%$, as measured in $Z\rightarrow\mu\mu$ decays.
As with electrons, muon candidates within $\Delta R=0.4$ of jets are
discarded.  

Jets are defined using the 
anti-$k_t$ algorithm~\cite{Cacciari:2008gp} with a radius parameter of 0.4,
starting from calorimeter energy clusters calibrated using the local weighting
method~\cite{Aad:2011he,Aad:2014bia}.
Jets are then calibrated using
a simulation-based energy- and $\eta$-dependent calibration scheme.
Jets with $\pT>25$~GeV and $|\eta|<2.5$ are considered for further analysis.
Contributions to the jet momentum from pileup interactions are suppressed
using a jet-area-based subtraction method~\cite{ATLAS-CONF-2013-083}.
Jets with $\pT < 50$~GeV and $|\eta|<2.4$ are further required to have a
jet vertex fraction (JVF) of at least 50\%, where JVF is defined as
the scalar sum of the $\pT$ of tracks associated with the
jet cone which originate from the selected primary vertex, divided by the
sum for all tracks associated with the jet.

Jets are ``tagged'' as $b$-quark jets using a multivariate discriminant
based on track impact parameters and reconstructed secondary
vertices~\cite{ATLAS-CONF-2011-102,ATLAS-CONF-2012-043}.
The discriminant threshold is set to correspond to approximately $70\%$
efficiency for $b$-quark jets from $t\overline{t}$ decays.  This threshold
achieves a rejection factor of approximately 140 against light-quark and
gluon jets and 5 against charm-quark jets.

In the lepton-plus-jets final-state topology studied, signal events should
include, in addition to the $W$ boson that decays leptonically and gives rise
to the charged electron or muon and \MET, some number of hadronically decaying $W$ 
and/or $Z$ bosons.  These hadronic decays of intermediate vector bosons can be
reconstructed using pairs of jets with a dijet invariant mass value that lies
within a window around the known $W$ and $Z$ masses. The masses of the 
$W$ and $Z$ bosons are sufficiently close in value so that, given the jet energy resolution,
it would be difficult to separate hadronic $W$ and $Z$ decays; instead, 
a dijet mass window is used that is wide enough to 
select with high efficiency either $W$ or $Z$ candidates.

In hadronic $W/Z$ decays, the typical angular separation 
between the two daughter jets is related to the mass ($m$) 
and the transverse momentum (\pT)
of the decaying $W/Z$ boson by $\Delta R \approx 2\times m/\pT$.
$W/Z$ bosons produced in the decays of massive VLQ $B$ quarks
typically possess large values of $\pt$, so that their
daughter jets lie relatively close to each other in the detector. 

To reconstruct hadronically decaying $W/Z$ candidates, all pairs of jets are
considered, and pairs are retained if they have a dijet mass 
within the range of 60--110~GeV, consistent with the $W$ and $Z$ boson masses.  
To reduce the combinatorial background in the
high jet-multiplicity events considered, the two jets must  be close to one another in the detector, 
separated by
$\Delta R<1.0$. In addition, the transverse momentum of the dijet system must satisfy 
$\pT(jj) > 120$~GeV.  To avoid double counting,
any individual jet may only be part of one selected dijet pairing. This condition is fulfilled by
considering dijets formed by selecting pairs of individual
jets from a list ordered from the highest \pT \ value to the lowest and removing
from further searching both jets of any pair that satisfies the requirements.
The number of hadronically decaying intermediate vector boson ($W/Z$)
candidates passing these requirements in a given event is denoted hereafter
by \NV \ and is subsequently used in the analysis.

The measurement of \met~\cite{Aad:2012re} is based on the energy deposits in the calorimeter
with $|\eta|<4.9$.
The energy deposits associated with reconstructed objects (jets, photons, and
electrons) are calibrated accordingly. Energy deposits not associated with a reconstructed object are
calibrated according to their energy sharing between the electromagnetic and hadronic calorimeters.
The energy associated with reconstructed muons, estimated 
using the momentum measurement of its reconstructed track, is taken into account in the calculation
of \met.

\section{Event Preselection}
\label{sec:selection}

The analysis searches for $\dfour\bar\dfour$ pair production, with the VLQ \dfour \
subsequently decaying via the modes $\dfour \rightarrow Wt/Zb/Hb$. 
The event 
preselection requires exactly one isolated charged lepton (electron or muon)
with high \pT \ and also a high value of \MET.  The selected electron or muon is
required to have $\pT > 25$~GeV and to pass the isolation and other requirements
described in Sec.~\ref{sec:objects}. 
The preselected data are divided into mutually exclusive 
electron and muon channels, according to the nature of the identified charged lepton.
As exactly one charged lepton candidate is required in each event, this analysis
shares no events with analyses of dilepton final states~\cite{Aad:2014efa}.

Additional requirements are applied to reduce the background contribution from
multijet events with a jet faking the leptonic signature. Events must
satisfy $\MET > 20$~GeV. In addition, a requirement is made that
$\MET + \MT > 60$~GeV, with \MT \ being the transverse mass of the
leptonic $W$ candidate, defined by

\begin{equation}
  \MT = \sqrt{2\pT \MET(1-\cos\Delta\phi)},
\end{equation}

\noindent where \pT \ is the transverse momentum of the lepton, and
$\Delta\phi$ is the azimuthal angle between the lepton and the 
 direction of the missing transverse momentum vector.

Signal $\dfour\bar\dfour$ events in the lepton-plus-jets final state should
include a large number of high-\pT \ jets. At least four jets, each with $\pT > 25$~GeV
and satisfying all the jet criteria outlined previously, must be present. 

The high jet multiplicities involved,
and resulting combinatorial background, complicate the task of
reconstructing the \dfour \ mass. Instead, the variable \HT, defined as the scalar
sum of \MET \ and of the \pT \ values of the lepton and of 
all selected jets, provides an effective measure of the mass scale
of the event. Given the large \dfour \ mass values probed, the event preselection requires
$\HT > 300$~GeV.

\section{Analysis Strategy}
\label{sec:optimize}

Signal events tend to have high values of jet multiplicity, with at least two
$b$ jets and higher values of \NV \ than background events.
In addition, due to the large \dfour \ mass values probed, \dfour \ signal
events are characterized by having higher energy jets than typical background events.
Two different analysis approaches are explored. The final results are derived using a multivariate analysis
technique, based on boosted decision trees 
(BDTs)~\cite{cart84-2,Friedman:2002:SGB:635939.635941,Freund1997119}, that
combines information from a number of input variables into a single discriminant.
A cuts-based approach, 
where requirements are imposed on individual final-state variables, is used as a cross-check.

 \subsection{Definition of signal and control regions}
\label{sec:regions}

The preselected data sample is divided into a set of mutually
exclusive subsamples: the signal region (SR) 
is used to perform the final search,
while five control regions (CRs) are used to validate the
background determination. As shown in Table~\ref{tab:srandcr}, the
variables used to define the various regions are the total
jet multiplicity (\Njets), the number of
hadronic $W/Z$ candidates (\NV), the number of $b$-tagged jets (\Nbjets), 
and the value of \HT \ in the event. The control region definitions ensure negligible
signal contamination of the control region samples. For example, 
the predicted contributions from a benchmark signal with a VLQ $B$ mass of 700~GeV 
are below 0.2\% in all control regions.

\begin{table}[htbp]
\begin{center}
\caption{Definitions of the SRs used in the
cuts-based and BDT analyses, in terms of jet multiplicity (\Njets), the number of
hadronic $W/Z$ candidates (\NV), the number of $b$-tagged jets (\Nbjets), 
and the \HT \ requirement.  The definitions of 
the various CRs used to validate the
background determination are also included.
Control regions WCR1 and WCR2 are dominated by $W$+jets events,
while TCR1, TCR2, and TCR3 are dominated by \ttbar\ events.
A dash in the \NV \ column means that no 
requirement is applied on that variable. A dash in the \HT \ column means
that no additional \HT \ selection is made, apart from the
$\HT > 300$~GeV requirement applied as part of the event preselection.}
\begin{tabular}{lcccc} \hline \hline
Data Region 	& \ \ \Njets \ \ & \NV & \Nbjets & \HT \ (GeV) \\ \hline
SR (cuts-based) \ \ &  $\ge 6$ & \ \ \ $\ge 1$ \ \ \  & \ \ \ $\ge 1$ \ \ \  & $> 800$ \\ 
SR (BDT) &  $\ge 6$ & \ \ \ $\ge 1$ \ \ \  & \ \ \ $\ge 1$ \ \ \  & $> 500$   \\  \hline
WCR1 &  $= 4,5$ & - &  $= 0$  & - \\ 
TCR1 &  $= 4,5$ & - & $\ge 1$    & -    \\  
WCR2 &  $\ge 6$ & - & $= 0$  &  -   \\ 
TCR2 &  $\ge 6$ &  -  & $\ge 1$       &  $< 500$  \\ 
TCR3 &  $\ge 6$ &  = 0  & $\ge 1$       & -  \\ \hline  \hline
\end{tabular} 
\label{tab:srandcr}
\end{center}
\end{table}

The SR is defined as those events with at least six jets, at least one of which
is tagged as a $b$ jet, and with at least one hadronic $W/Z$ candidate. Signal-region events  
must also have a high \HT \ value:  for the cuts-based analysis, \HT \ must exceed
800~GeV, while the BDT analysis requires $\HT > 500$~GeV. 
The analysis using the BDT discriminant uses a
lower \HT \ threshold since it has greater discriminating
power between the signal and background.
For both analysis methods, the expected signal contributions are rather small compared to
the overall SR sample sizes. As described in Secs.~\ref{sec:tmva} and \ref{sec:cuts}, additional
information is used to provide further separation between signal and background.

The signal selection efficiency in the SR is roughly constant as a function of the mass of
the heavy particle being pair produced,
decreasing only slightly for higher mass values.  Given the
selection requirement of exactly one isolated lepton, the efficiency is sensitive to
the value of the branching ratio  of the heavy quark for the decay to $Wt$.
For the case of a 100\% branching ratio,
as is true for the $T_{5/3}$ and for the VLQ $B$ in the chiral limit,
the efficiency is highest, with values in the range of 16\%--19\%.
For branching ratio values 
expected for the
cases where the VLQ $B$ is an $SU(2)$ singlet or part of an $SU(2)$
doublet~\cite{AguilarSaavedra:2009es,Aguilar-Saavedra:2013pxa},
the efficiencies are
lower; the efficiency for an $SU(2)$ singlet VLQ $B$
is approximately 8\%, while that for the $SU(2)$ doublet case, for which the branching
ratio for the decay to $Wt$ is zero, is less than 2\%.

Three control regions, denoted TCR1 through TCR3,  are used to validate the modeling and 
estimation of the  $t\bar{t}$ background, which dominates in these control regions as well
as in the SR. Two additional control regions,
denoted WCR1 and WCR2,  are defined to select samples that are dominated by
the $W$+jets background, and are used to validate the prediction of this background source.
WCR1 and TCR1 each require either four or five jets and therefore
have lower jet multiplicity than in SR events; these control regions differ from each other in
that TCR1 requires at least one $b$-tagged jet, whereas WCR1 requires that
none of the jets is $b$ tagged. Background events with higher jet multiplicity
are selected in WCR2 and TCR2, which each require at least six jets, as in the SR.
Again, these control regions differ from each other in that TCR2 requires at least one $b$-tagged 
jet, whereas WCR2 requires that none of the jets is $b$ tagged. To preserve
the orthogonality of TCR2 and the SR, TCR2 has the additional requirement
of $\HT < 500$~GeV. 
No requirement on the number of hadronic $W/Z$ candidates is made
on the four control regions TCR1, TCR2, WCR1 and WCR2.

The final background control region is TCR3. 
As with TCR2, TCR3 requires at least six jets, at least one of which
is $b$ tagged.  However, instead of requiring $\HT < 500$~GeV, TCR3 maintains
orthogonality with the SR by requiring zero hadronic $W/Z$ candidates. 
Since \HT \ is found to be an important discriminating variable in the BDT,
the relatively
low range of \HT \ values selected in TCR2 eliminates background events that would populate the 
signal-like region of high BDT values, thereby limiting the ability to validate the
performance of the BDT on background events.  The TCR3 definition selects
background events that are kinematically more similar to signal events, and thereby
allows a more sensitive validation of the BDT, while still ensuring negligible
signal contamination of the TCR3 sample.

As expected, the \ttbar\ background
dominates in the SR, as well as in the three \ttbar\ control regions, namely TCR1 through TCR3, used
to validate the modeling of the \ttbar\ background, contributing  80\% or more
of the total background in these samples. 
The $W$+jets background contributes 
typically 10\% or less of the total background in those regions.
It dominates the
backgrounds in WCR1 and WCR2, contributing 70\% and 55\% to the total background, respectively. 
The other backgrounds are in all cases small, and
their sum contributes typically about 10\% or less of the total background in any of
the regions.

\subsection{Multivariate discriminant analysis}
\label{sec:tmva}

Decision trees~\cite{cart84-2} recursively partition a data sample into
multiple regions where signal or background purities are enhanced. Boosting is a method
that improves the performance and stability of decision trees and involves the combination
of many trees into a single final discriminant~\cite{Friedman:2002:SGB:635939.635941,Freund1997119}.
After boosting, the final score
undergoes a transformation to map the scores on the interval $-1$ to $+1$, with the most signal-like
(background-like) events having BDT scores near $+1$ ($-1$).

The BDT implementation uses the TMVA tool in ROOT~\cite{Speckmayer:2010zz}.
Initial studies considered a long list of $\approx 50$ possible input variables. The list was reduced
by choosing variables with a high BDT ranking, which measures their ability to
separate signal from background, and removing variables that
have a high degree of correlation with a higher-ranked variable. The final BDT uses the
following twelve input variables, ordered from highest to lowest according to their rankings in the BDT training:

\begin{itemize}
\item \HT, defined previously;
\item $\Delta R(\ell,{\mathrm bjet1})$, the angular separation between the lepton and the
leading $b$-tagged jet;
\item \MT, the transverse mass of the leptonically decaying $W$ boson candidate;
\item $\pT(W_{{\mathrm lep}})$, the \pT \ value of the leptonically decaying $W$ boson candidate;
\item Min[$\Delta R(\ell,W_{{\mathrm had}})$], the minimum angular separation between the lepton and a 
hadronic $W/Z$ candidate;
\item $E_{{\mathrm bjet1}}$, the energy of the leading \pT\ $b$-tagged jet;
\item Average $\Delta R(j,j)$ for the jets of dijet hadronic $W/Z$ candidates;
\item \NV, the number of hadronic $W/Z$ candidates;
\item \Njets, the total jet multiplicity;
\item \Nbjets, the number of $b$-tagged jets;
\item $\pT(\ell)$, the \pT \ value of the lepton; and
\item \MET, the magnitude of the missing transverse momentum.
\end{itemize}

The signal sample used for training the BDT is made by combining 
large samples of simulated chiral $d_4$ decays
at four different masses (600, 700, 800, and 900~GeV).
The background samples used in the BDT training include  $t\bar{t}$ and $W+$jets; 
these two background processes together account for over 90\% of
the total background contribution in the signal region.  
Both the signal and background samples were divided into separate training and test samples to verify
 that there was no over-training of the BDT.

Figure~\ref{fig:tmva_vars} shows unit-area-normalized distributions of the twelve BDT variables,
again ordered by their BDT rankings, in the signal 
region for two VLQ $B$ signal masses (700 and 800~GeV), which are in the vicinity of the
expected sensitivity of the analysis, and also for the combined background contributions.
Figure~\ref{fig:tmva_vars} shows that each variable has some discriminating power between background and signal. 
The BDT technique combines these individual discriminants to produce an improved separation 
between background and signal.  

\begin{figure*}[hbpt]
\centering
\includegraphics[width=0.32\textwidth]{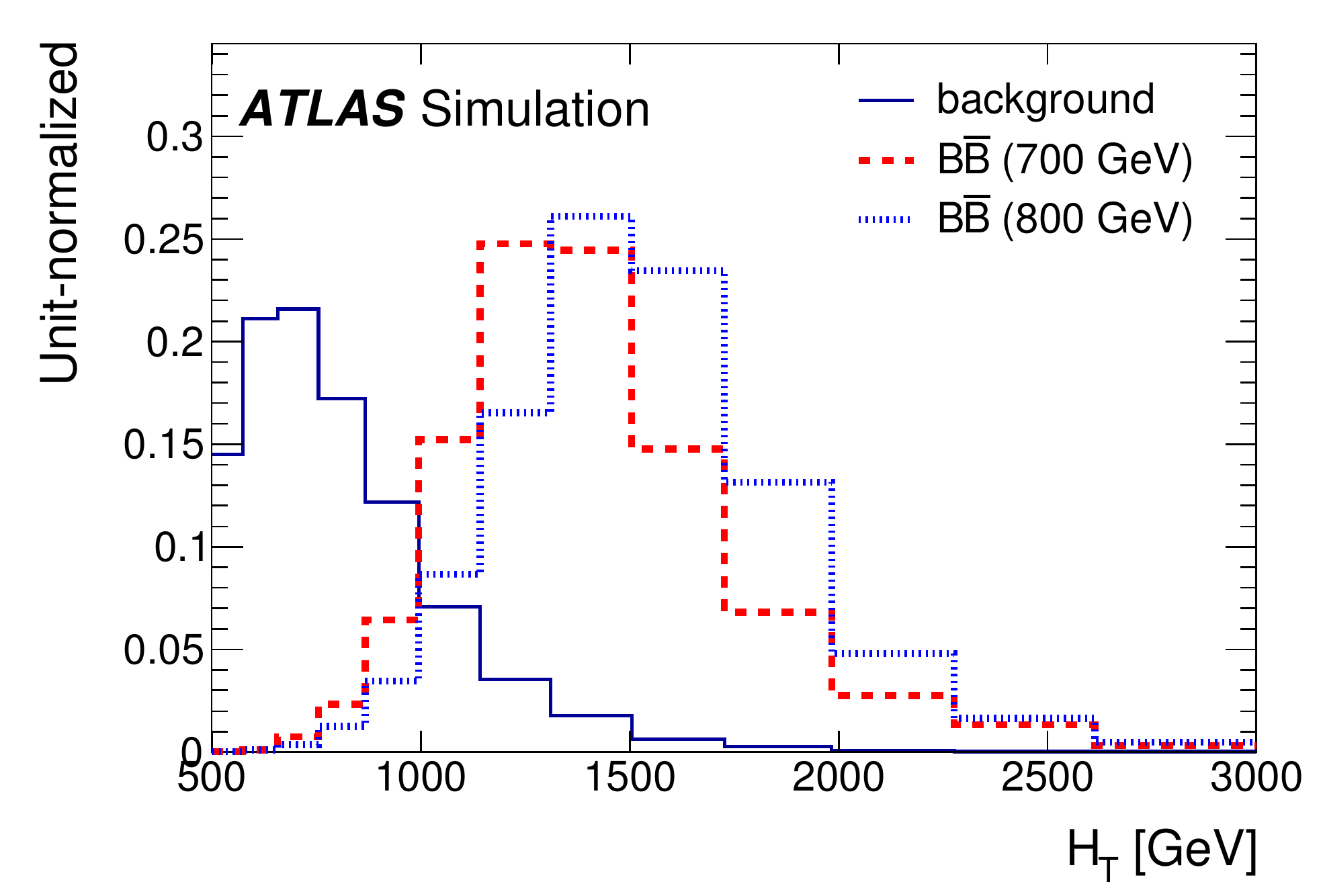}
\includegraphics[width=0.32\textwidth]{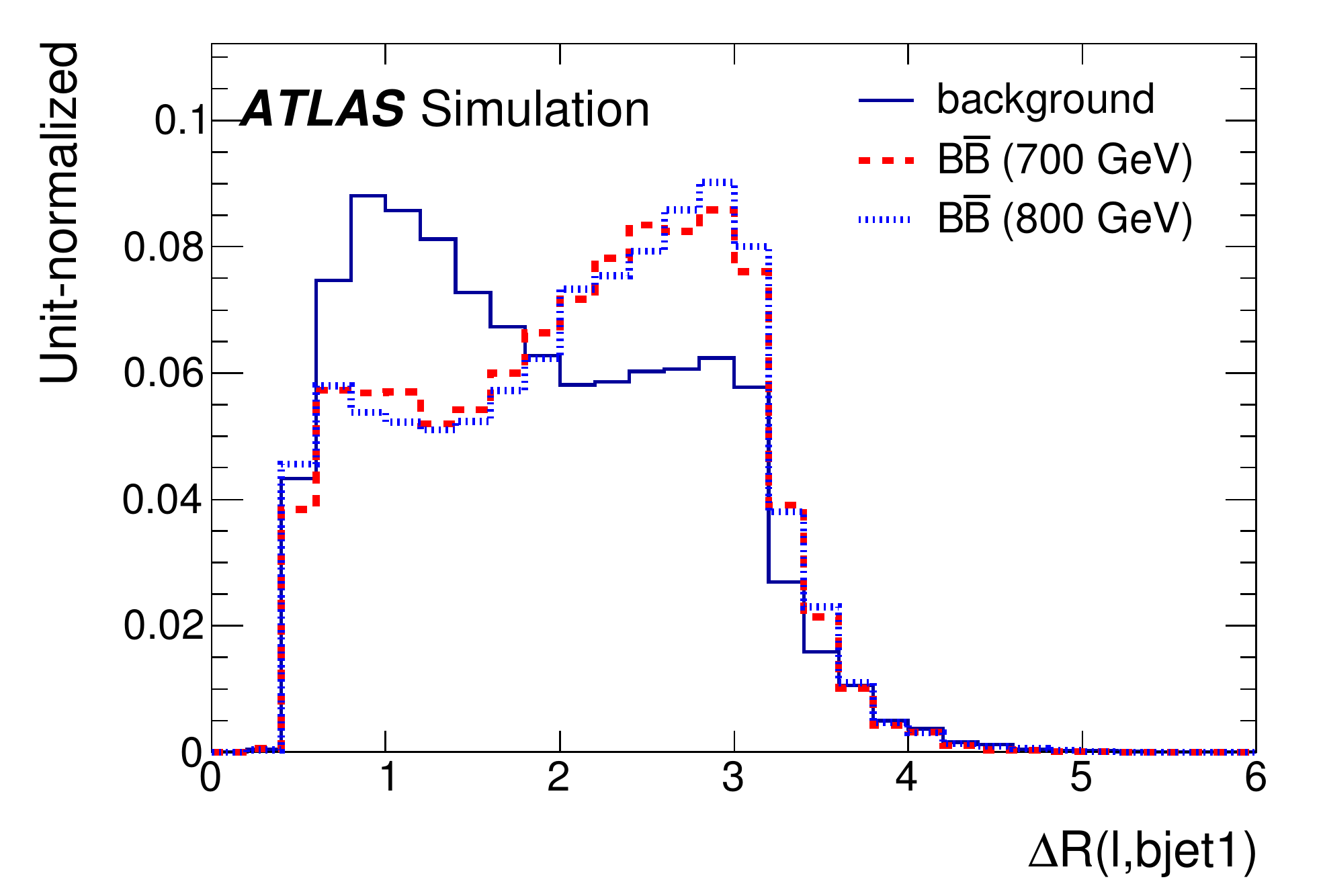}
\includegraphics[width=0.32\textwidth]{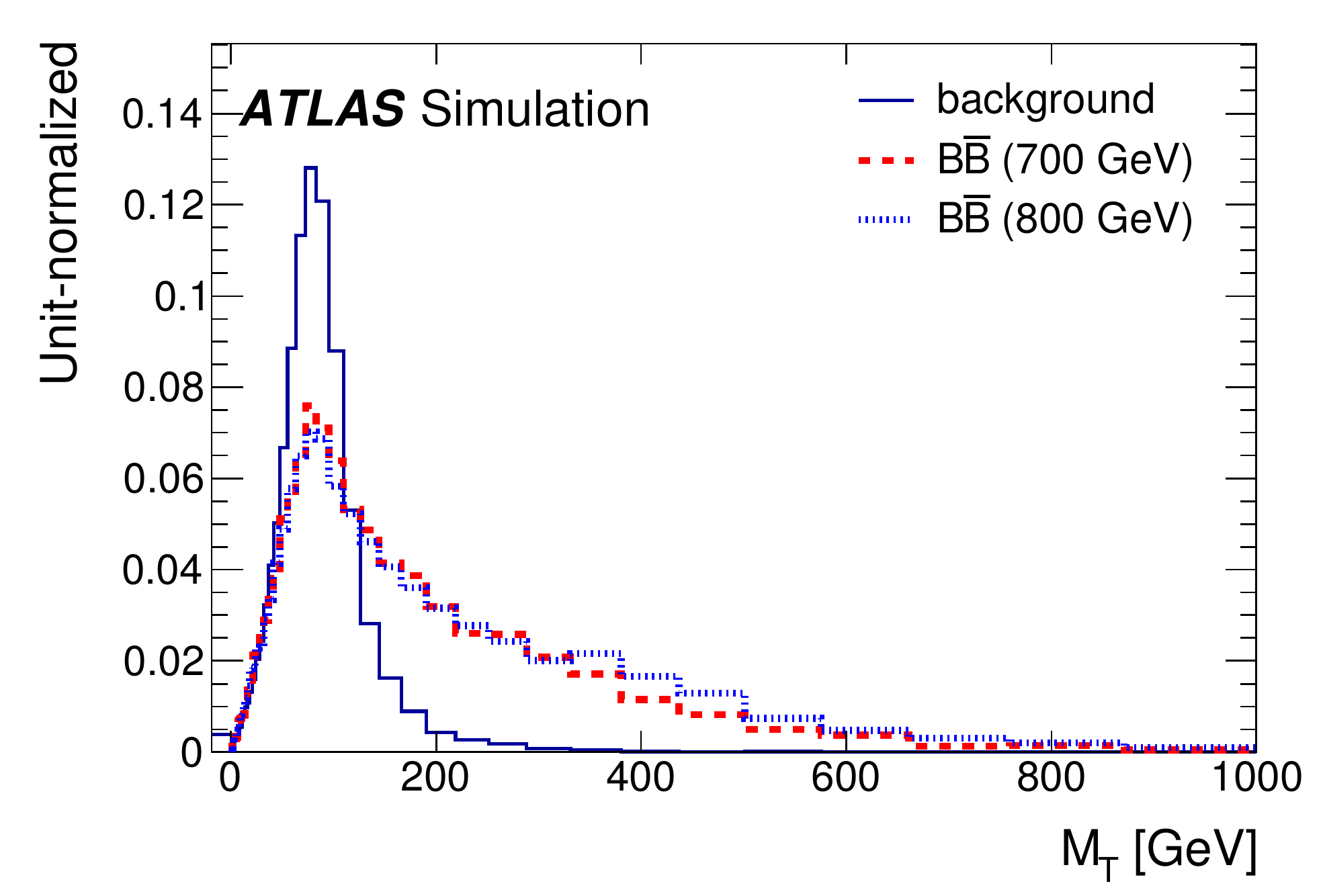}
\includegraphics[width=0.32\textwidth]{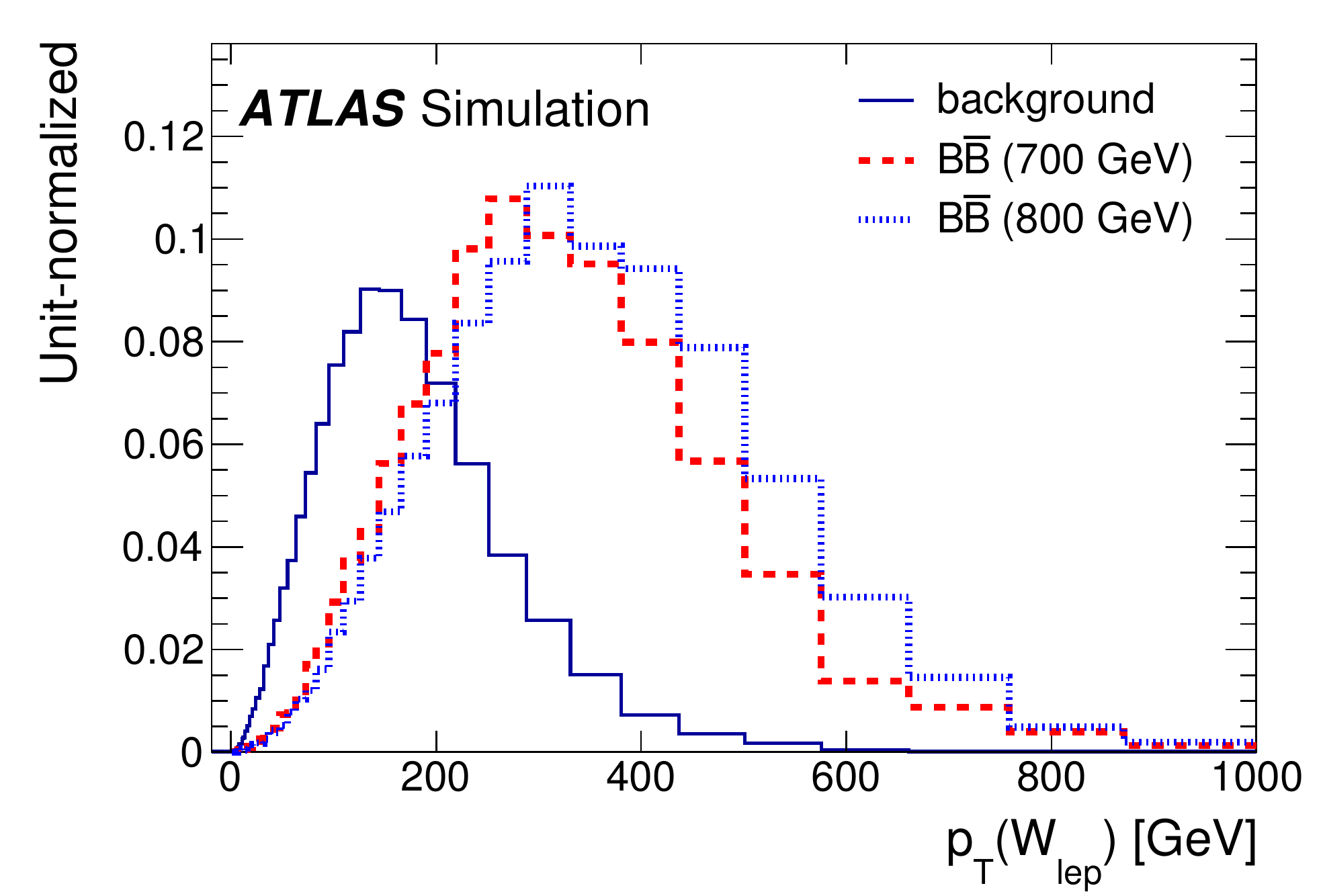}
\includegraphics[width=0.32\textwidth]{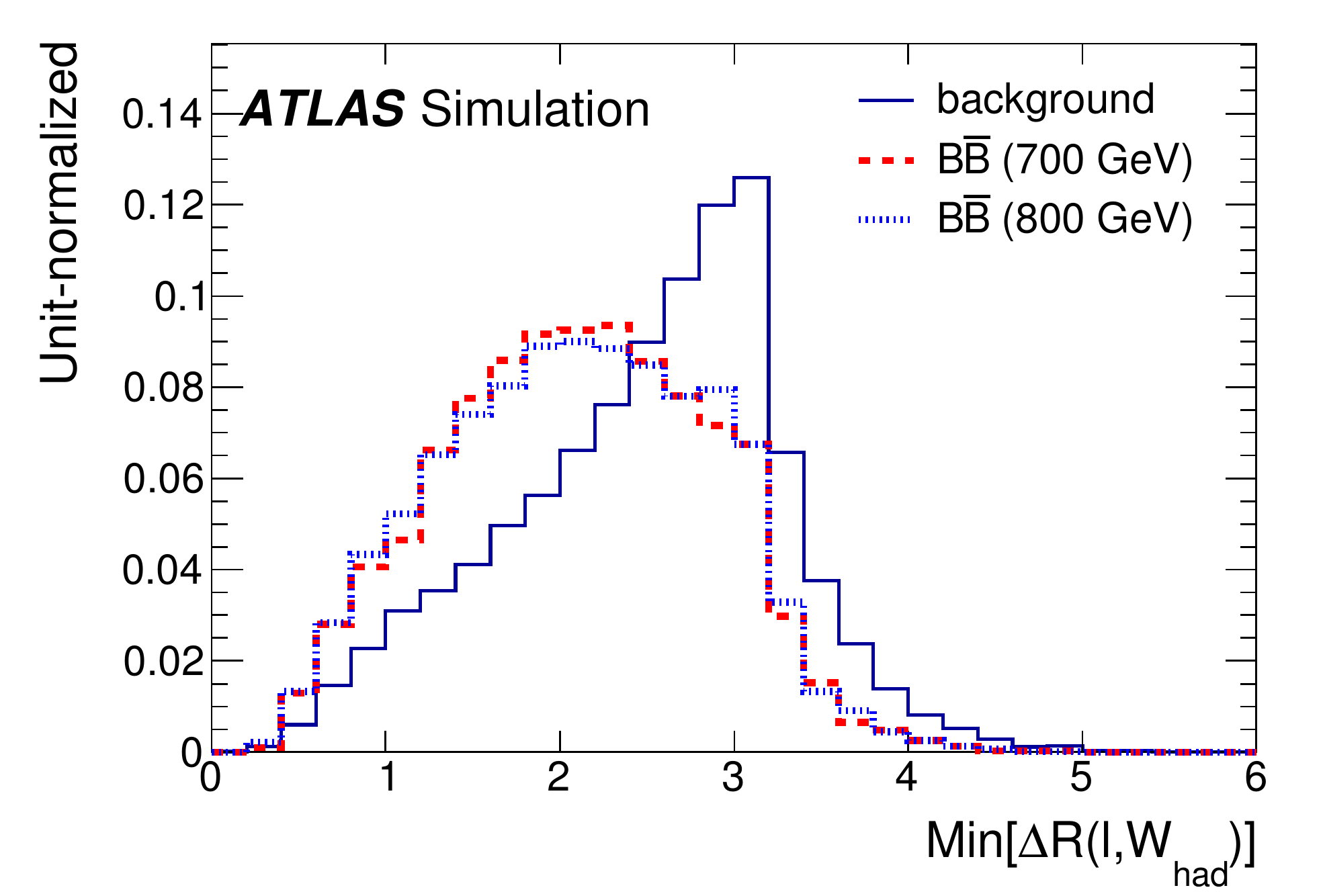}
\includegraphics[width=0.32\textwidth]{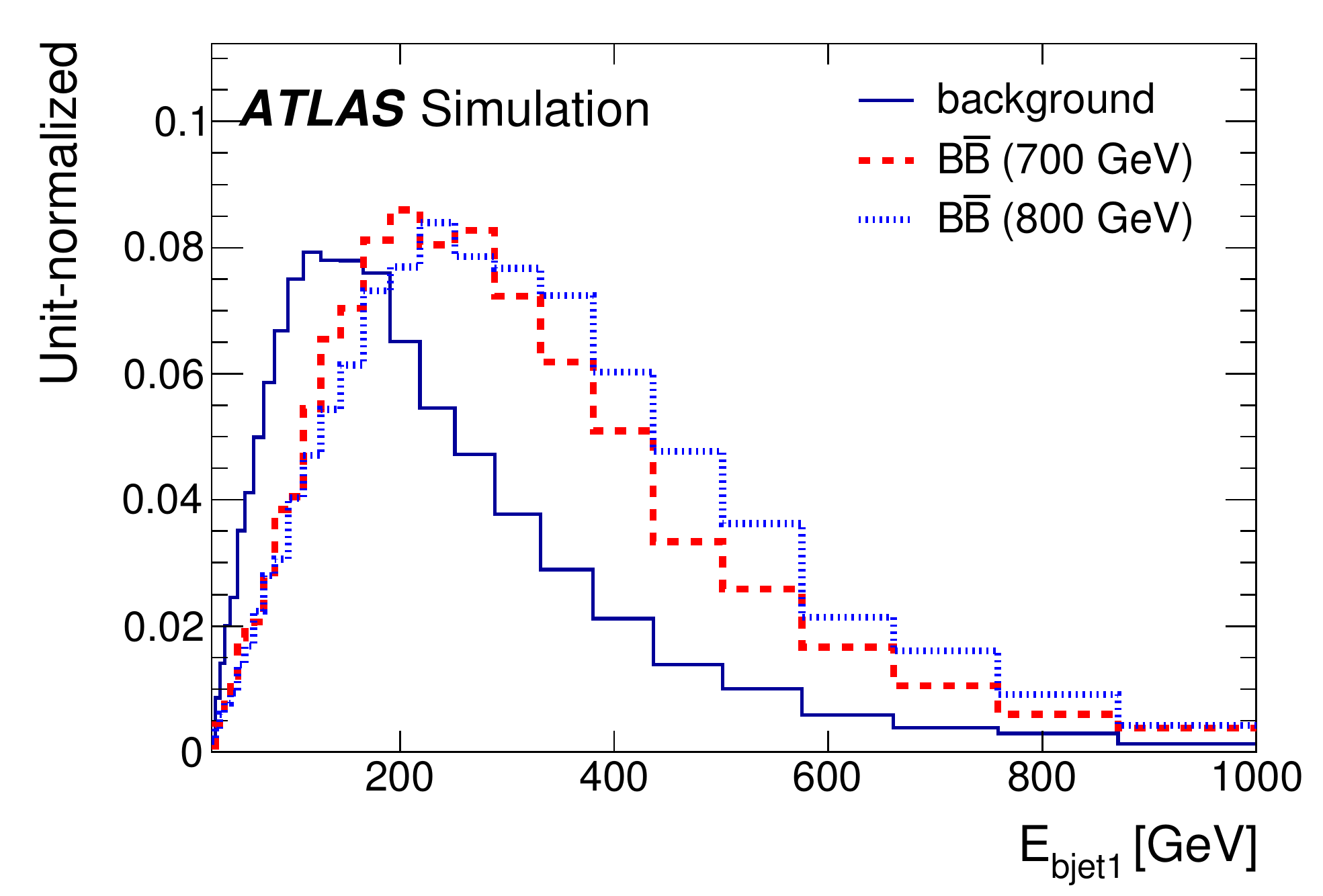}
\includegraphics[width=0.32\textwidth]{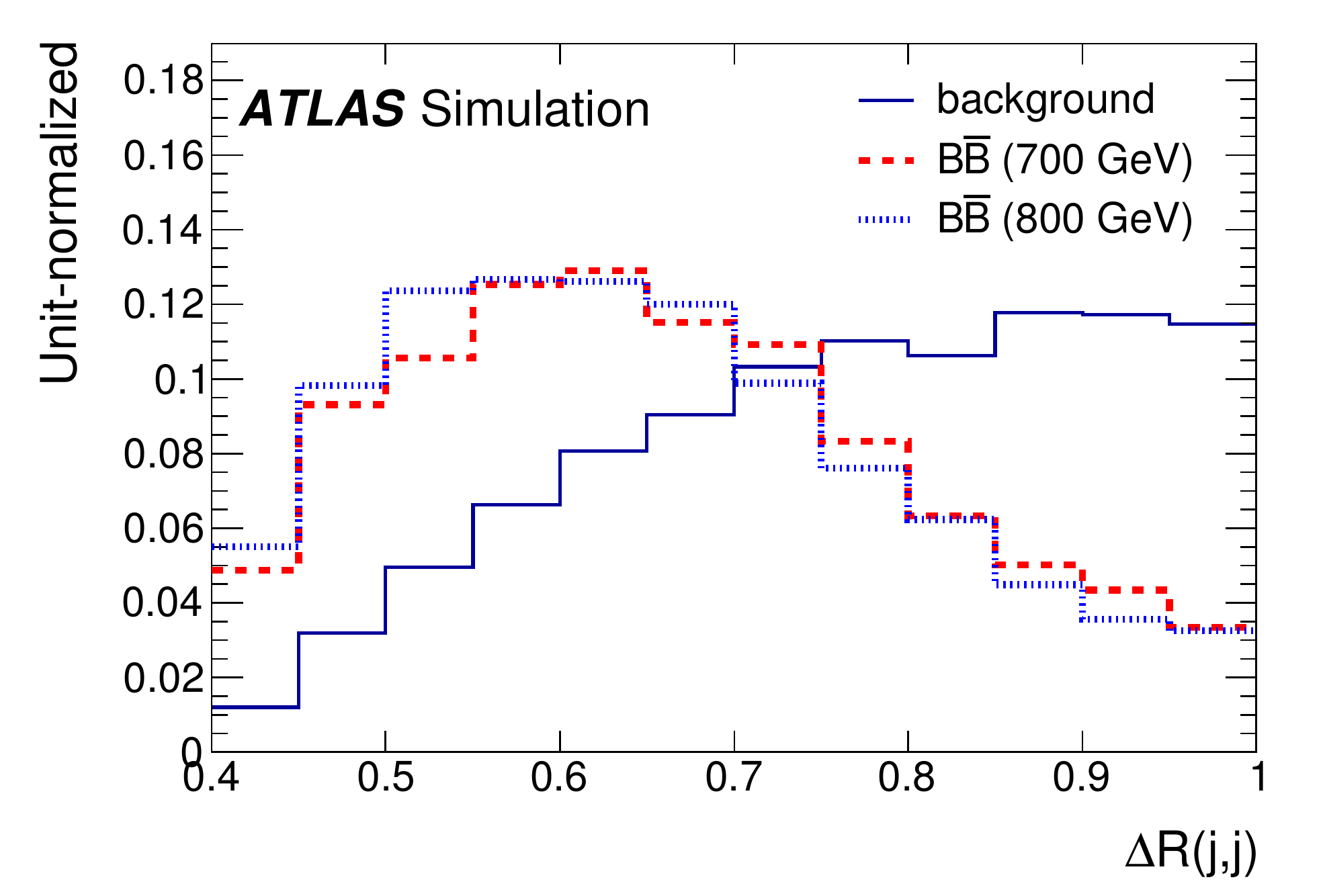}
\includegraphics[width=0.32\textwidth]{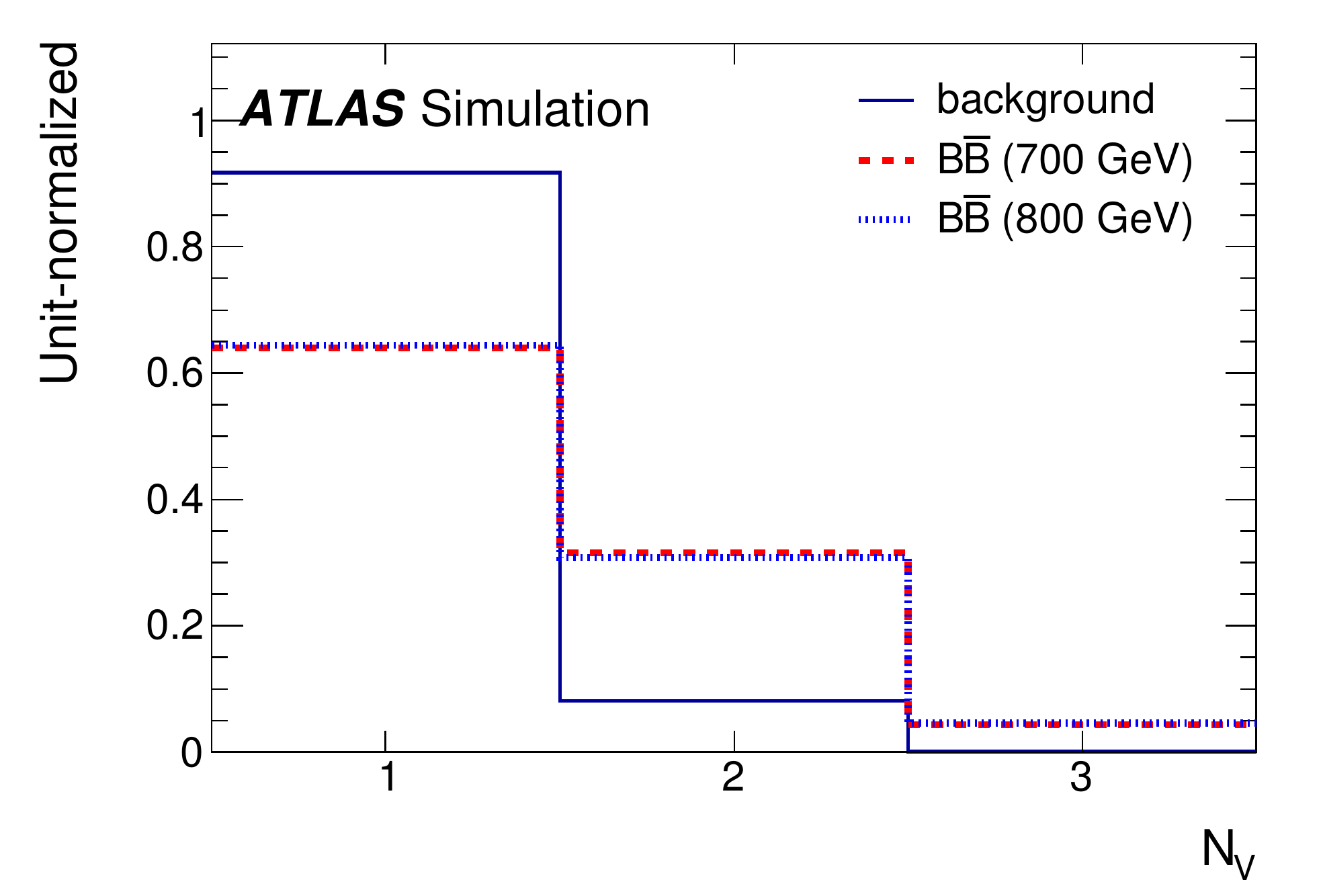}
\includegraphics[width=0.32\textwidth]{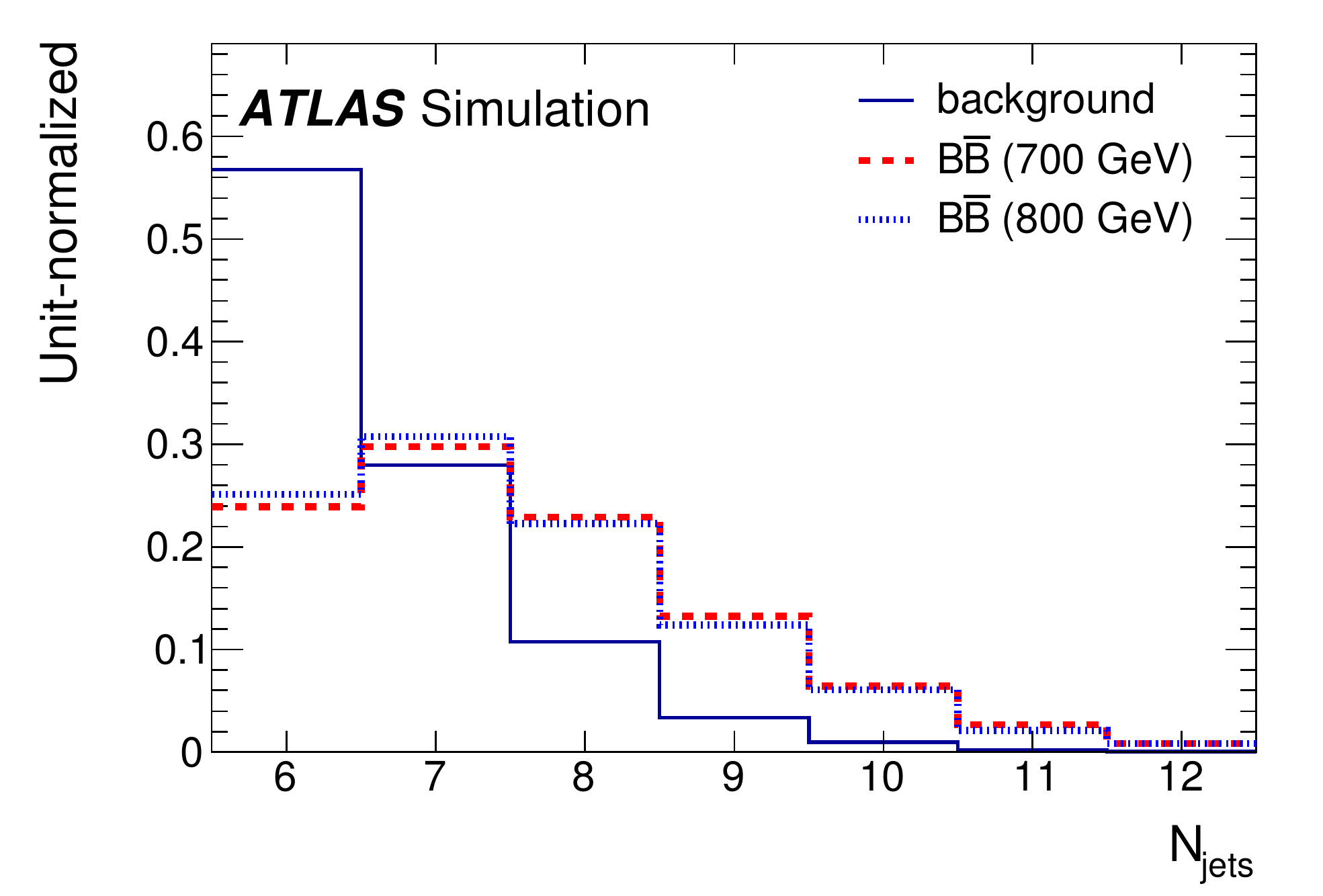}
\includegraphics[width=0.32\textwidth]{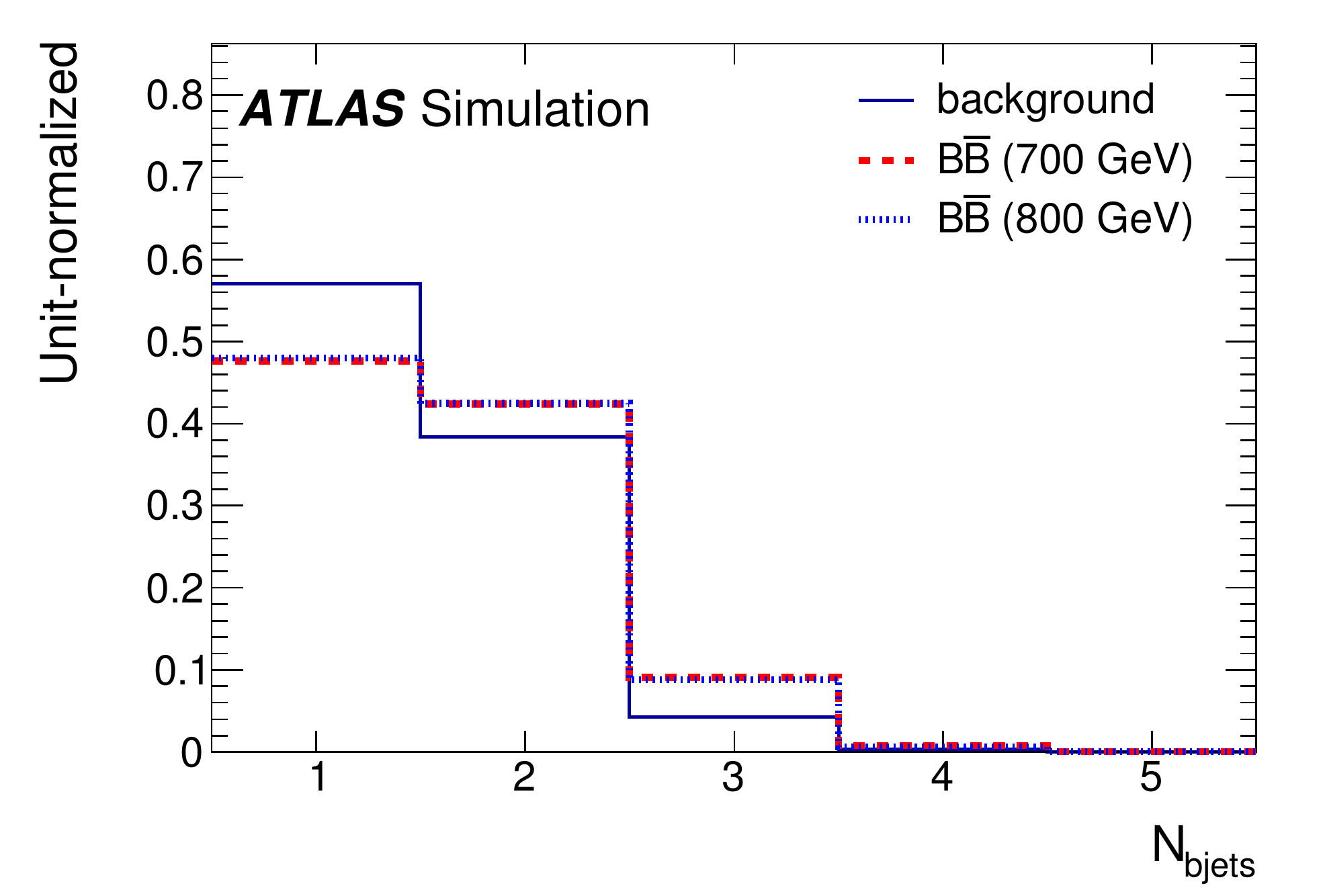}
\includegraphics[width=0.32\textwidth]{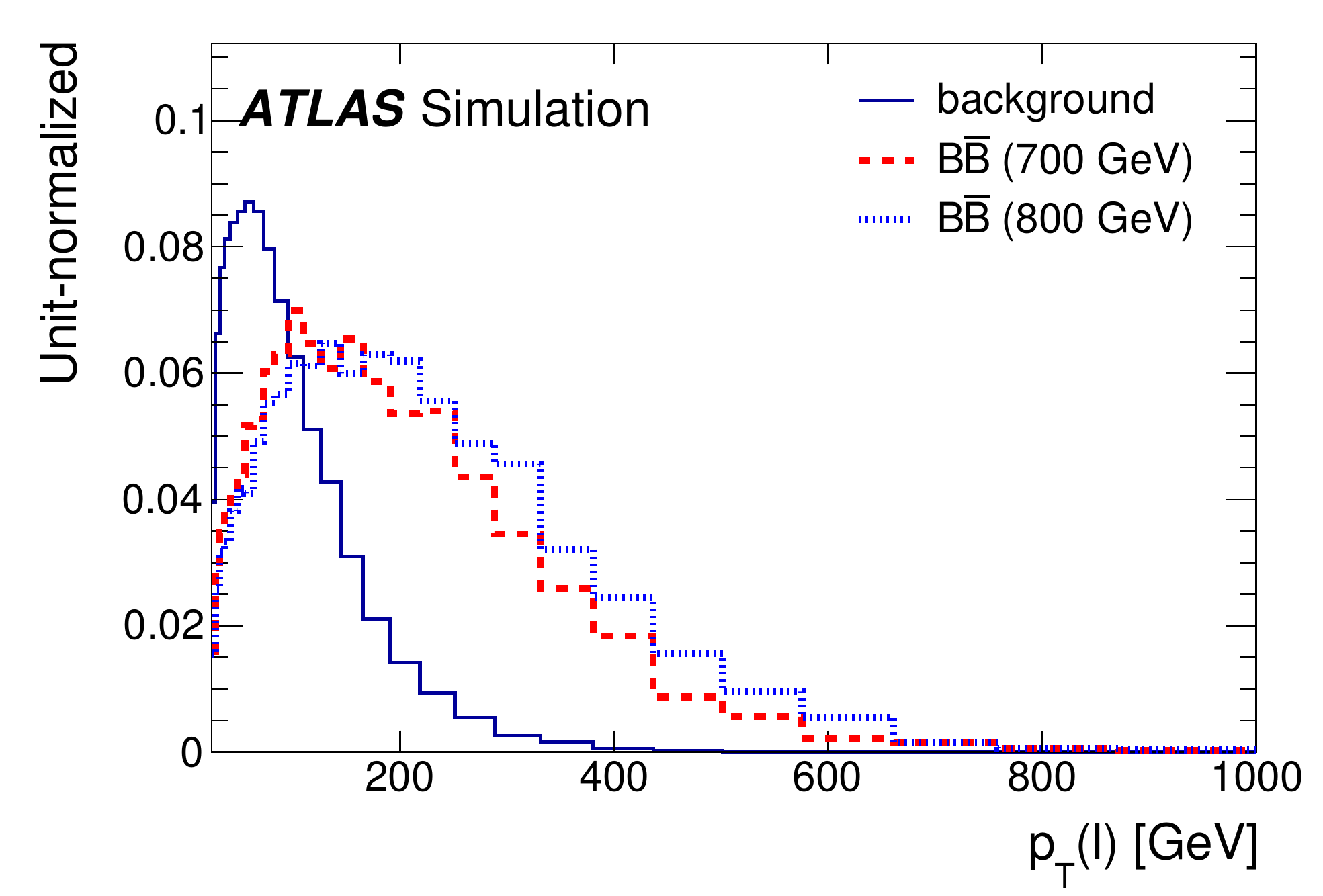}
\includegraphics[width=0.32\textwidth]{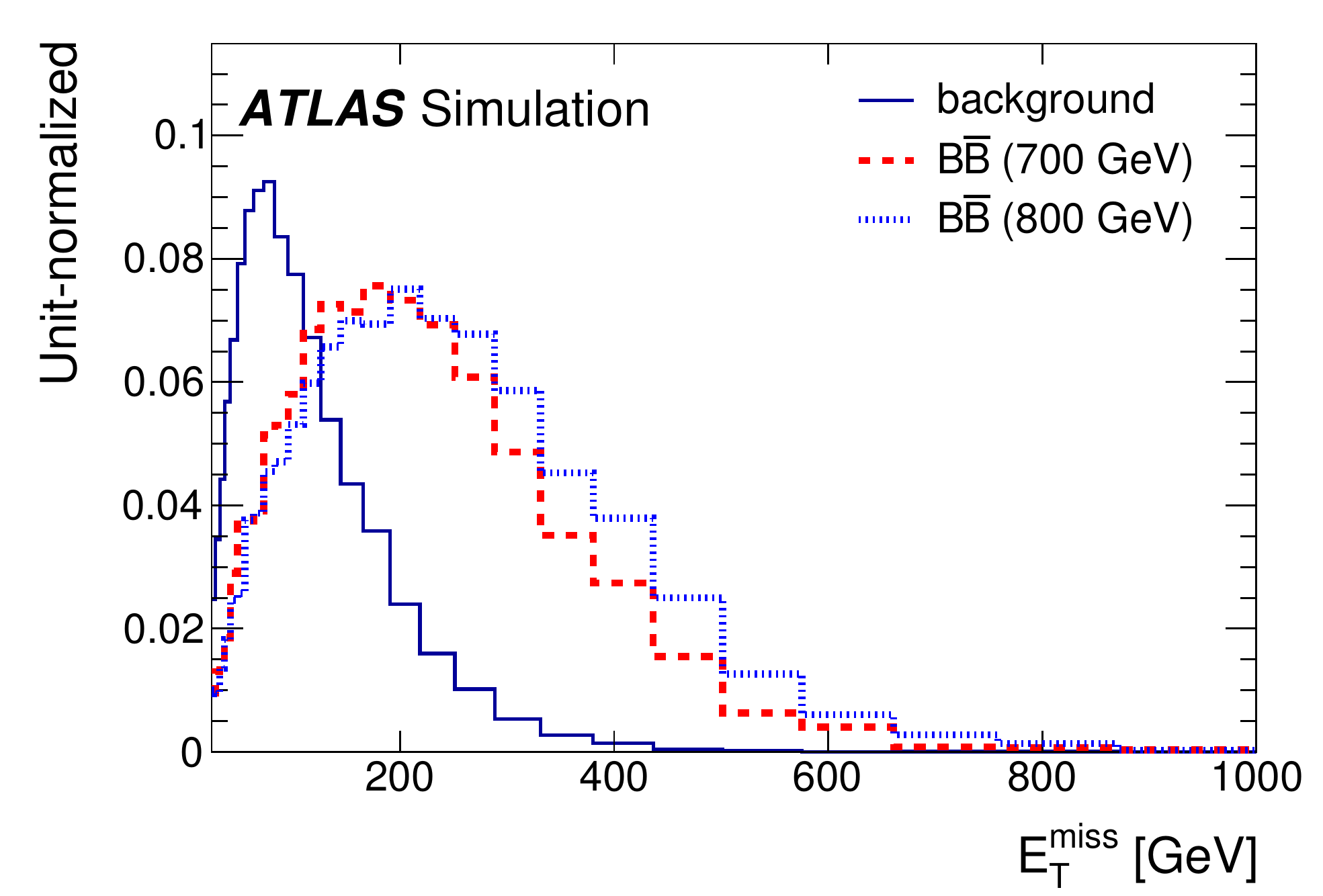}
\caption{Unit-normalized distributions of the twelve variables used in the BDT discriminant. 
The background contribution ($t\bar t$ combined
with $W$+jets) is shown with a dark blue solid line, while signal distributions are shown for 700 and 
800~GeV VLQ $B$ with BR($B \rightarrow Wt) = 100\%$ with red dashed and light blue dotted lines, 
 respectively. The following selection requirements, which define the signal region for the BDT analysis,
are applied: $\Njets \ge 6$, $\Nbjets \ge 1$, $\NV \ge 1$, and $\HT>500$~GeV.}
\label{fig:tmva_vars}
\end{figure*}  

\subsection{Cuts-based analysis}
\label{sec:cuts}

As a cross-check of the BDT analysis results, a cuts-based approach is used.
As shown in Table~\ref{tab:srandcr}, the signal region for the cuts-based analysis is almost
identical to that of the BDT analysis: in addition to the preselection procedure described in Sec.~\ref{sec:selection},
the requirements that $\Njets \ge 6$, $\NV \ge 1$, and $\Nbjets \ge 1$ are applied. As mentioned earlier, given the lower
sensitivity of the cuts-based approach, a more restrictive requirement is made
on \HT, namely $\HT > 800$~GeV, for the cuts-based signal region definition.
Given the sizable background, particularly from $t\bar t$ events, that passes
these requirements, simply comparing the total number of events in the signal region with the number
expected from background processes would not be very sensitive.  Instead, the cuts-based
analysis exploits the fact that signal events tend to have higher values of \HT \
and \NV \ than do background events, as shown in Fig.~\ref{fig:tmva_vars}.  The
cuts-based SR data are divided into two exclusive
subcategories, one with $\NV = 1$ and the second with $\NV \ge 2$. The \HT \
spectra in the two subcategories are then used to search for a signal excess
over the predicted background.

\section{Background determination}
\label{sec:bg}

In this section, the background models used
for both the BDT and cuts-based analyses are described.

\subsection{Multijet background}
\label{sec:mjetbg}

The normalization and shape of the multijet background contribution, with a jet being misidentified as a lepton,
are determined directly from data using the so-called
matrix method~\cite{matrix_method}. 
This method makes use of samples of
events that are kinematically similar to the signal but enriched in multijet events:
these samples are obtained by relaxing
lepton identification criteria such as isolation requirements. The yields and kinematic distributions
of the multijet background contribution in the signal region or a given control region can then be derived
by applying the efficiency and false-identification
rate of the relaxed selection. The efficiency is estimated from MC
samples of prompt lepton sources, and validated against data. The false-identification rate is
estimated using data in a multijet-enriched control sample selected by requiring low \MET \ and \MT \ values.

Multijet events contribute only a small component of the total background in this analysis.  
Different methods were compared for obtaining the multijet background
rate.  No significant differences were found, and a conservative
$\pm 50\%$ uncertainty on the normalization is used~\cite{Aad:2014fwa}.

\subsection{$W$+jets background}

The shape of the expected $W$+jets background contribution is obtained using MC samples.
The overall yield of $W$+jets events is verified by exploiting the lepton charge
asymmetry measured in data~\cite{ATLAS:2012an,Aad:2012hg}.
The method uses the fact that the production of
$W$ bosons at the LHC is charge asymmetric, and the 
theoretical prediction of the ratio of the numbers of events with different lepton 
charges has an uncertainty of only a few percent.
Charge-symmetric contributions from $t\bar t$, $Z$+jets and multijet processes cancel in the ratio.
Slightly charge-asymmetric contributions from the remaining backgrounds such as single top are estimated using MC
simulation. The procedure is performed without any $b$-tagging requirement and for different lepton flavors
and jet multiplicities. The resulting yield is consistent with the MC calculation within systematic uncertainties,
and therefore the unscaled MC predictions are used for the total $W$+jets yield.

To correct for mismodeling of the $W$ boson kinematics in the
simulation, the $W$ boson \pT\ distribution is reweighted in accord with
the difference between expected and observed $Z$ \pT\
spectra~\cite{Aad:2013ysa}.

The modeling of the $W$+jets background is validated using data in
control regions WCR1 and WCR2, in which both the signal and $t\bar t$
contributions are suppressed by the $\Nbjets = 0$ requirement.

\subsection{$t\bar t$ background}
\label{sec:ttbkg}

The $t\bar t$ background shape is studied using
{\sc Powheg-box} samples
and cross-checked using samples generated with
{\sc Alpgen} (version 2.14)
with up to five additional partons.
The simulated \ttbar \ events are reweighted in order to correct for
the observation that there are
more events with high $t$ or $t\bar t$ \pt \ in MC
simulation than in data~\cite{ATLAS-CONF-2014-011}.
The uncertainty in the reweighting is included in the overall
systematic uncertainty, with more details available in Ref.~\cite{ATLAS-CONF-2014-011}.
The $t\bar t$ background normalization in the plots and tables is
taken from the NNLO+NLL prediction as mentioned in 
Sec.~\ref{sec:mc:simulation}.
For the final results, however, the background normalization is
determined from data (see Sec.~\ref{sec:stats}).

The modeling of the $t\bar t$ background is validated using data in
control regions TCR1, TCR2 and TCR3.
In each control region, the $t\bar t$ reweighting in terms of the $t$ and $t\bar t$ \pt \
spectra improves the agreement in other kinematic variables, especially \HT.

\subsection{Other backgrounds}

Other background sources, including electroweak single top production,
$Z$+jets events, $t\bar t$ production in association with a $W$ or $Z$ boson
(denoted $t\bar t+V$), and diboson ($WW, WZ, ZZ$) production, are modeled using MC simulation.
These backgrounds are small,
with their sum contributing less than 10\% of the total background in the
signal region as well as in most of the control regions.

\subsection{Validation of background modeling}

The various background-dominated control regions (WCR1, WCR2, TCR1, TCR2, and
TCR3, as defined in Table~\ref{tab:srandcr}) are used to validate the
background prediction.

The data distributions for each of the twelve variables used in the BDT analysis are well described
by the background expectation in each of the five control regions, demonstrating that the
variables chosen are well modeled in the MC simulation. For example, Fig.~\ref{fig:bdtvars_tcr3}
shows  the distributions for nine of the twelve BDT input variables for the data in
control region TCR3, which is the most signal-like control region. The other three BDT input
variables are all identically zero for this control region, given the requirement in the definition
of TCR3 that $\NV = 0$. These three variables are instead shown separately in Fig.~\ref{fig:otherbdtvars}
for control region TCR2.

In both Figs.~\ref{fig:bdtvars_tcr3} and \ref{fig:otherbdtvars},
the predicted distributions for the background expectation are shown superimposed on the
data.  The panel beneath each distribution shows the bin-by-bin
ratio of the data to the background expectation. Within the
total uncertainties on the background prediction (shown as the shaded bands), the data
are in good agreement with the total expected background.

\begin{figure*}[hbpt]
\centering
\includegraphics[width=0.32\textwidth]{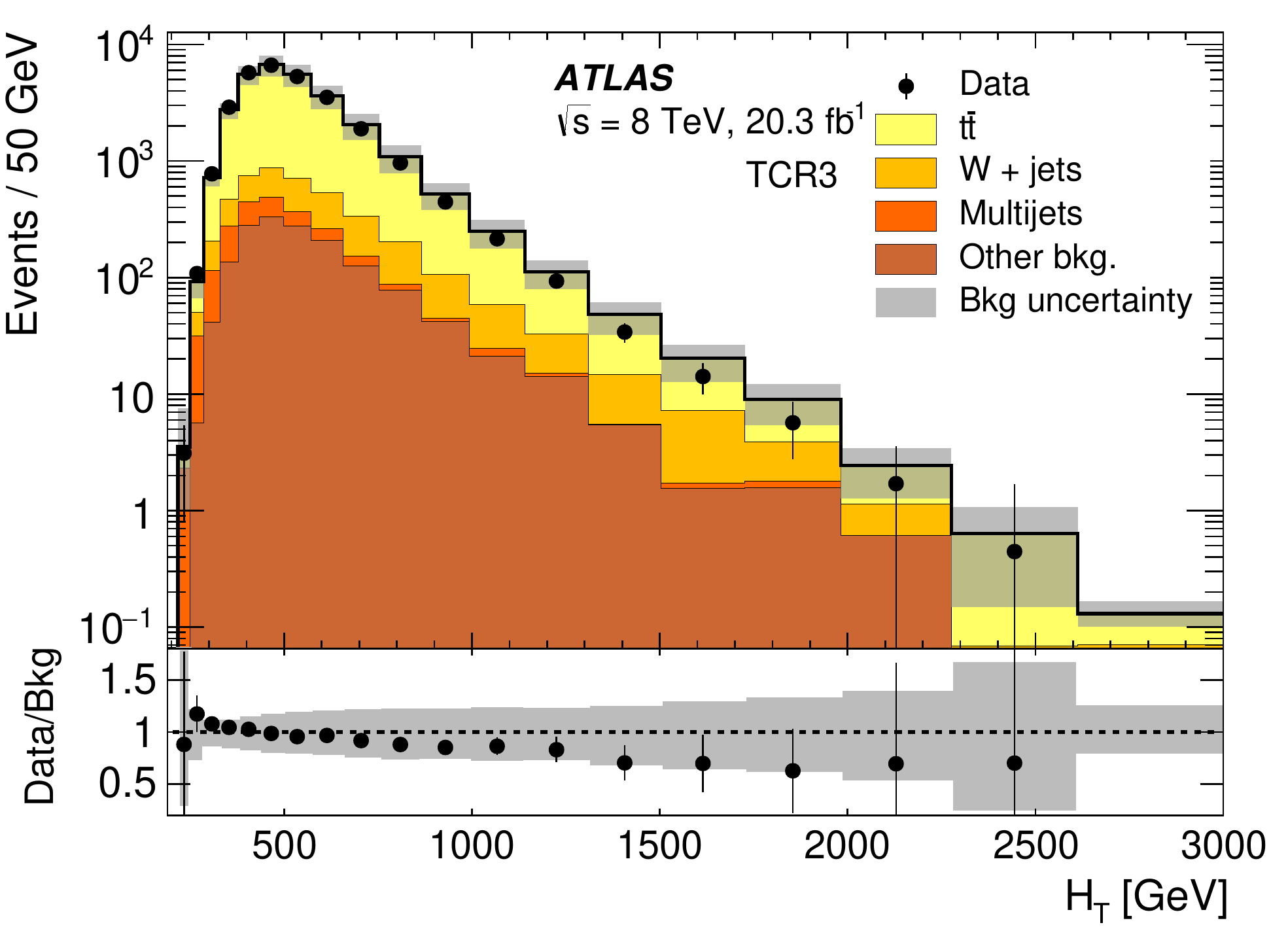}
\includegraphics[width=0.32\textwidth]{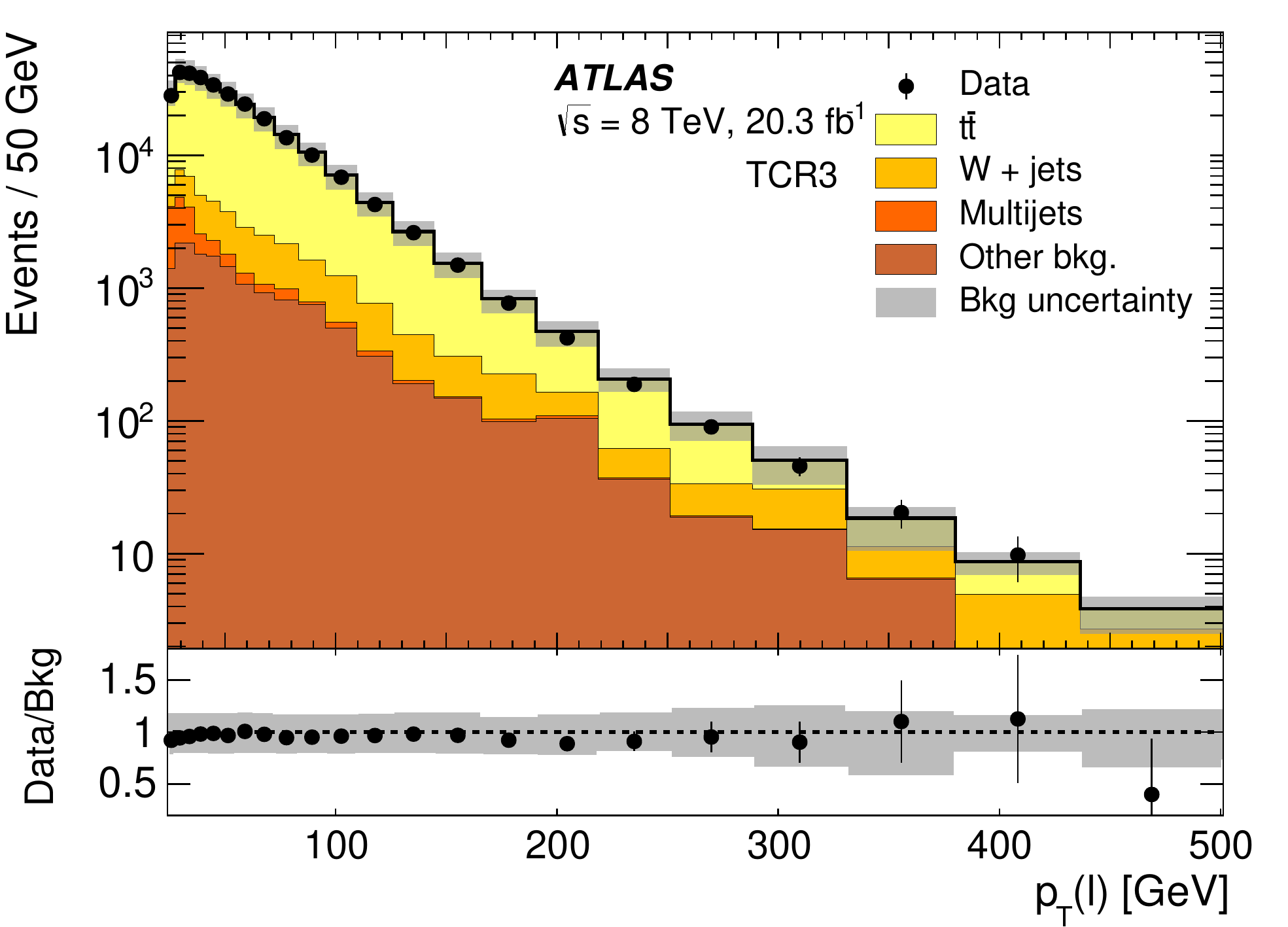}
\includegraphics[width=0.32\textwidth]{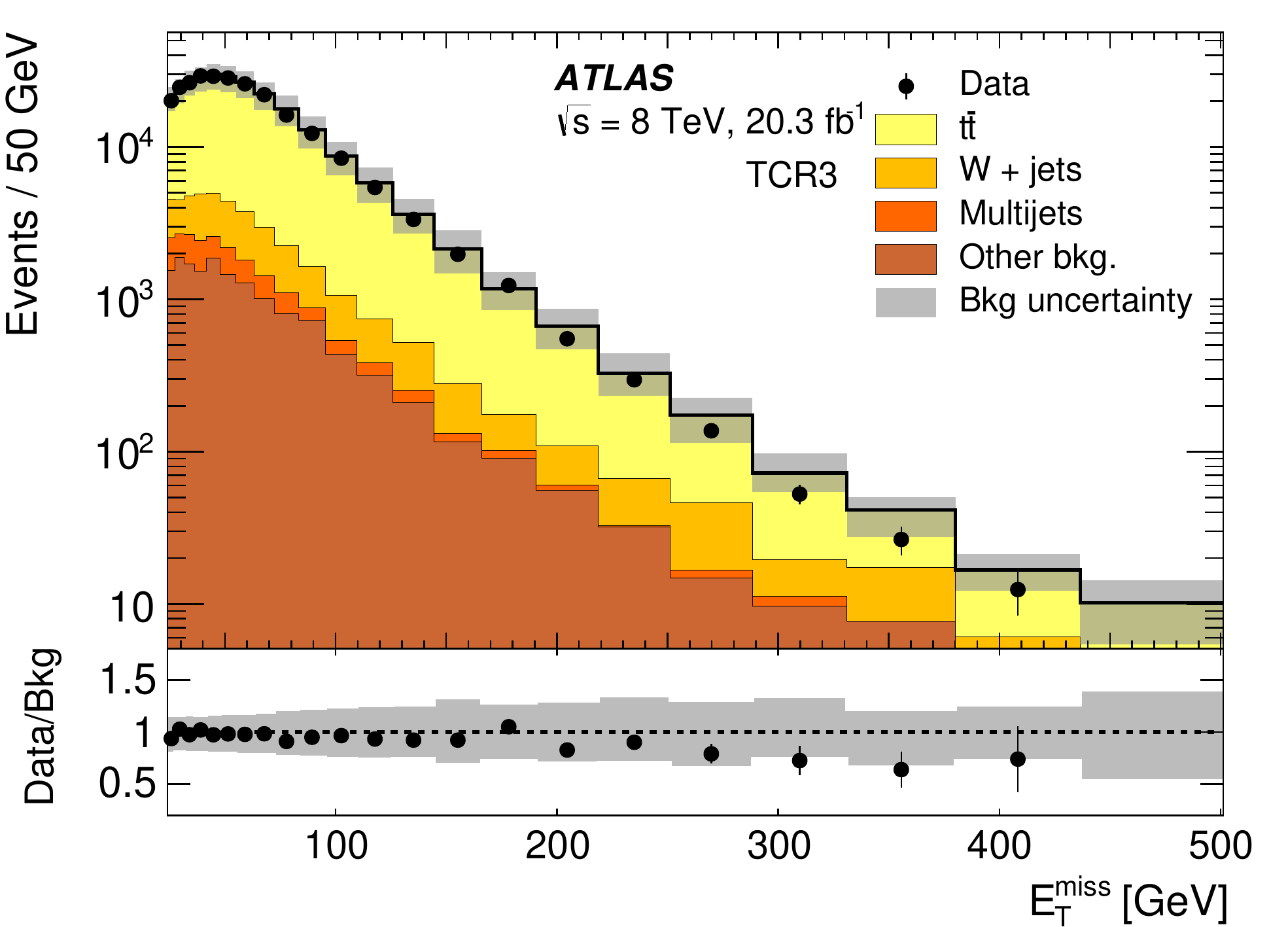}
\includegraphics[width=0.32\textwidth]{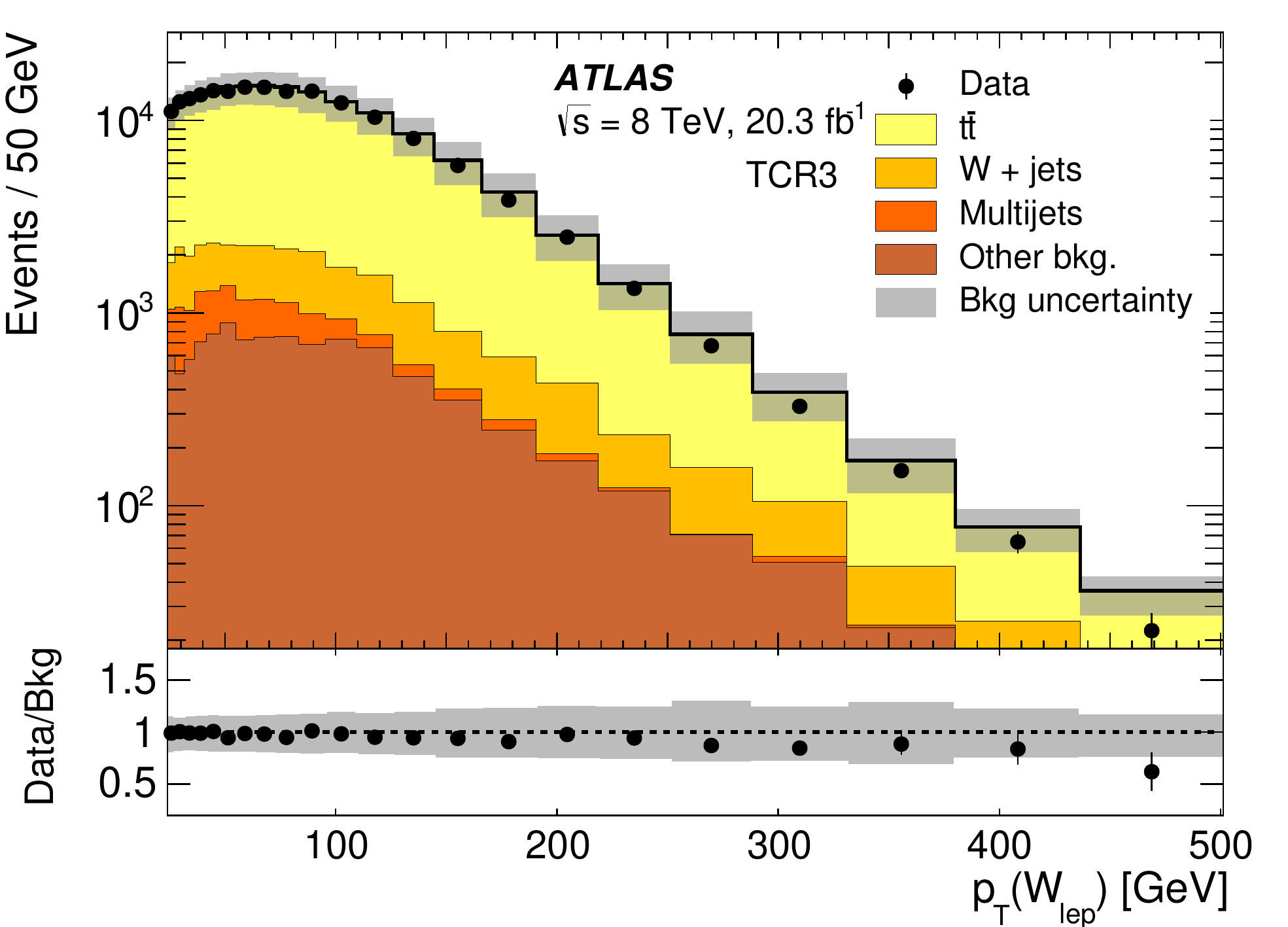}
\includegraphics[width=0.32\textwidth]{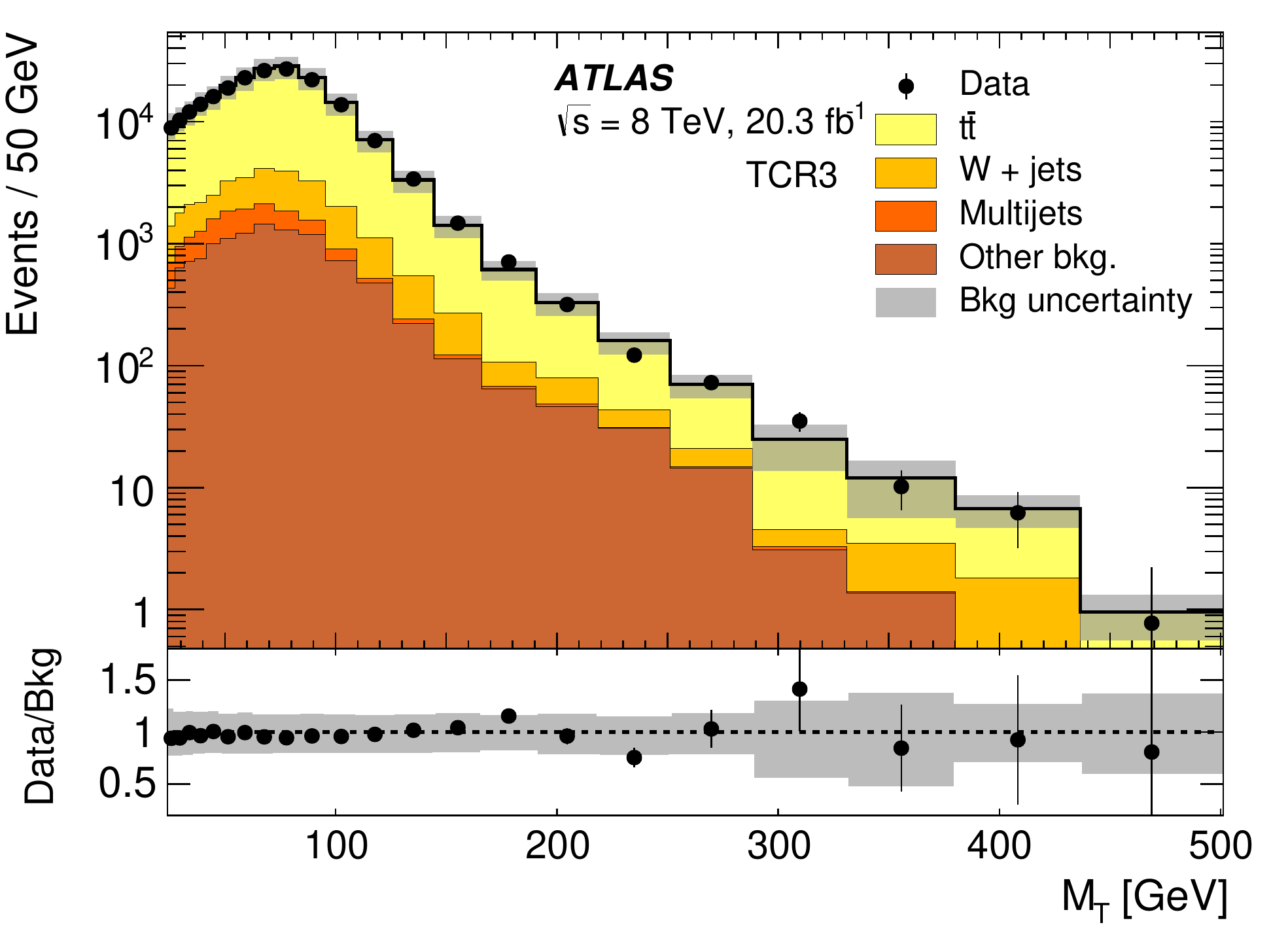}
\includegraphics[width=0.32\textwidth]{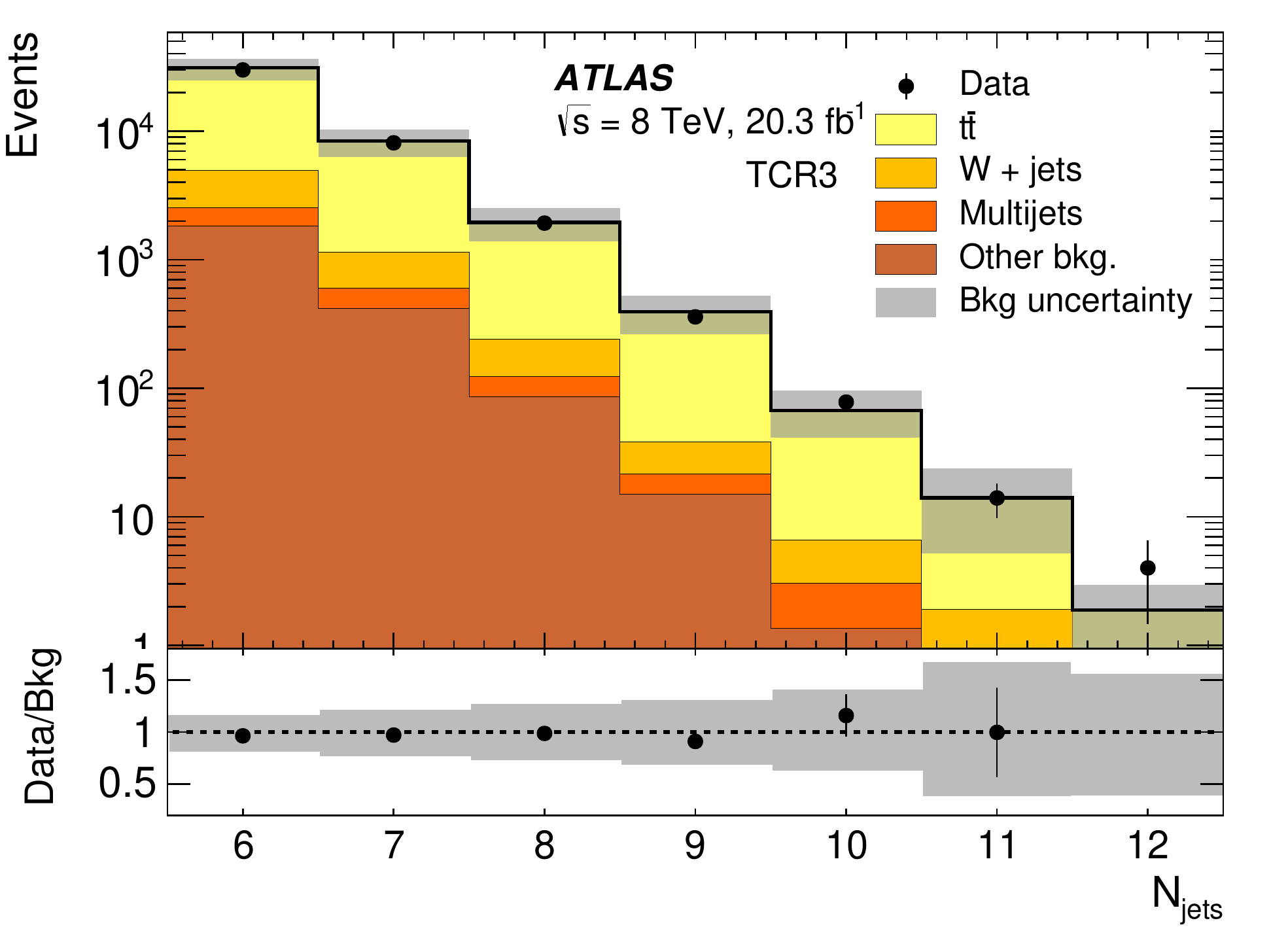}
\includegraphics[width=0.32\textwidth]{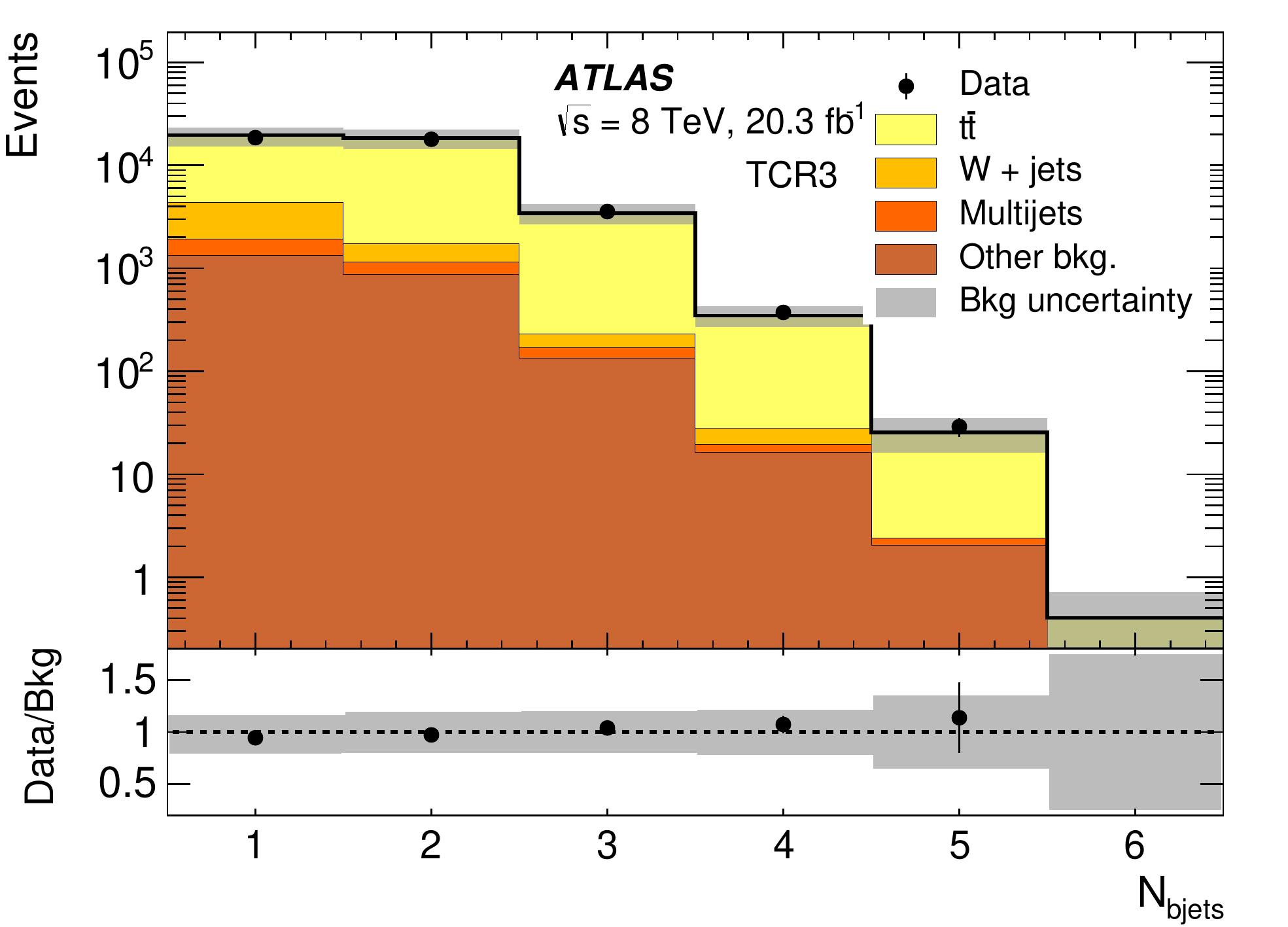}
\includegraphics[width=0.32\textwidth]{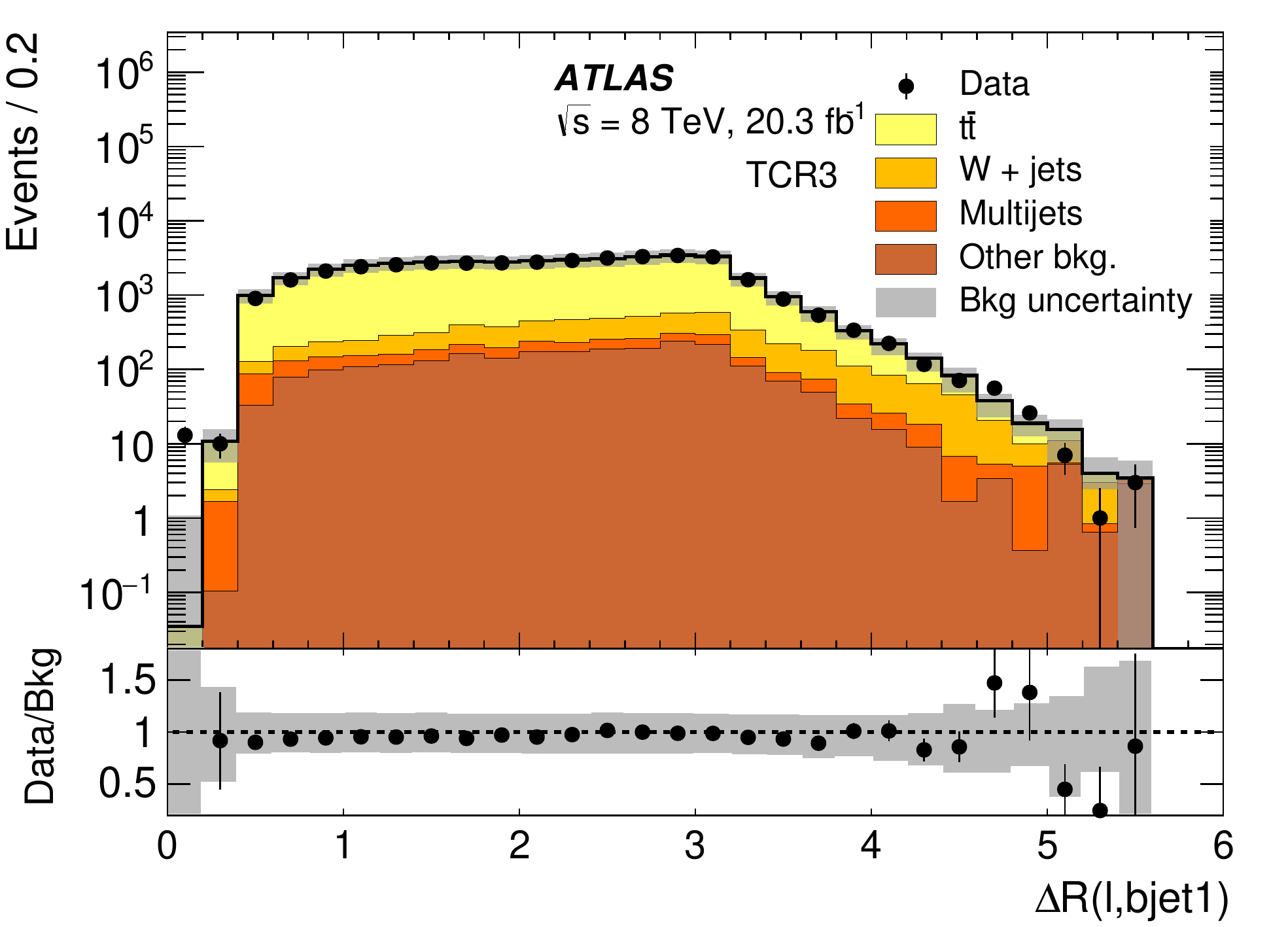}
\includegraphics[width=0.32\textwidth]{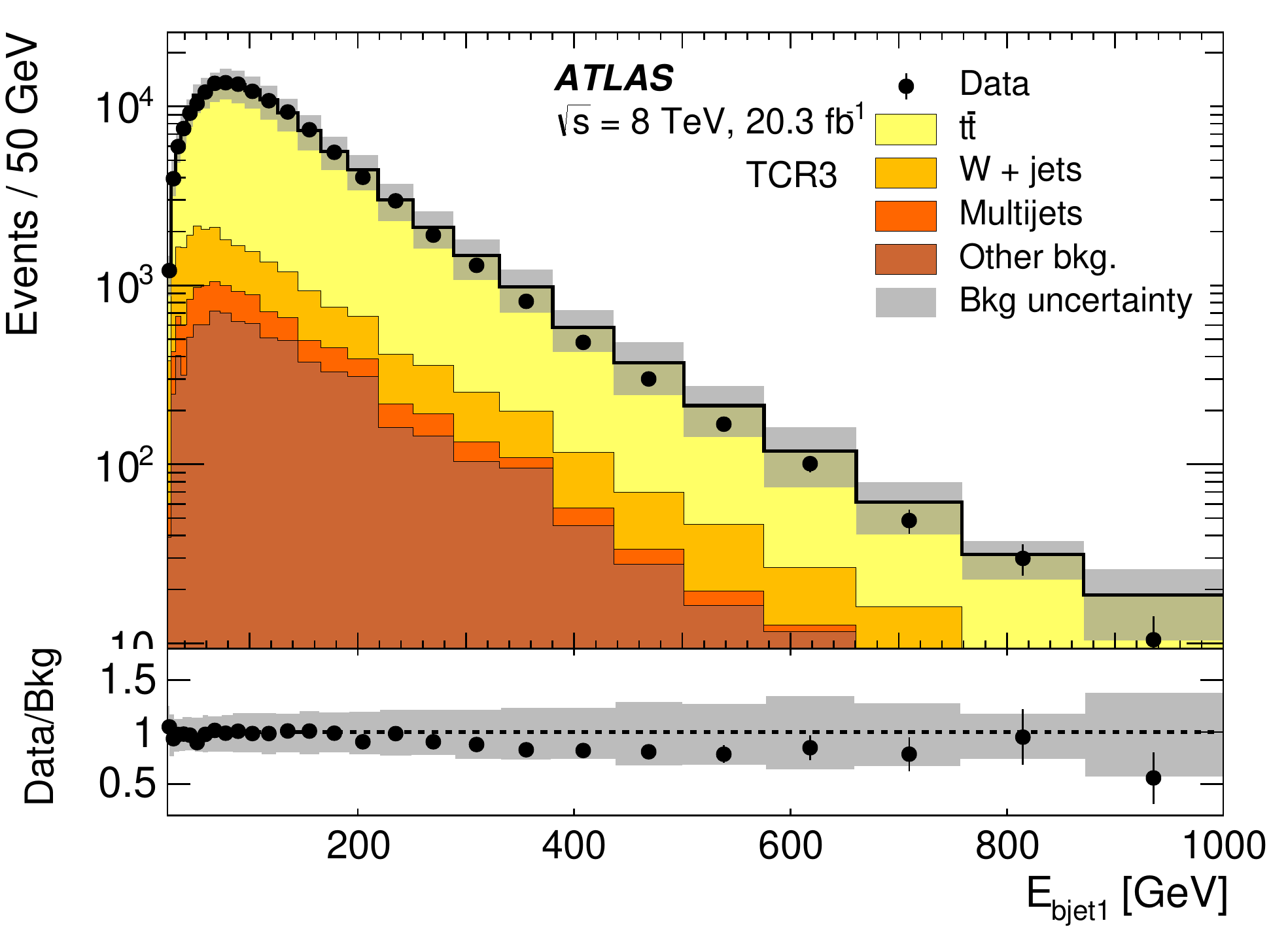}
\caption{Distributions of nine of the twelve BDT input variables for data and for the background
expectation, for control region TCR3.
The lower panels show the
bin-by-bin ratio of the data to the total background expectation. The data are shown
with statistical error bars. The shaded bands
show the total uncertainties on the background expectation, 
including both the statistical and systematic uncertainties.}
\label{fig:bdtvars_tcr3}
\end{figure*}  

\begin{figure*}[hbpt]
\centering
\includegraphics[width=0.32\textwidth]{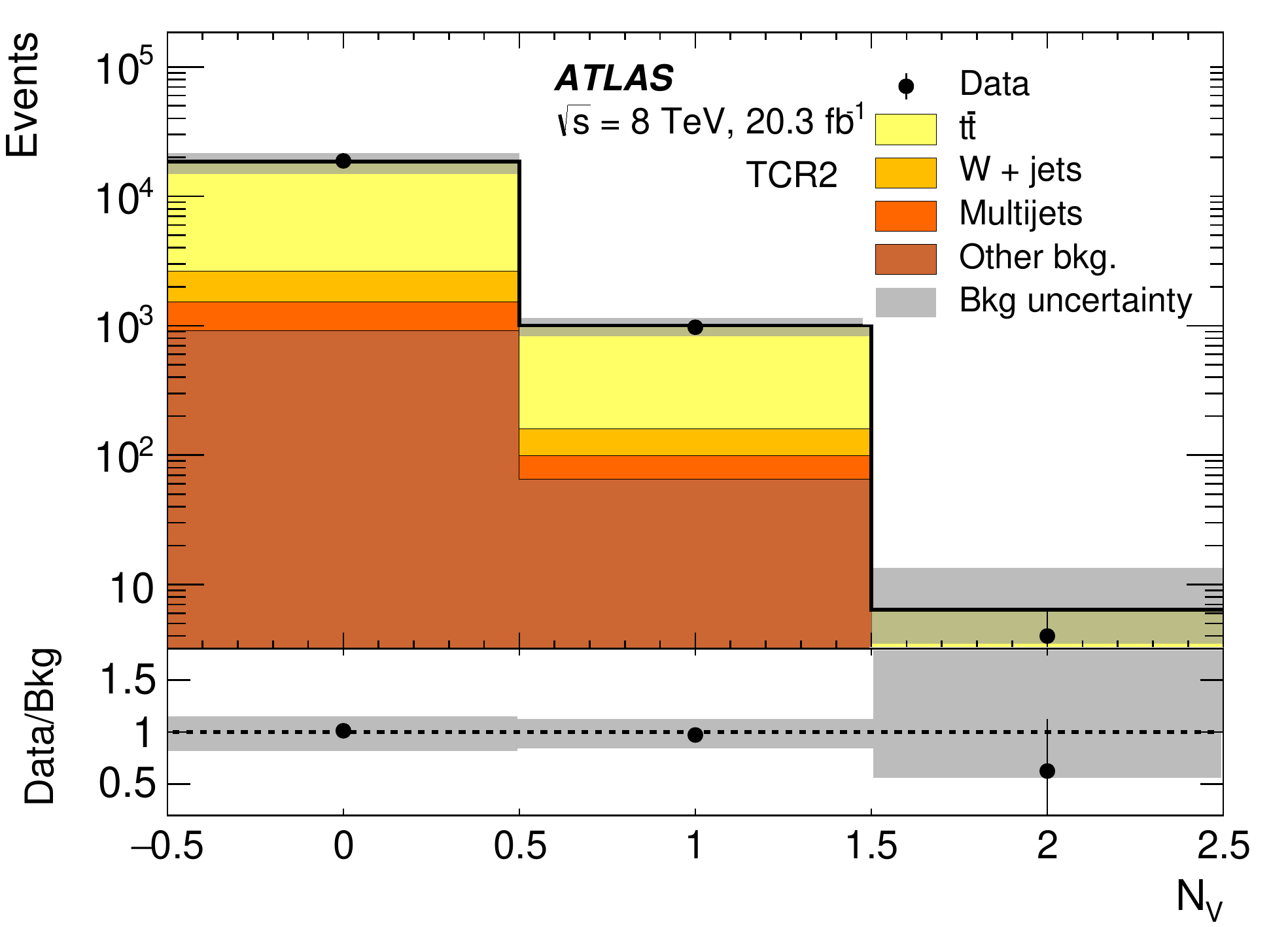}
\includegraphics[width=0.32\textwidth]{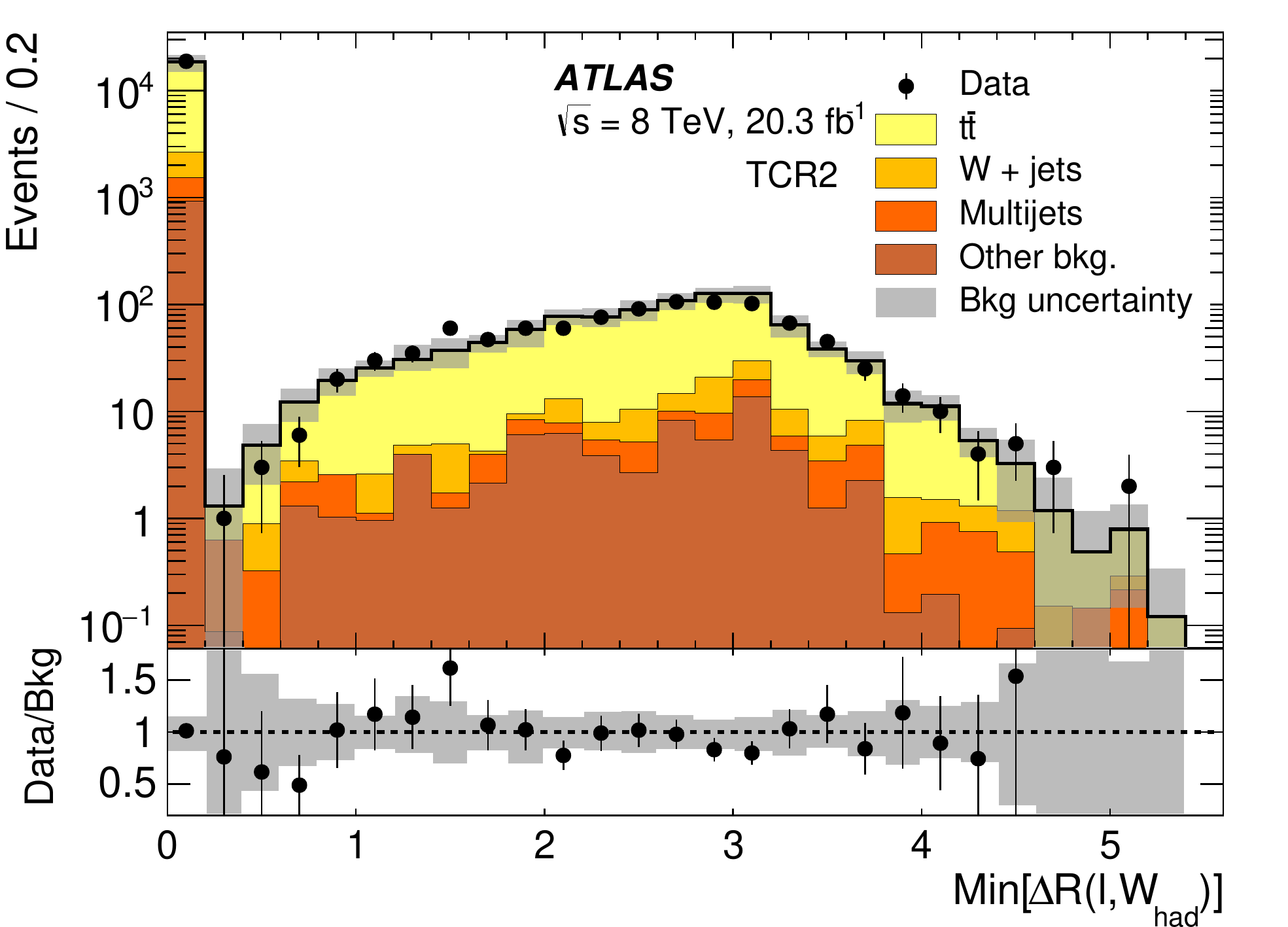}
\includegraphics[width=0.32\textwidth]{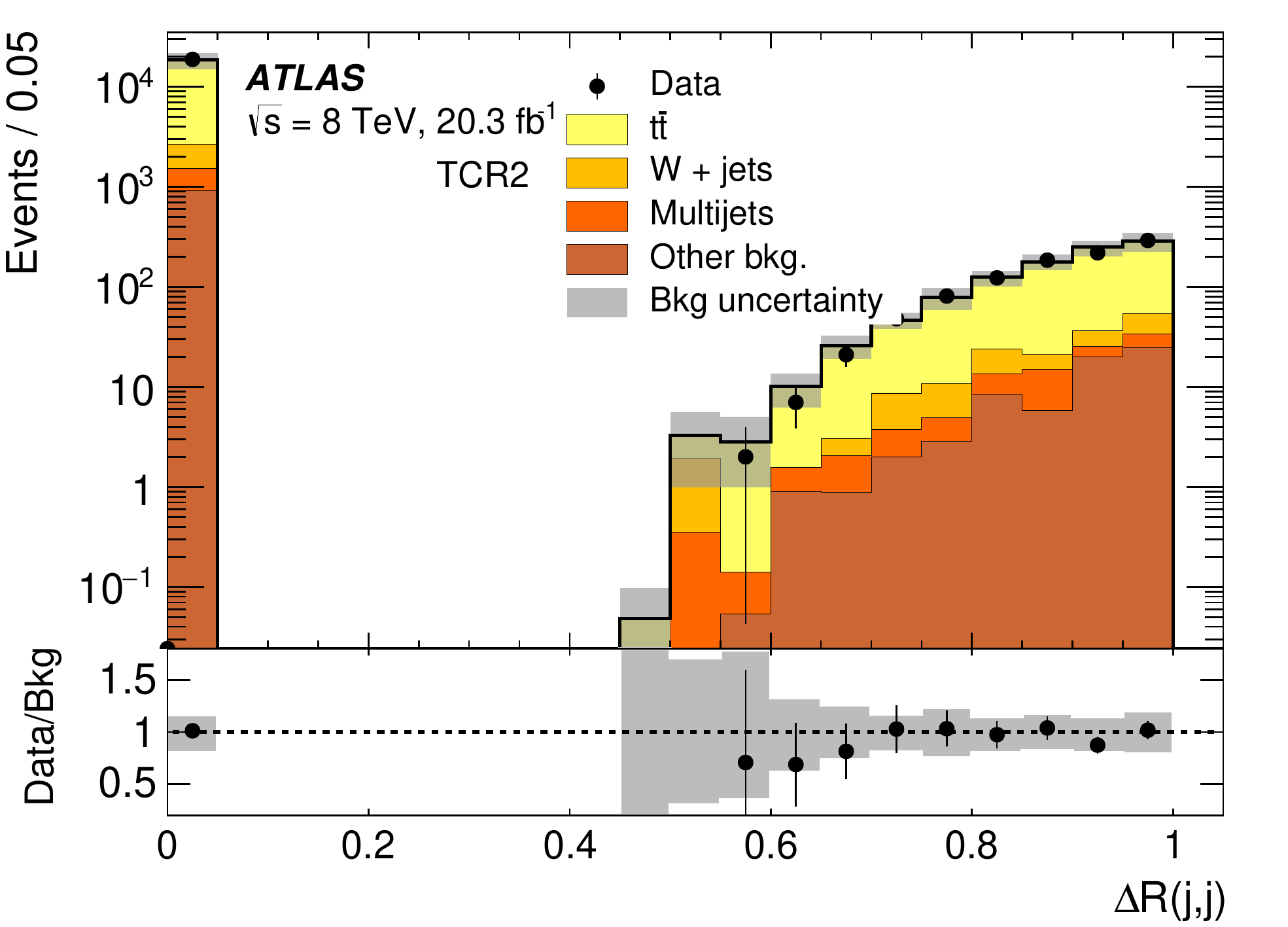}

\caption{Distributions of three of the twelve BDT input variables for data and for the background
expectation, for control region TCR2. For the two rightmost plots, events with zero reconstructed
hadronically decaying $W/Z$ boson candidates are included in the first bin, at zero. The lower panels show the
bin-by-bin ratio of the data to the total background expectation. The data are shown
with statistical error bars. The shaded bands
show the total uncertainties on the background expectation, 
including both the statistical and systematic uncertainties.}
\label{fig:otherbdtvars}
\end{figure*}  

Figure~\ref{fig:bdt_cr_plots} shows the distributions of the BDT discriminant observed in data 
in each of the five background-dominated control regions. Superimposed are the distributions expected for
background. Good agreement is observed between the data distribution and the background expectation in all
five control regions, supporting the validity of the background modeling. Additional
checks reveal that
the pairwise correlations between the BDT input variables, and also the BDT output, all show
reasonable agreement between data and the background predictions.

\begin{figure*}[hbpt]
\centering
\includegraphics[width=0.48\textwidth]{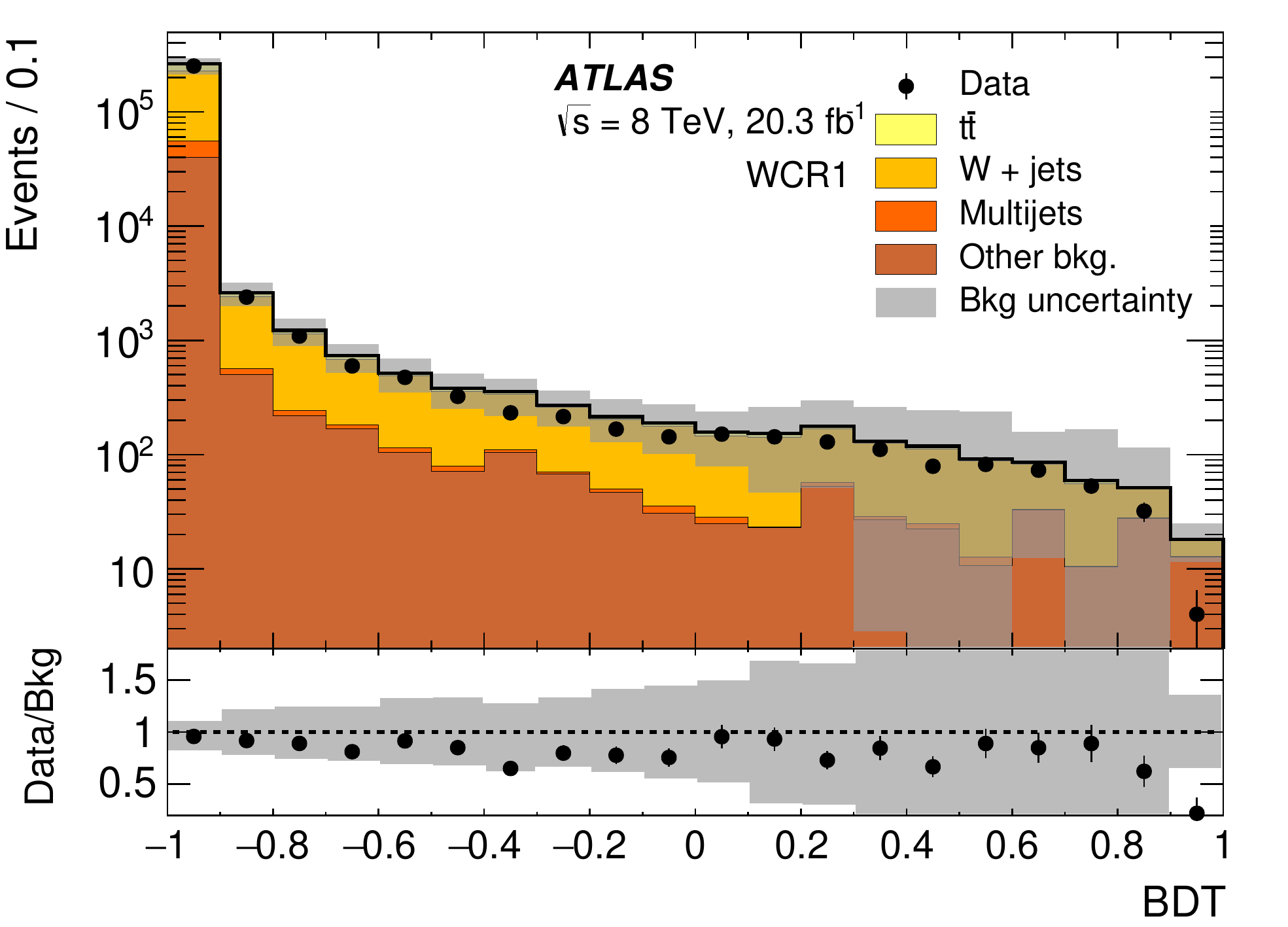}
\includegraphics[width=0.48\textwidth]{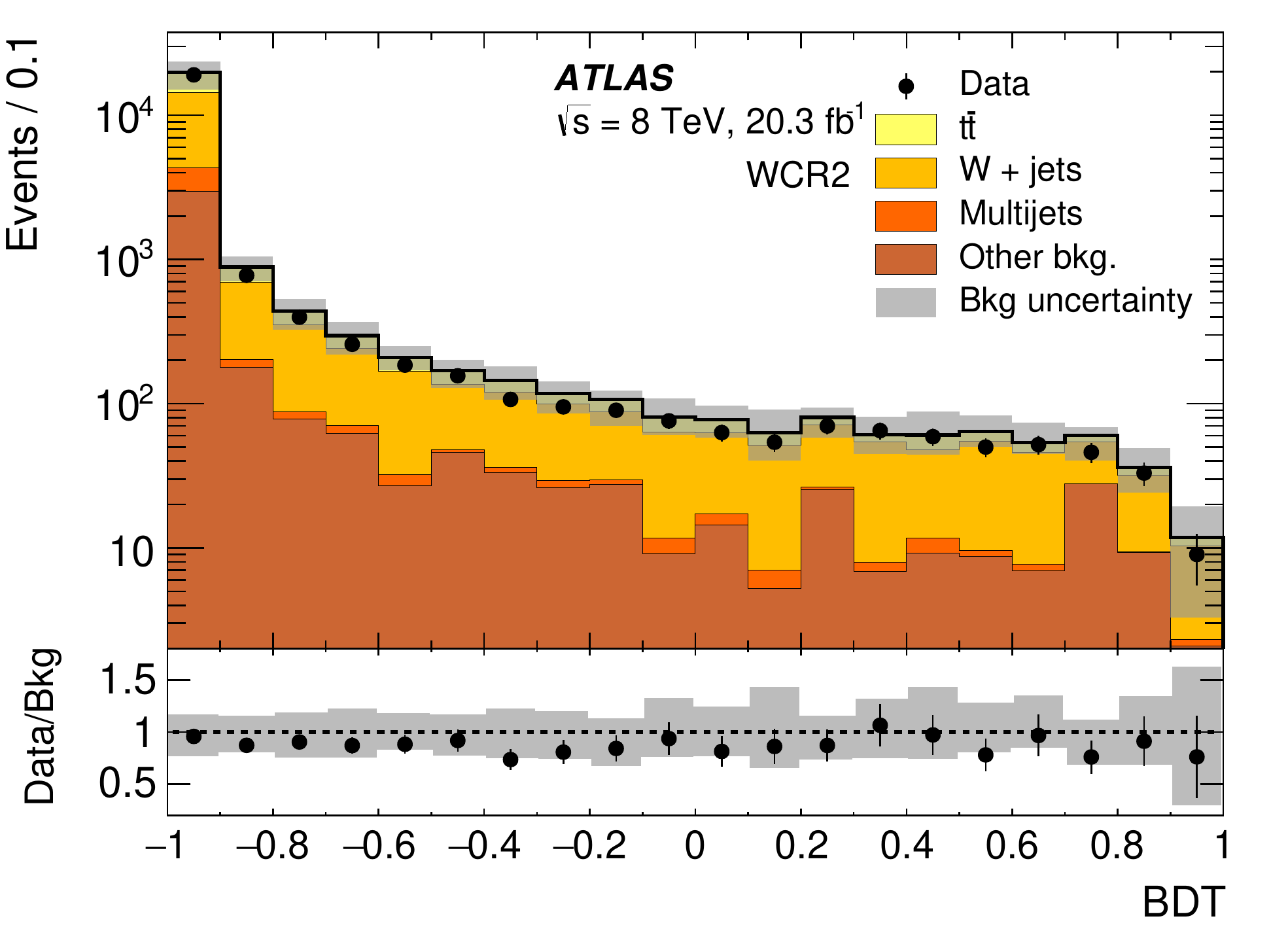}
\includegraphics[width=0.48\textwidth]{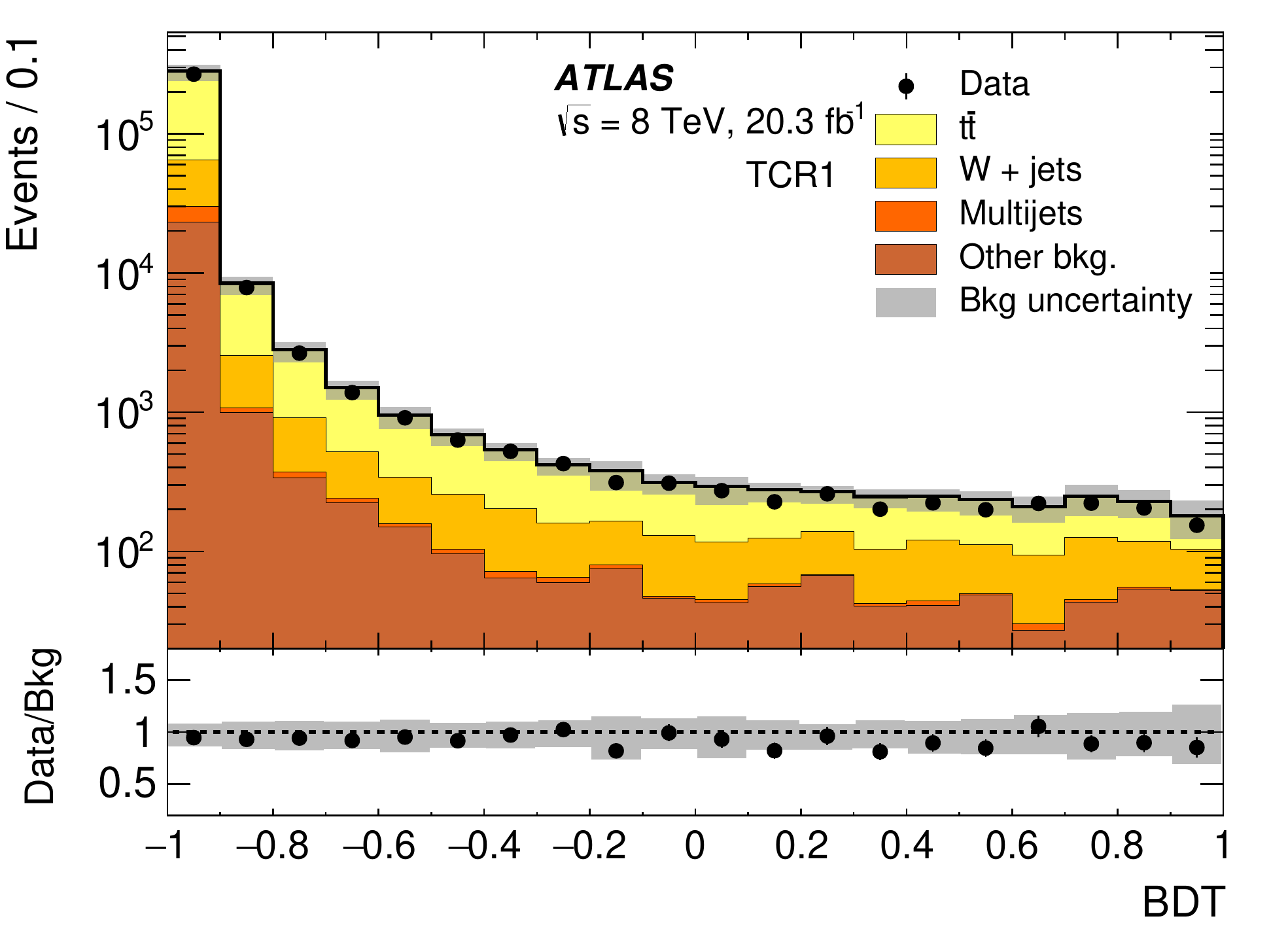}
\includegraphics[width=0.48\textwidth]{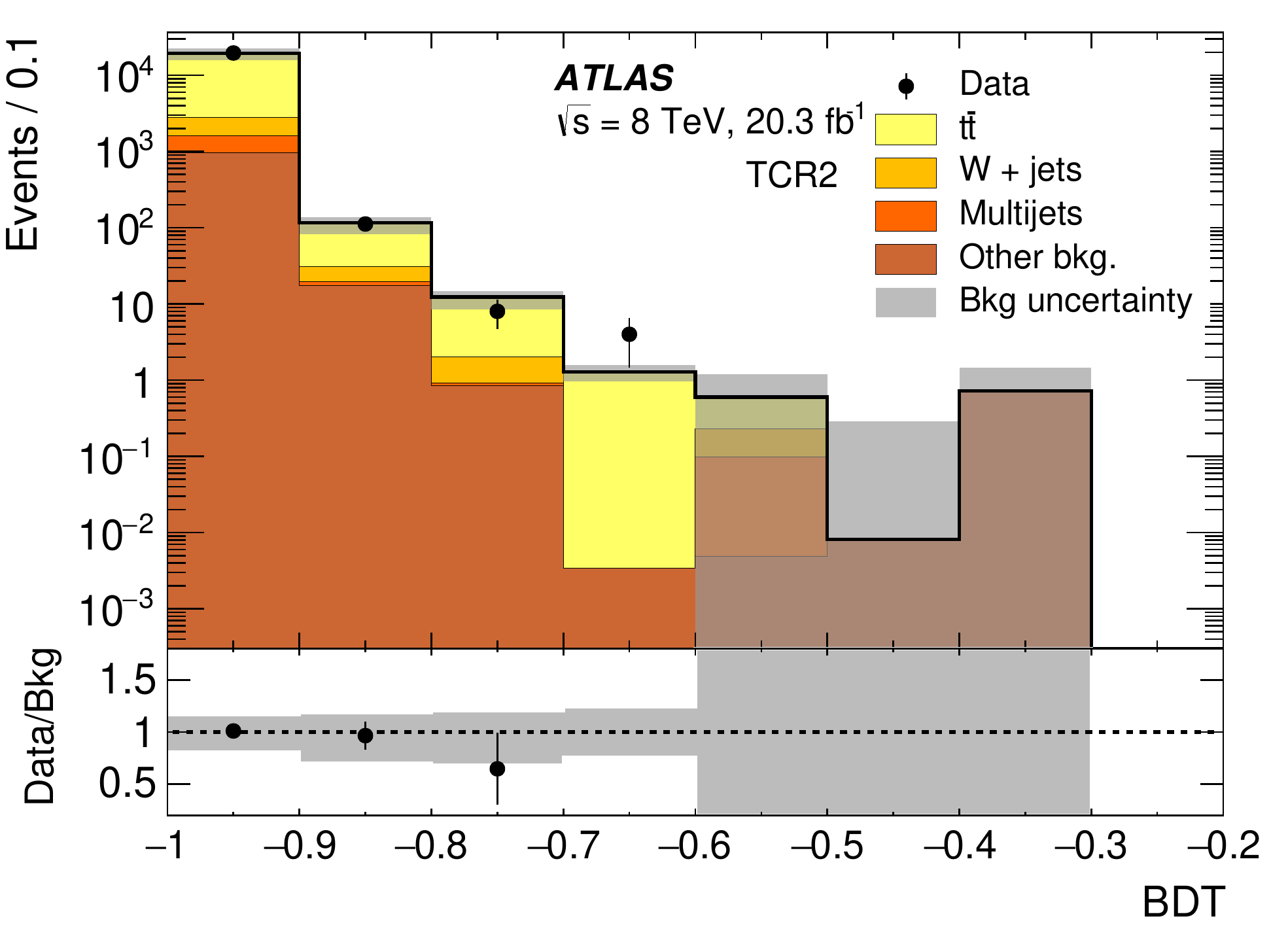}
\includegraphics[width=0.48\textwidth]{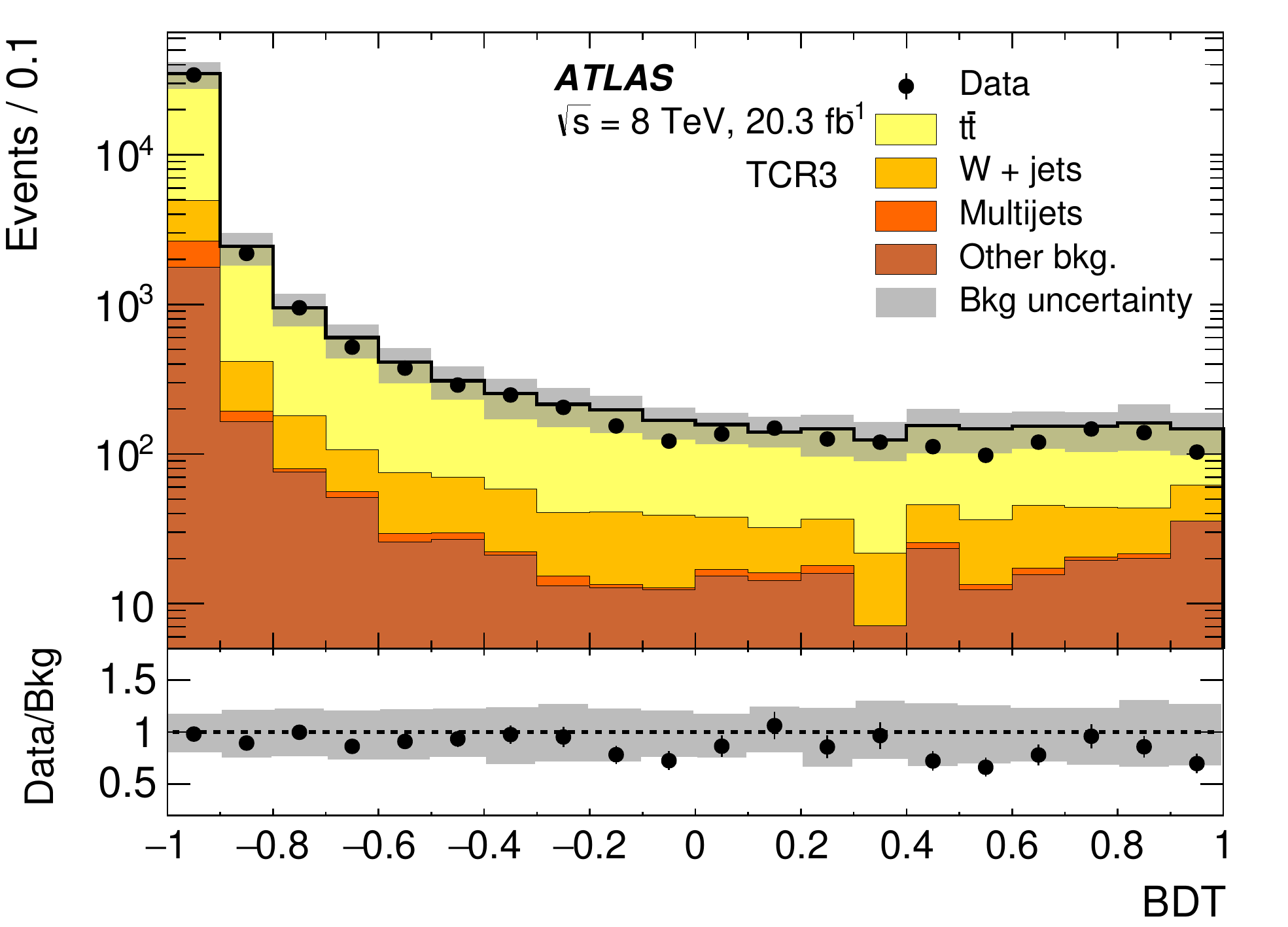}
\caption{Distributions of the BDT discriminant for data and for the background
expectation, for the various control regions: WCR1 (top left), WCR2 (top right),
TCR1 (middle left), TCR2 (middle right), and TCR3 (bottom). The lower panels show the
bin-by-bin ratio of the data to the total background expectation. The data are shown
with statistical error bars. The shaded bands
show the total uncertainties on the background expectation, 
including both the statistical and systematic uncertainties.}
\label{fig:bdt_cr_plots}
\end{figure*}  

\section{Systematic Uncertainties}
\label{sec:syst}

Tables~\ref{tab:systobj}, \ref{tab:systttbar}, and \ref{tab:systbkg} list
the sources of systematic uncertainty considered in this analysis.
The different effects change both the size and shape of the
signal and background contributions to the discriminant distribution;
the tables show the overall effect on the expected numbers
in the BDT SR.

Table~\ref{tab:systobj} lists those uncertainties that affect all the samples.
The uncertainty on the integrated luminosity, to which all
non-data-driven samples are
 normalized, is $2.8\%$~\cite{Aad:2013ucp}.  
 The jet energy resolution is measured by studying dijet
events in data and simulation. The simulation is found to agree with data to better than 10\%~\cite{Aad:2012ag};
the difference in resolutions between data and simulation is used to further smear the simulation. 
The effect of this additional smearing is taken as a systematic uncertainty.
The largest
uncertainty arises from the jet energy scale (JES), which
is varied in simulation by amounts derived from test beam and $Z/\gamma$+jet
collision data along with simulation~\cite{Aad:2011he,Aad:2014bia}.
The jet reconstruction efficiency is estimated using track-based jets,
and is well described by data, with the effect of small (approximately 0.2\%)
differences for jets below 30~GeV  assessed
by discarding randomly selected jets in simulated events.
The effect of the JVF requirement is evaluated
in data using events with a $Z$ boson produced in association with a single jet.
The lepton identification uncertainties include those on the electron energy and
muon momentum scale and resolution and trigger efficiencies,
evaluated in data using leptonic decays of $W$ and $Z$
bosons~\cite{Aad:2014zya,Aad:2014fxa,Aad:2011mk}.
The systematic uncertainties on the \ETM \ reconstruction include
the uncertainties on the constituent objects, as well as an additional uncertainty on the unclustered 
energy originating mainly from the pileup modeling.  
Finally, the tagging efficiency of $b$ jets, as well as of $c$ jets and
light-flavor jets, is derived from data and parametrized as a function
of jet \pT\ and $\eta$~\cite{ATLAS-CONF-2014-004,ATLAS-CONF-2014-046}.
The corresponding efficiencies in simulation are corrected to match
those observed in data, and the uncertainties in the calibration are
propagated through this analysis.

\begin{table*}[hbpt]
\begin{center}
\caption{Relative uncertainty (\%) on the expected number of events
due to uncertainties in luminosity determination and physics object reconstruction 
in the BDT signal region defined in Table~\ref{tab:srandcr}.
The signal column is for a VLQ $B$ mass of 700~GeV and for BR$(B \rightarrow Wt) = 1$.
The rightmost column indicates the corresponding uncertainty on the total background
in each row, taking into account the different background fractions.}
\begin{small}
\begin{tabular}{l|c|cccccc|c}
\hline\hline
 Source of uncertainty & Signal & $t\bar{t}$ & $W$+jets & Single top & $Z$+jets & $t\bar{t}+V$ & Diboson & Total background \\ \hline
Integrated luminosity & $\pm 2.8$ & $\pm 2.8$ & $\pm 2.8$ & $\pm 2.8$ & $\pm 2.8$ & $\pm 2.8$ & $\pm 2.8$ & $\pm 2.8$\\
Jet energy resolution & $\pm 1.0$ & $\pm 3.0$ & $\pm 6.0$ & $\pm 2.0$ & $\pm 9.0$ & $\pm 1.0$ & $\pm 14.0$ & $\pm 3.3$\\
JES & $\,^{+2.4}_{-3.1}$ & $\,^{+15.5}_{-13.5}$ & $\,^{+18.4}_{-15.0}$ & $\,^{+18.6}_{-16.4}$ & $\,^{+17.9}_{-19.2}$ & $\,^{+8.8}_{-8.7}$ & $\,^{+17.6}_{-10.3}$ & $\,^{+15.6}_{-13.5}$\\
JVF & $\pm 1.0$ & $\pm 4.0$ & $\pm 6.0$ & $\pm 6.0$ & $\pm 4.0$ & $\pm 2.0$ & $\pm 5.0$ & $\pm 4.2$\\
Lepton identification & $\,^{+1.2}_{-1.3}$ & $\pm 1.3$ & $\,^{+1.3}_{-1.9}$ & $\,^{+1.1}_{-2.0}$ & $\,^{+4.1}_{-8.5}$ & $\,^{+1.3}_{-1.4}$ & $\,^{+2.0}_{-2.1}$ & $\,^{+1.2}_{-1.5}$\\
\ETM & $\pm 1.0$ & $\pm 2.0$ & $\pm 2.0$ & $\pm 1.0$ & $\pm 8.0$ & $\pm 0.1$ & $\pm 7.0$ & $\pm 2.0$\\
$b$-tagging efficiency & $\,^{+6.3}_{-6.1}$ & $\pm 4.5$ & $\pm 1.6$ & $\,^{+5.0}_{-4.9}$ & $\,^{+4.3}_{-10.3}$ & $\,^{+4.9}_{-4.8}$ & $\pm 1.2$ & $\,^{+4.1}_{-4.3}$\\ \hline\hline
\end{tabular}
\end{small}
\label{tab:systobj}
\end{center}
\end{table*}

Table~\ref{tab:systttbar} summarizes systematic uncertainties related
to reweighting the {\sc Powheg-box} $t\bar{t}$ simulation for better agreement
with data.  As noted in Sec.~\ref{sec:ttbkg}, the reweighting improves
agreement in the control regions TCR1--TCR3, especially in the
\HT \ variable.  The largest uncertainties in the measurement of top quark
and $t\bar t$ \pT \ are taken as independent contributions to the systematic
uncertainty.  Of these, the largest contribution arises from the
modeling of initial- and final-state radiation.  The uncertainty due
to PDF choice on the acceptance is estimated by comparing $t\bar t$ events generated
with the HERAPDF 1.5 NLO PDF set~\cite{CooperSarkar:2011aa}
with those using the nominal CT10 PDFs.
The uncertainty due to the choice of the parton shower model is estimated by replacing {\sc Pythia}
with {\sc Herwig}~\cite{herwig} (version 6.520)
in the $t\bar t$ simulation.  For both uncertainties,
the comparison samples are corrected to match top quark \pT \ and $t\bar t$ \pT \
distributions in data, as done with the nominal simulation.
In addition, the {\sc Powheg-box+Pythia} sample is compared with
a \pT-reweighted {\sc Alpgen} (version 2.14) sample, with {\sc Herwig}
parton showering,
to check its behavior at high
jet multiplicity, and is found to be consistent within statistical uncertainties.

\begin{table}
\begin{center}
\caption{Further relative systematic uncertainties (\%) due to $t\bar{t}$ modeling.
Included are the relative uncertainties on the yields for $t\bar{t}$
background events, as well as the corresponding uncertainties on the total
background prediction.}
\begin{tabular}{lcc}
\hline\hline
Source of & \ \ Uncertainty on  \ \ & Uncertainty on \\
uncertainty & $t\bar{t}$ yield (\%) & total background (\%) \\ \hline
$\pT$ reweighting & $\pm 14$ & $\pm 12$ \\
PDF acceptance & $\pm 13$ & $\pm 11$\\
Parton shower model & $\pm 6.4$ & $\pm 5.4$\\
\hline\hline
\end{tabular}
\label{tab:systttbar}
\end{center}
\end{table}

Further relative uncertainties on the expected numbers of background
events are summarized in Table~\ref{tab:systbkg}.
The inclusive $t\bar t$ production cross-section uncertainty at NNLO+NNLL is
taken to be +5\%/--6\%.
A 4\% overall theoretical uncertainty (5\% for diboson) is assigned to the 
production rates  of $W$+jets, $Z$+jets, single top, and diboson
backgrounds, with an additional 24\% per jet
(estimated from variations in
the predicted cross-section ratio of $W+n$-jets to $W+(n-1)$-jets
production~\cite{Berger:2010zx,Alwall:2007fs})
added in quadrature.
An additional uncertainty is included for the shape of the $W$+jets background contribution, 
based on variations of the
matching scale and the functional form of the factorization
scale in {\sc Alpgen}.
A conservative uncertainty of $30\%$ is assigned to the
$t\bar t+V$ rate, based upon the NLO results of Ref.~\cite{Garzelli:2012bn}. 
As mentioned previously, a $50\%$ systematic uncertainty
is assigned to the normalization of the small multijet background contribution.

\begin{table}[hbpt]
\begin{center}
\caption{Relative yield uncertainties (\%) due to normalization.}
\begin{tabular}{lcc}
\hline\hline
Source of & \ \ \ \ Uncertainty on \ \ \ \  & Uncertainty on \\
uncertainty & background process (\%) & total background (\%) \\ \hline
$t\bar t$ rate & $\,^{+5.0}_{-6.0}$ & $\,^{+4.2}_{-5.0}$\\
$W$+jets rate & $\pm 59$ & $\pm 4.6$\\
$W$+jets shape & $\pm 5.7$ & $\pm 0.4$\\
$Z$+jets rate & $\pm 59$ & $\pm 0.8$\\
Single-top rate & $\pm 48$ & $\pm 2.2$\\
$t\bar{t}+V$ rate & $\pm 30$ & $\pm 0.3$\\
Diboson rate & $\pm 48$ & $\pm 0.3$\\
Multijet rate & $\pm 50$ & $\pm 0.7$\\
\hline\hline
\end{tabular}
\label{tab:systbkg}
\end{center}
\end{table}


\section{Results}
\label{sec:result}

With the background model determined, and validated using the
various background-dominated control regions, the data in the
BDT and cuts-based signal regions are examined and
compared with the expected background contributions in order
to search for any evidence of a signal-like excess.

\subsection{Signal-region distributions}

Table~\ref{tab:systtotal} lists the predicted event yields for the BDT signal region, including
both the statistical and systematic
uncertainties, for the various background contributions, as well as
the total background expectation. The number of events observed
in data in the BDT signal region is 12235, which is in good agreement with
the total expected background of $12900 \pm 100 \pm 3100$~events.
The first row of Table~\ref{tab:systtotal} shows the expected signal yield of
$164 \pm 2 \pm 13$ events, for the specific case of  a VLQ $B$ with a
mass of 700~GeV and a 100\% branching ratio to $Wt$. The
signal-to-background ratio is only about 1.3\% in this case, supporting the usage
of the BDT to obtain additional discrimination power. 

\begin{table}[hbpt]
\begin{center}
\caption{Expected signal and background yields for the BDT signal region with
$\Njets\ge 6$, $\Nbjets\ge 1$, and $\HT>500$~GeV, with associated
statistical and systematic uncertainties.
The signal row is for a VLQ $B$ mass of 700~GeV,
with a 100\% branching ratio to $Wt$. The last row provides the
numbers of events observed in data.}
\begin{tabular}{lc}
\hline\hline
  &           Event yield \\
Physics process  & [$\pm$(stat)$\pm$(syst)] \\ \hline
700~GeV VLQ $B$; BR$(B \rightarrow Wt) = 1$ & $164 \pm 2 \pm 13$ \\ \hline
$t\bar{t}$ &  \ \ $10800 \pm 100 \pm 2800$ \ \ \\
$W$+jets &  $1020 \pm 30 \pm 630$ \\
Single top & $490 \pm 20 \pm 300$ \\
$Z$+jets &  $180 \pm 30 \pm 120$ \\
$t\bar{t}+V$ &  $147 \pm 1 \pm 47$ \\ 
Diboson &  $66 \pm 5 \pm 42$ \\
Multijets &  $183 \pm 9 \pm 92$ \\ \hline
Total background &  $12900 \pm 100 \pm 3100$ \\ \hline
Observed in data & 12235 \\ \hline\hline
\end{tabular}
\label{tab:systtotal}
\end{center}
\end{table}

Figure~\ref{fig:sr_bdtvariables} shows the distributions of several of the BDT input 
variables for data in the signal region of the BDT analysis, with
$\Njets \ge 6$, $\Nbjets \ge 1$, $\NV \ge 1$, and $\HT>500$~GeV.
Within the total uncertainties on the background prediction, shown in the figures as the shaded bands, the data
are in good agreement with the total expected background.

\begin{figure*}[hbpt]
\centering
\includegraphics[width=0.48\textwidth]{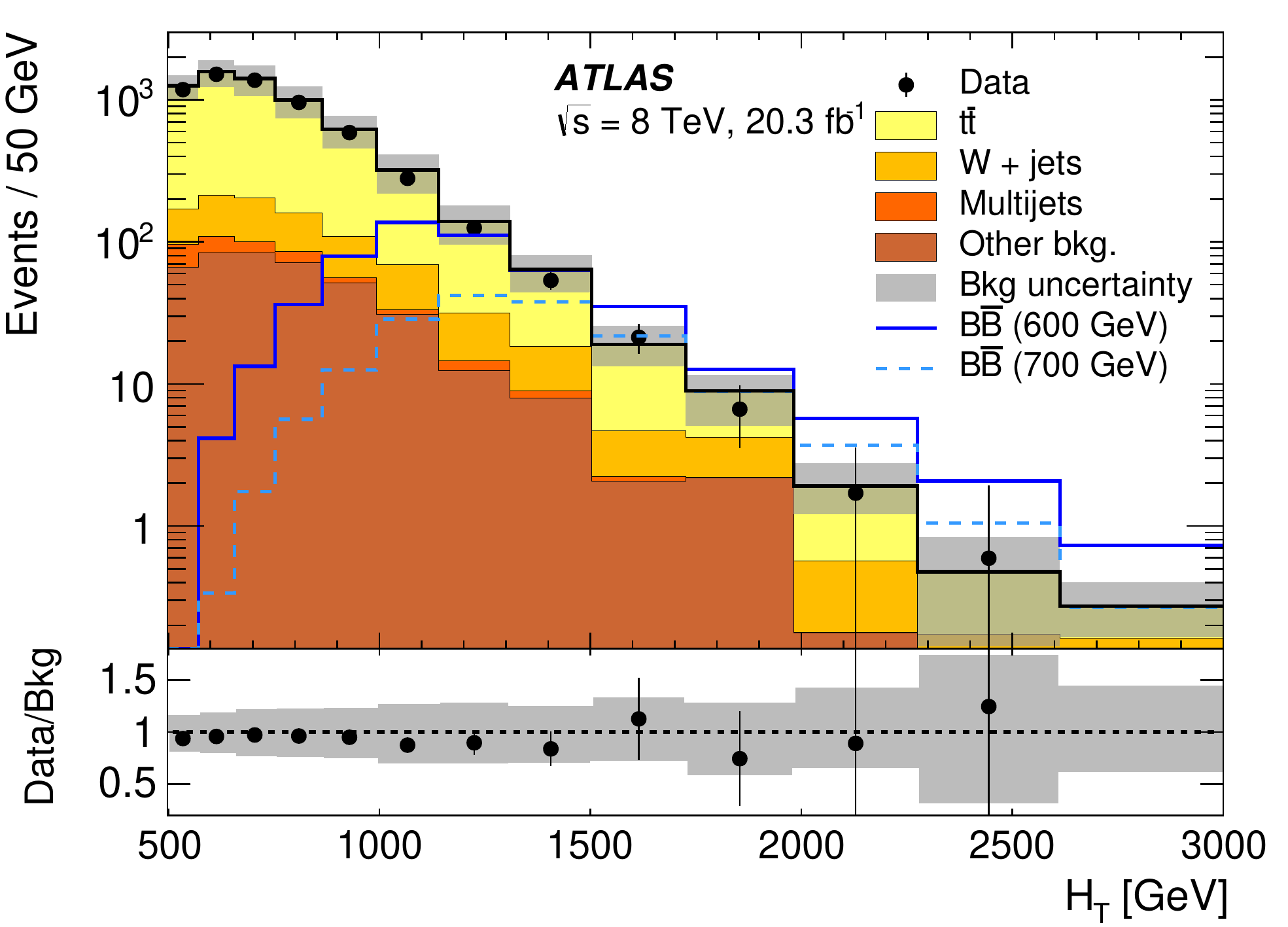}
\includegraphics[width=0.48\textwidth]{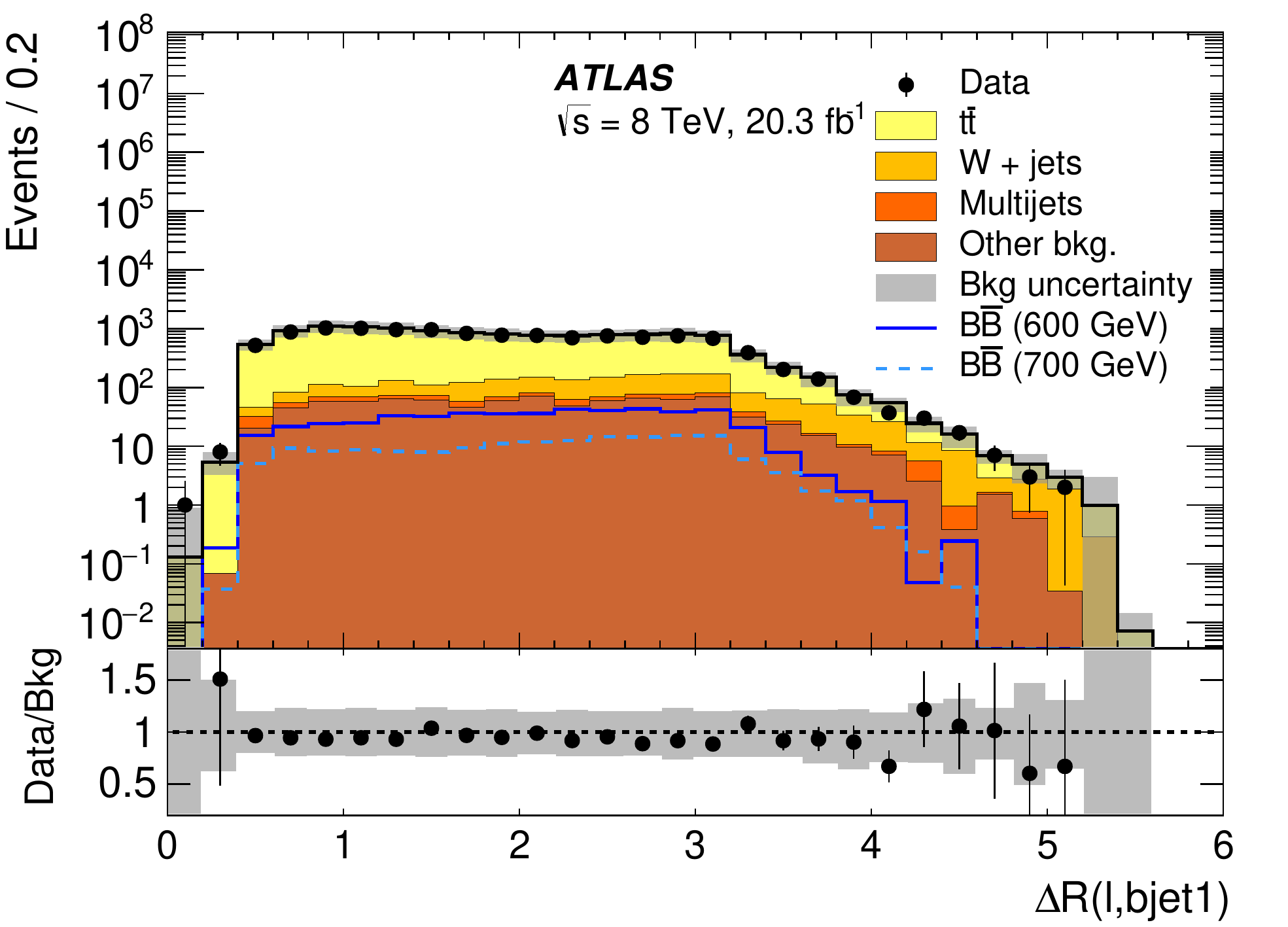}
\includegraphics[width=0.48\textwidth]{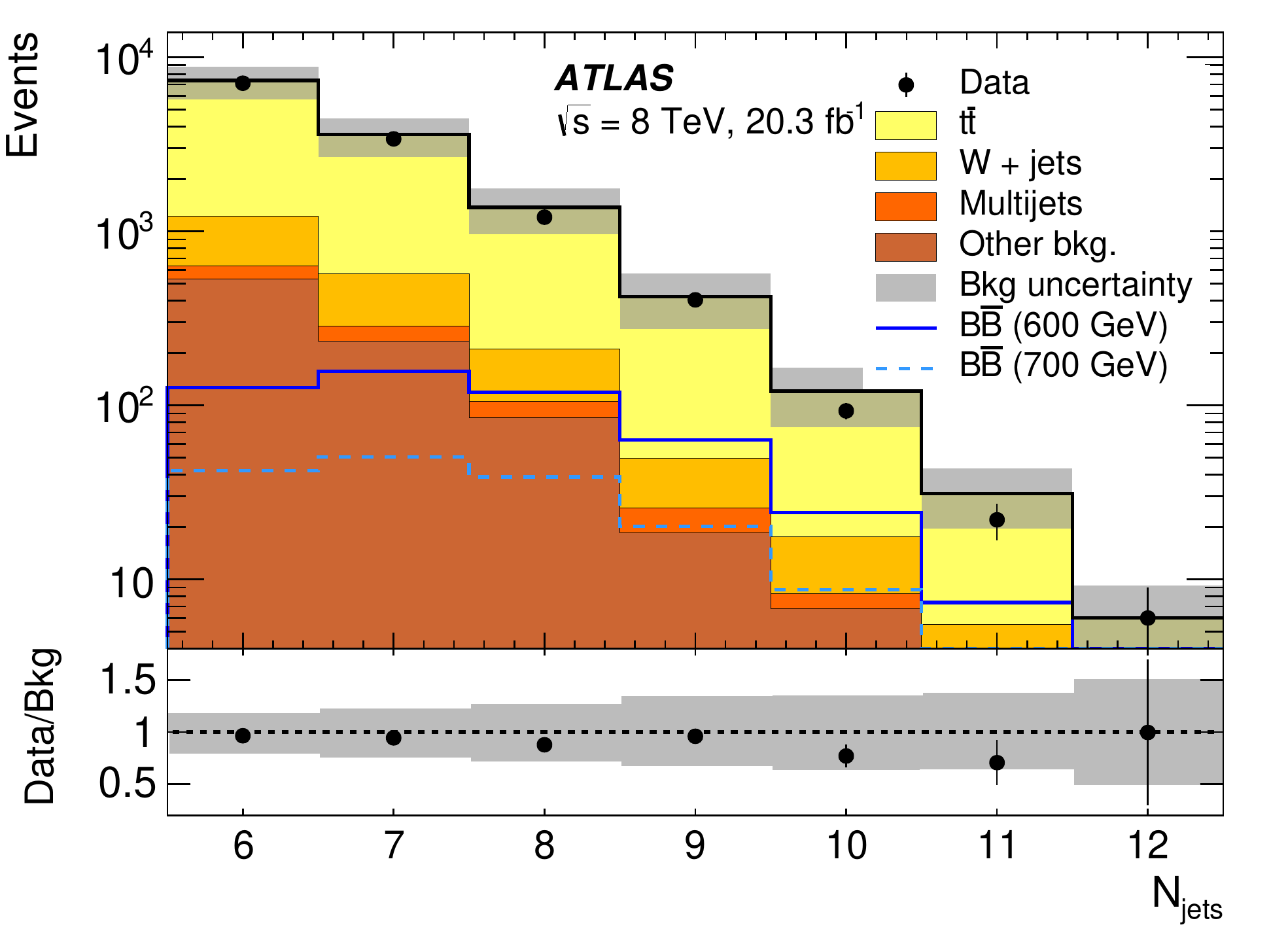}
\includegraphics[width=0.48\textwidth]{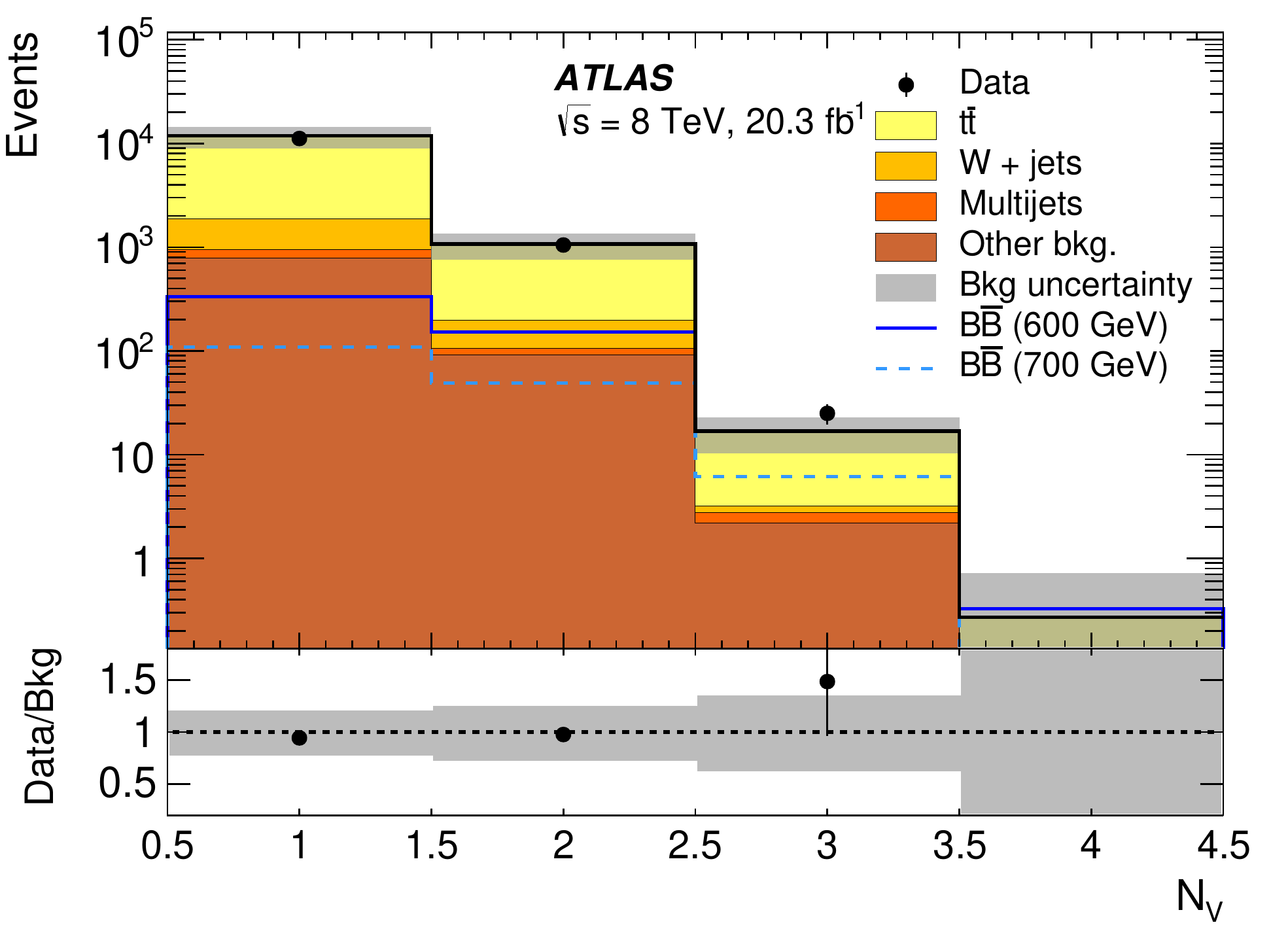}
\caption{Distribution of four BDT input variables for data in the BDT signal region, with
$\Njets \ge 6$, $\Nbjets \ge 1$, $\NV \ge 1$, and $\HT>500$~GeV.  Superimposed is
the expectation for the total background. The variables shown in the upper row are the two 
variables with the highest sensitivity in the BDT training, namely \HT \ (left),
and $\Delta R(\ell,{\mathrm bjet1})$, the angular separation between the lepton and the
leading $b$-tagged jet (right). The lower row shows two of the multiplicity-related variables,
namely the number of jets (left) and the number of hadronic $W/Z$ candidates (right).
Also shown are the expected signal contributions for VLQ $B$ masses of 600 and
700~GeV, assuming BR$(B \rightarrow Wt) = 1$.
The data are shown with statistical error bars, and the shaded band
shows the total uncertainty on the background expectation.
The lower panels show the bin-by-bin
ratio of the data to the background expectation.}
\label{fig:sr_bdtvariables}
\end{figure*} 

The final result of the BDT analysis exploits the increased sensitivity obtained by
combining the twelve input variables into the final BDT discriminant.
Figure~\ref{fig:limit_input_bdt} shows the distribution of the BDT discriminant 
for data in the signal region of the BDT analysis.  The data
are in good agreement, within the uncertainties, with the total expected background contribution. 
Given the lack of
evidence for a signal-like excess, the BDT distribution is used to set upper limits on
VLQ production, as described in Sec.~\ref{sec:limits}. Figure~\ref{fig:limit_input_bdt}(a)
shows the entire range of the BDT discriminant with uniform
binning.  Figure~\ref{fig:limit_input_bdt}(b) shows the same data in the nonuniform binning optimized  for the
determination of the final exclusion limits, with the background-dominated
region of BDT values from $-1$ to $+0.95$ combined in a single bin. 

\begin{figure*}[hbpt]
\centering
\includegraphics[width=0.48\textwidth]{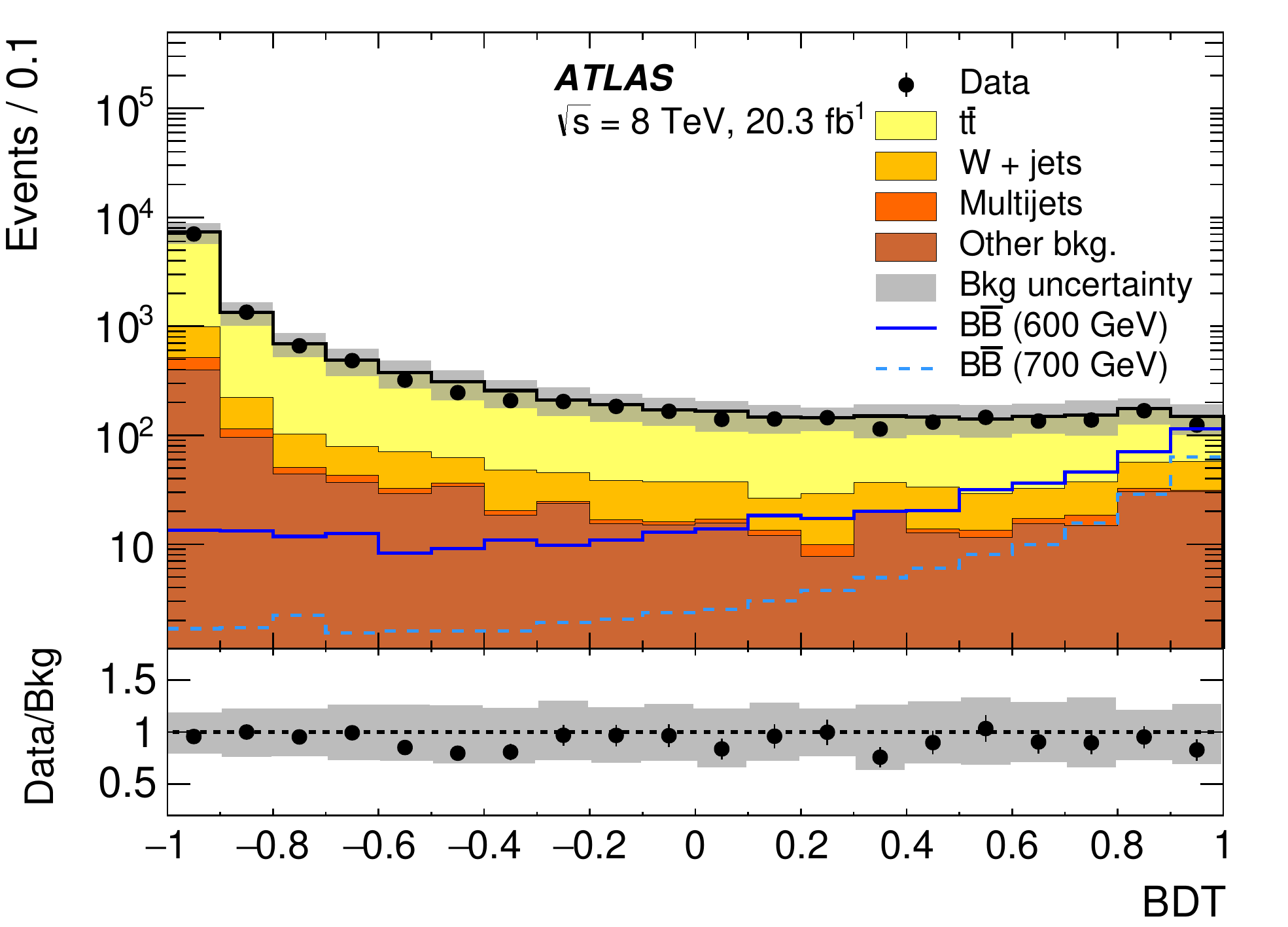}
\includegraphics[width=0.48\textwidth]{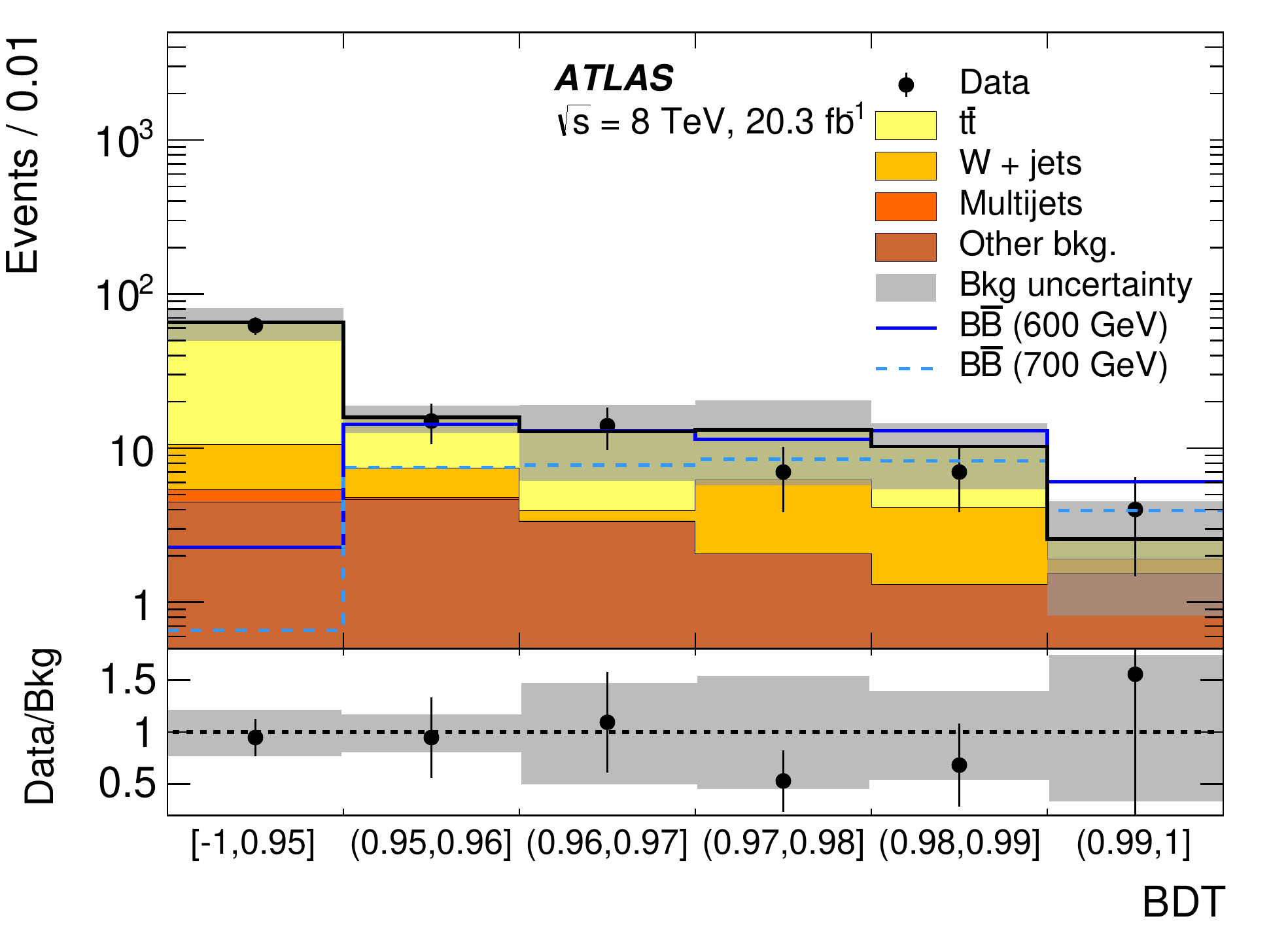}
\caption{Distribution of the BDT discriminant for data in the signal region of
the BDT analysis, with
$\Njets \ge 6$, $\Nbjets \ge 1$, $\NV \ge 1$, and $\HT>500$~GeV.  Superimposed is
the expectation for the total background, as well as signal contributions that would be expected
for VLQ $B$ masses of 600 and 700~GeV, in both cases for a branching ratio of unity for
the decay $B \rightarrow Wt$. The data are shown with statistical error bars, and the shaded band
shows the total uncertainty on the background expectation.
Uniform binning is used in the left plot, and the right one shows
the same data in the non-uniform binning used to
determine the final exclusion limits, with the background-dominated
region of BDT values from $-1$ to $+0.95$ combined in a single bin. 
The lower panels show the bin-by-bin
ratio of the data to the background expectation. }
\label{fig:limit_input_bdt}
\end{figure*}  

The cuts-based analysis is used to cross-check the BDT results. 
Figure~\ref{fig:cutsbasedistributions} shows the \HT \ distributions for 
events in the signal region for the cuts-based analysis, with
$\Njets \ge 6$, $\Nbjets \ge 1$, $\NV \ge 1$, and $\HT>800$~GeV. The data after the
cuts-based signal selection are divided into two exclusive subsamples in order to
improve the sensitivity of the analysis by exploiting the different signal-to-background
ratios in the two subsamples. 
Figure~\ref{fig:cutsbasedistributions}(a) shows the \HT \ distribution for the
subsample of events with  $\NV=1$, while
Fig.~\ref{fig:cutsbasedistributions}(b) shows the data for events
with $\NV\ge 2$. For both subsamples, the data are in good agreement with the background
expectation, and there is no evidence for any excess. These two \HT \ distributions are
used to set upper limits on VLQ production using the cuts-based analysis, as described below.

\begin{figure*}[hbpt]
\centering
\includegraphics[width=0.48\textwidth]{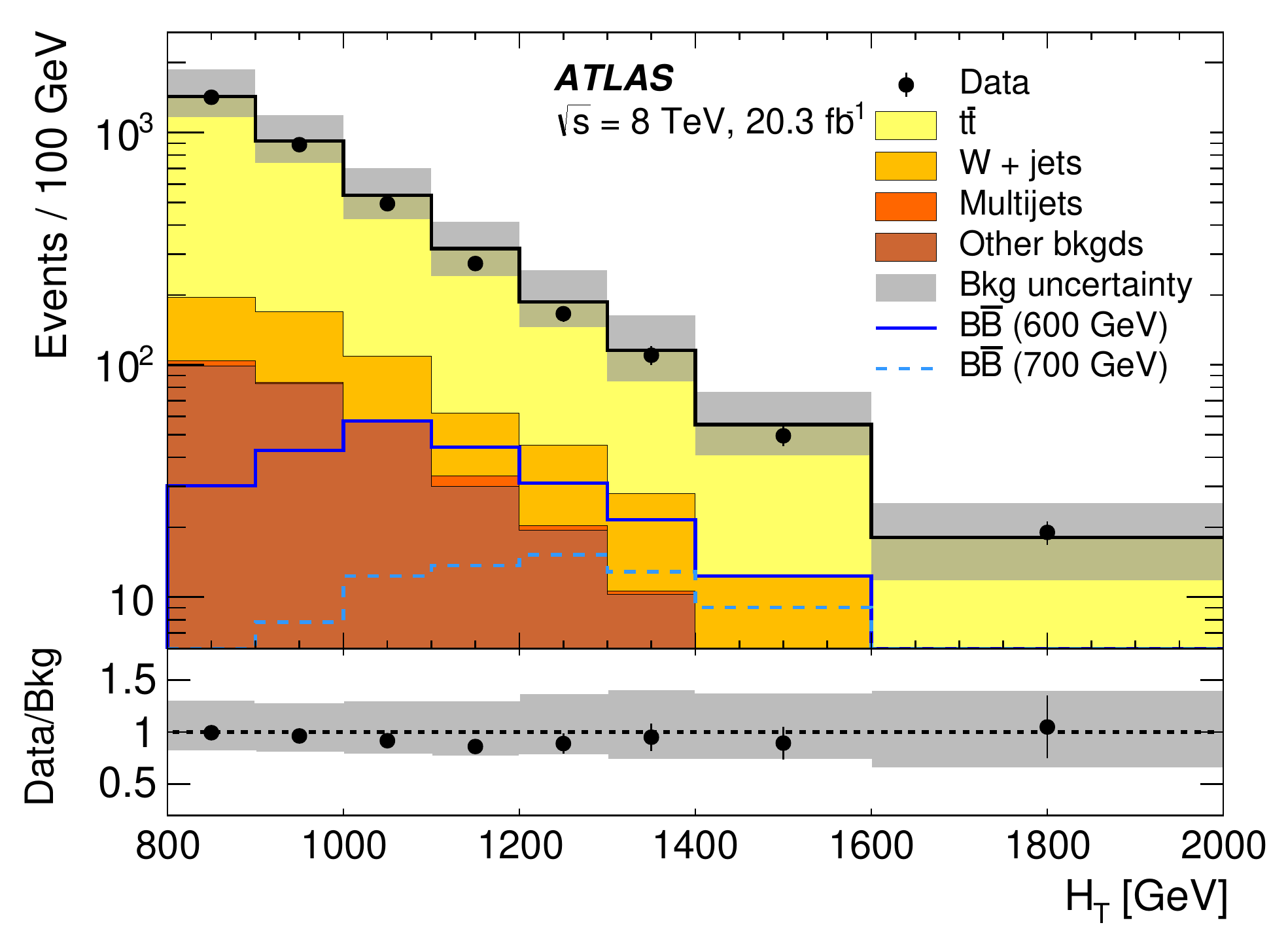}
\includegraphics[width=0.48\textwidth]{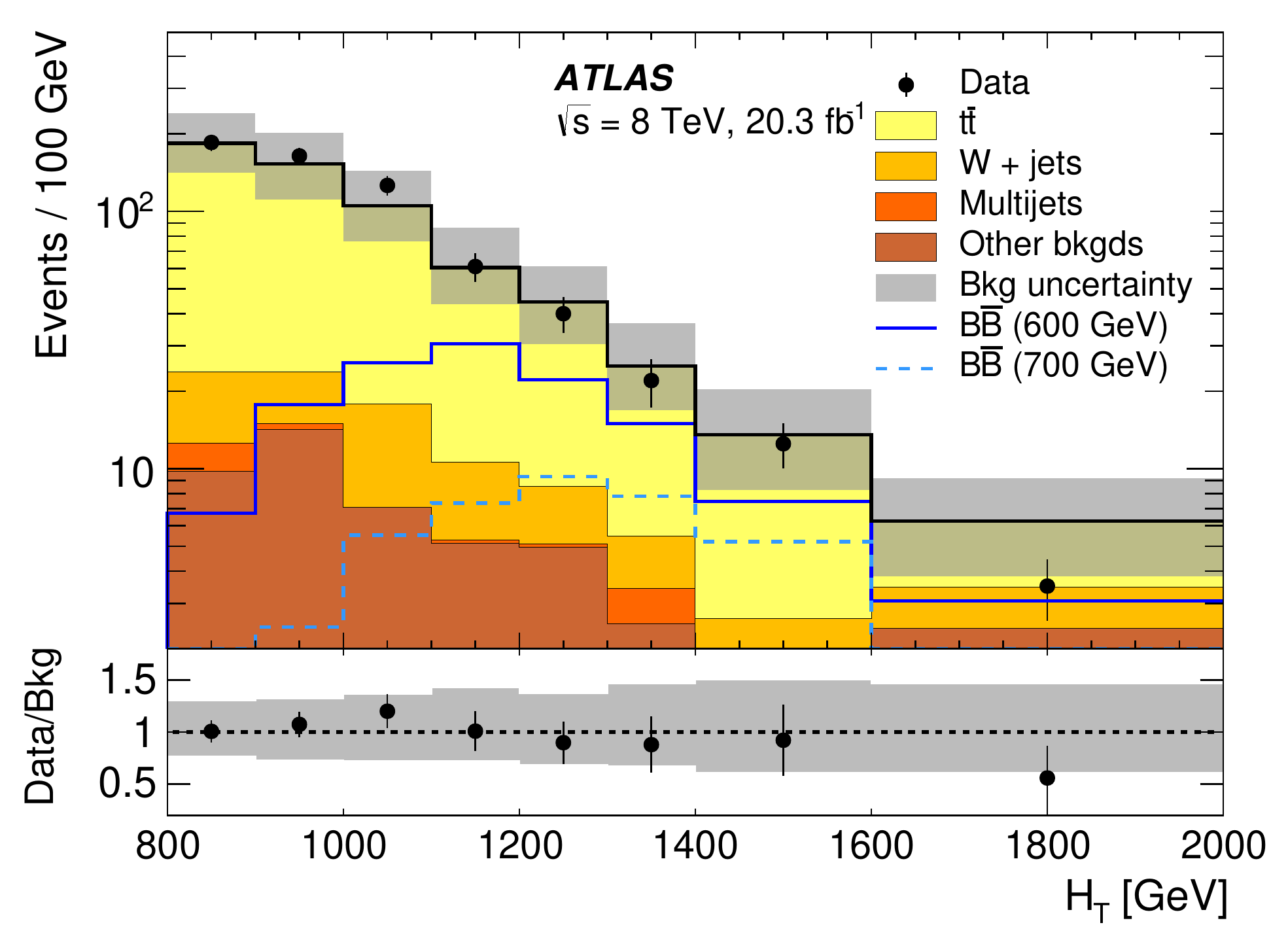}
\caption{\HT \ distributions for events in the signal region for the cuts-based analysis, with
$\Njets \ge 6$, $\Nbjets \ge 1$, $\NV \ge 1$, and $\HT>800$~GeV. 
The figure shows events with (left) $\NV=1$ and (right) $\NV\ge 2$.
Superimposed is
the expectation for the total background, as well as signal contributions that would be expected
for VLQ $B$ masses of 600 and 700~GeV, in both cases for a branching ratio of unity for
the decay $B \rightarrow Wt$.
The data are shown with statistical error bars, and the shaded band
shows the total uncertainty on the background expectation.
The lower panels show the bin-by-bin
ratio of the data to the background expectation.}
\label{fig:cutsbasedistributions}
\end{figure*}

\subsection{Statistical procedure}
\label{sec:stats}

A binned likelihood test is performed, assuming Poisson statistics for the distributions of the final discriminating
variables, to assess the compatibility of the observed data with the background-only
and signal-plus-background hypotheses. The test employs a log-likelihood ratio function,
$R_{\mathrm LL} = -2$log$(L_{\mathrm s+b}/L_{\mathrm b})$, where $L_{\mathrm s+b}$ ($L_{\mathrm b}$) is the likelihood to observe 
the data under the signal-plus-background (background-only) hypothesis.
Pseudoexperiments assuming Poisson statistics are generated
for the two hypotheses, using the predicted signal and background distributions and including the
impact of each systematic uncertainty. The latter are evaluated for their impact on both
the normalization and the shape of the final discriminating variable distributions, and are varied during
the generation of the pseudoexperiments assuming a Gaussian distribution as the prior
probability distribution function.

To reduce the impact of the acceptance effects of the \ttbar{} modeling uncertainties, the 
likelihood is parametrized as a function of an overall \ttbar\ normalization factor.
The likelihood is then minimized with respect to this normalization factor. 
The likelihood minimization thus constrains the absolute number of \ttbar{} events.  
This constraint comes from the low region of the BDT, which is dominated by \ttbar\ events.
The uncertainties on the shape of the \ttbar\ distribution, and therefore on the extrapolation of the number of 
\ttbar\ events in the low BDT region to the region populated by signal, are not constrained with this method.

The final discriminating variable for the BDT analysis is the distribution of the BDT discriminant,
using the binning in Fig.~\ref{fig:limit_input_bdt}(b). For the cuts-based analysis,
the two \HT \ distributions of Fig.~\ref{fig:cutsbasedistributions} are used in a combined fit as the
final discriminating variables.

The data are found to be consistent with the background-only hypotheses for both analysis
methods, and limits are subsequently derived according to the CL$_{\mathrm s}$
prescription~\cite{Junk:1999kv,Nakamura:2010zzi}
using the above likelihood-based test statistic.
Upper limits at the 95\% CL are set on the pair production 
cross sections of both the VLQ $B$ and $T_{5/3}$ scenarios. 

\subsection{Limits on VLQ $B$ production}
\label{sec:limits}

The values of the branching ratios for the various VLQ $B$ decay modes are model dependent. 
For the case of a $SU(2)$ singlet VLQ $B$, Fig.~\ref{fig:limit_vlb_singlet_bdt} shows 
the observed and expected 95\% CL limits on the pair-production cross section vs VLQ $B$ mass, 
as obtained with the BDT analysis.  Also shown
are the $\pm 1\sigma$ and $\pm 2\sigma$ uncertainty bands for the expected limit.
The observed limit curve is slightly lower than the expected limit curve due to the small deficit of
observed events in Fig.~\ref{fig:limit_input_bdt}(b), compared to the background expectation, 
in the signal-enriched bins of the BDT discriminant near a value of 1.0.
Figure~\ref{fig:limit_vlb_singlet_bdt} shows that the deficit is about $-1\sigma$, and therefore
is not significant. The uncertainty on the theoretical cross section of the signal,
from varying the
renormalization and factorization scales, 
as well as the PDF set and the value of $\alpha_{\mathrm s}$, is
indicated by the width of the theory band in the figure. Compared with the theoretical prediction
of the cross section, the results correspond to an observed (expected) 95\% CL lower limit on the
$SU(2)$ singlet VLQ $B$ mass of  \obslimitBsinglet~GeV (\explimitBsinglet~GeV).

\begin{figure}[hbpt]
\centering
\includegraphics[width=0.48\textwidth]{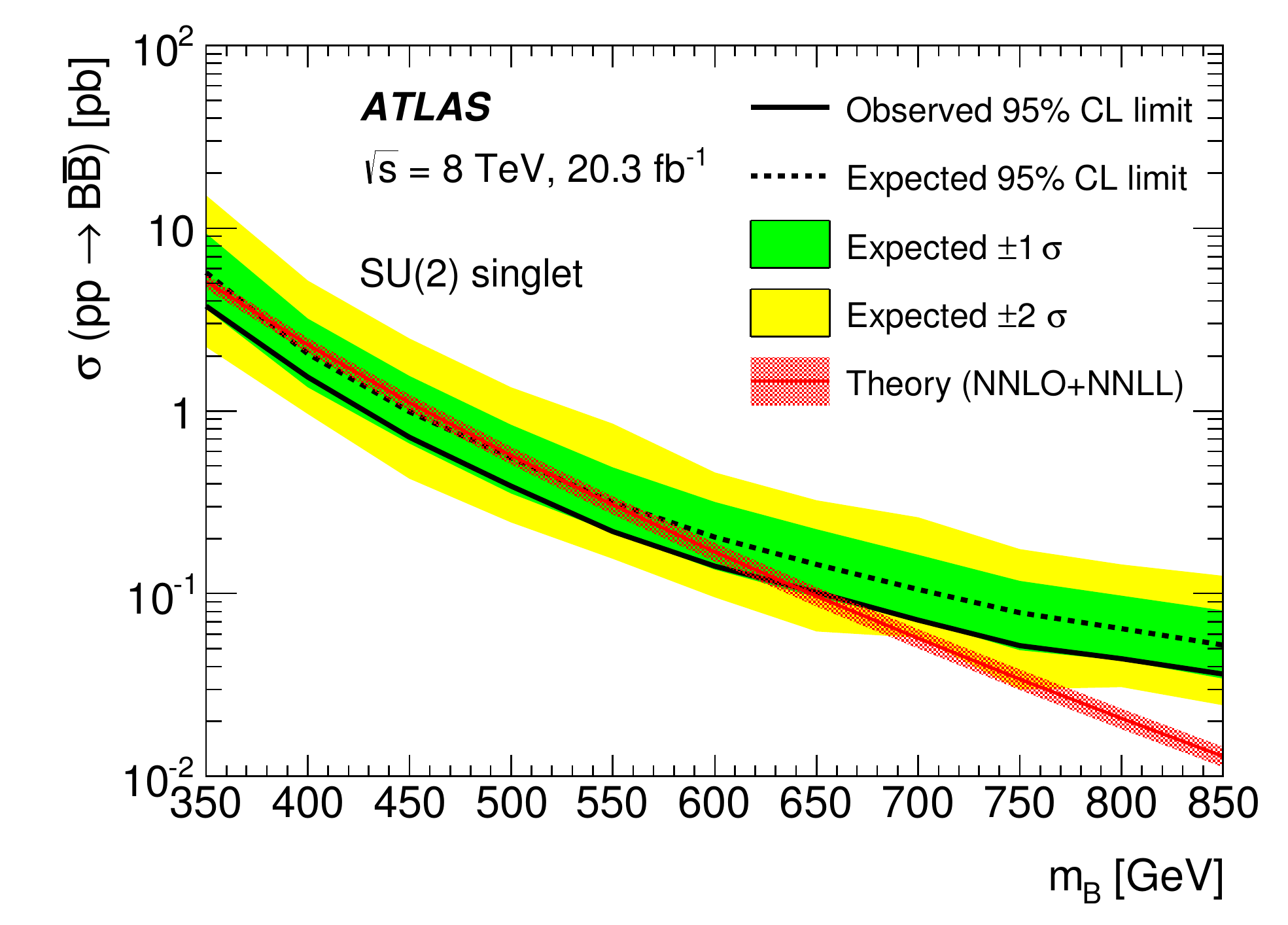}
\caption{Predicted pair production cross section as a function of the VLQ $B$ mass
and 95\% CL observed and expected upper limits for an $SU(2)$ singlet VLQ $B$, 
as obtained with the BDT analysis. The uncertainty on the theoretical cross section is
indicated by the width of the theory band.}
\label{fig:limit_vlb_singlet_bdt}
\end{figure} 

In addition to lower limits on the VLQ $B$ mass for this benchmark $SU(2)$ singlet
scenario, limits are also derived for all sets of VLQ $B$ branching ratios consistent with the three decay
modes ($B \rightarrow Wt/Zb/Hb$) summing to unity.
Figure~\ref{fig:limit_vlb_bdt} shows, for a variety of VLQ $B$ mass values,
the observed and expected $95\%$ CL exclusions of the BDT analysis. The results are shown 
in terms of the decay branching ratios, with BR($B \rightarrow Hb$) plotted on the vertical axis and
BR($B \rightarrow Wt$) on the horizontal axis.  Superimposed on Fig.~\ref{fig:limit_vlb_bdt} are 
two particular benchmark models, the case discussed above where the VLQ $B$ 
is an $SU(2)$ singlet (shown as a filled circle) and the case where it
is part of a $(B,Y)$ $SU(2)$ doublet (shown as a star). The
analysis is not sensitive to the $SU(2)$ doublet case, which predicts BR$(B \rightarrow Wt) = 0$,
while the results for the $SU(2)$ singlet case were shown in Fig.~\ref{fig:limit_vlb_singlet_bdt}.

An alternative
representation of the same results is shown in Fig.~\ref{fig:temp_plot_bdt}, which displays
 the observed and expected 95\% CL lower limits on the VLQ $B$ mass in the 
same plane of branching ratio values. The analysis is most sensitive in the bottom right 
corner of the plane of branching ratios, where BR$(B \rightarrow Wt) = 1$.
In that case,  the observed and expected 
95\% CL lower limits on the VLQ $B$ mass are \obslimitB \ and \explimitB~GeV, respectively.

\begin{figure*}[hbpt]
\centering
\includegraphics[width=1\textwidth]{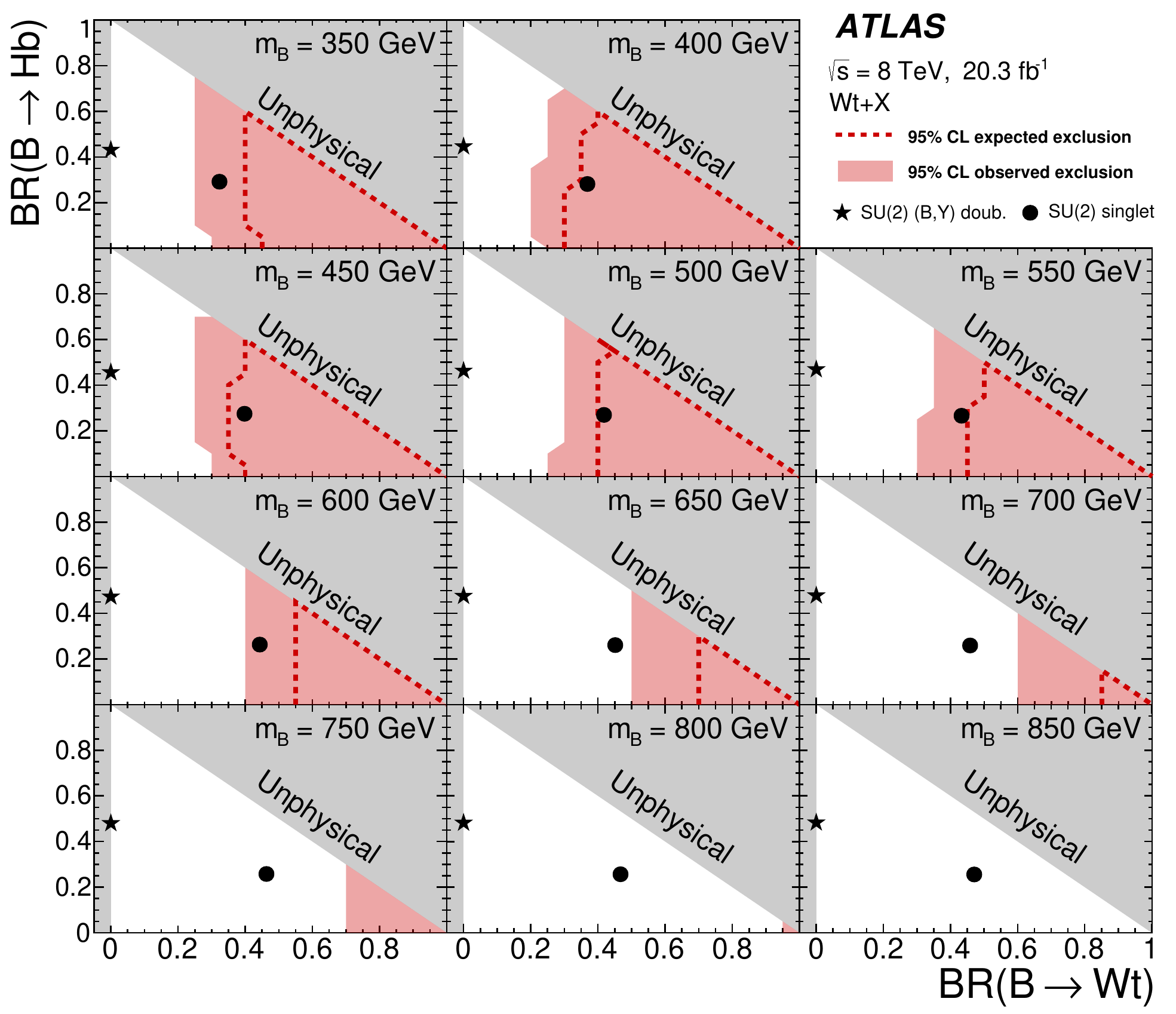}
\caption{For a variety of VLQ $B$ mass values,
the expected and observed $95\%$ CL exclusion limits of the BDT analysis. The results are shown 
in terms of the decay branching ratios, with BR($B \rightarrow Hb$) plotted on the vertical axis,
 BR($B \rightarrow Wt$) on the horizontal axis, and the branching ratios of the three decay
 modes considered, namely $B \rightarrow Wt/Hb/Zb$, assumed to sum to unity. Shown with a
 closed star (circle) are the branching ratio values for the particular case of an $SU(2)$
 doublet (singlet) implementation of the VLQ $B$ model.}
\label{fig:limit_vlb_bdt}
\end{figure*}

\begin{figure*}[hbpt]
\centering
\includegraphics[width=0.48\textwidth]{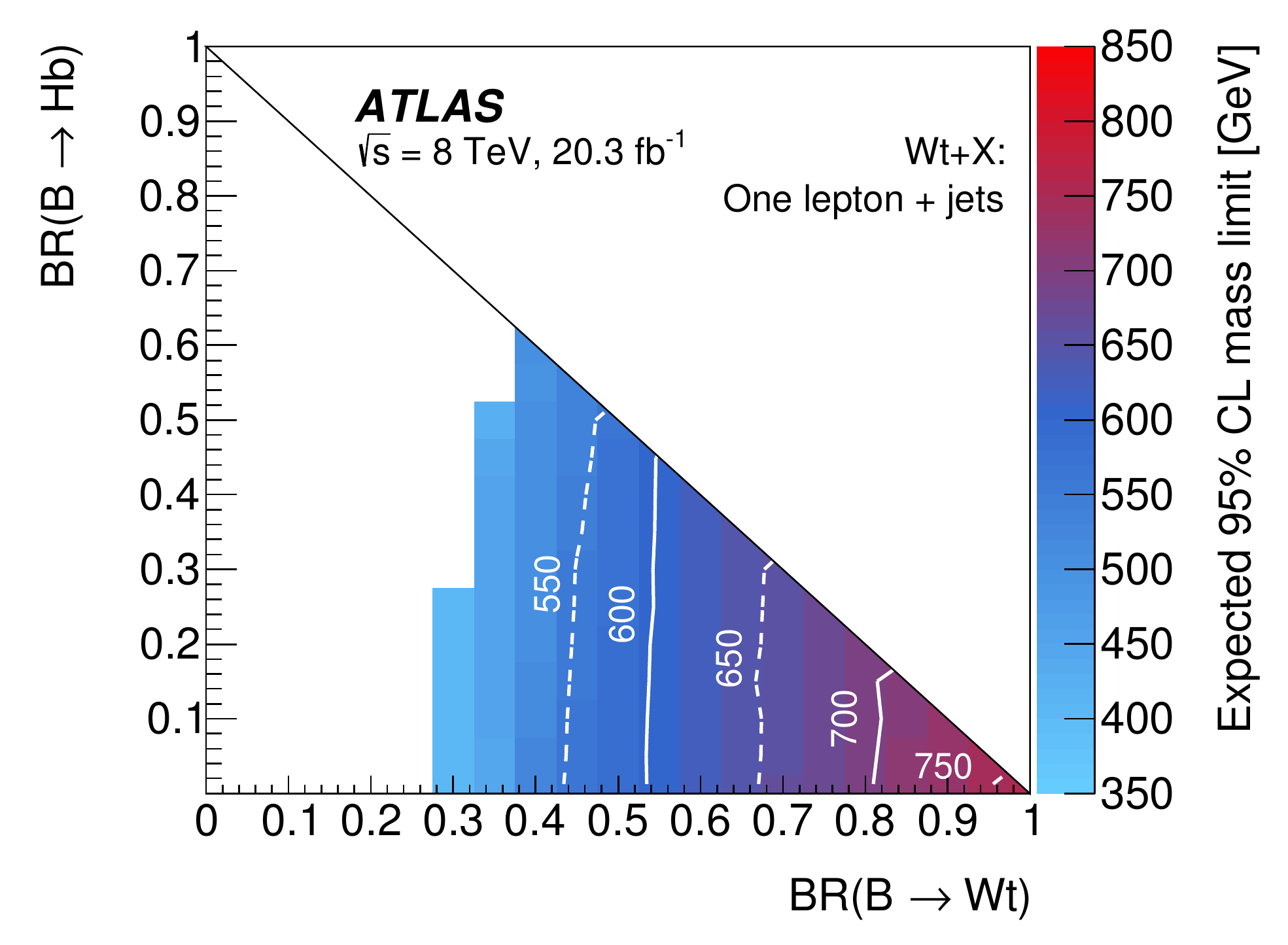}
\includegraphics[width=0.48\textwidth]{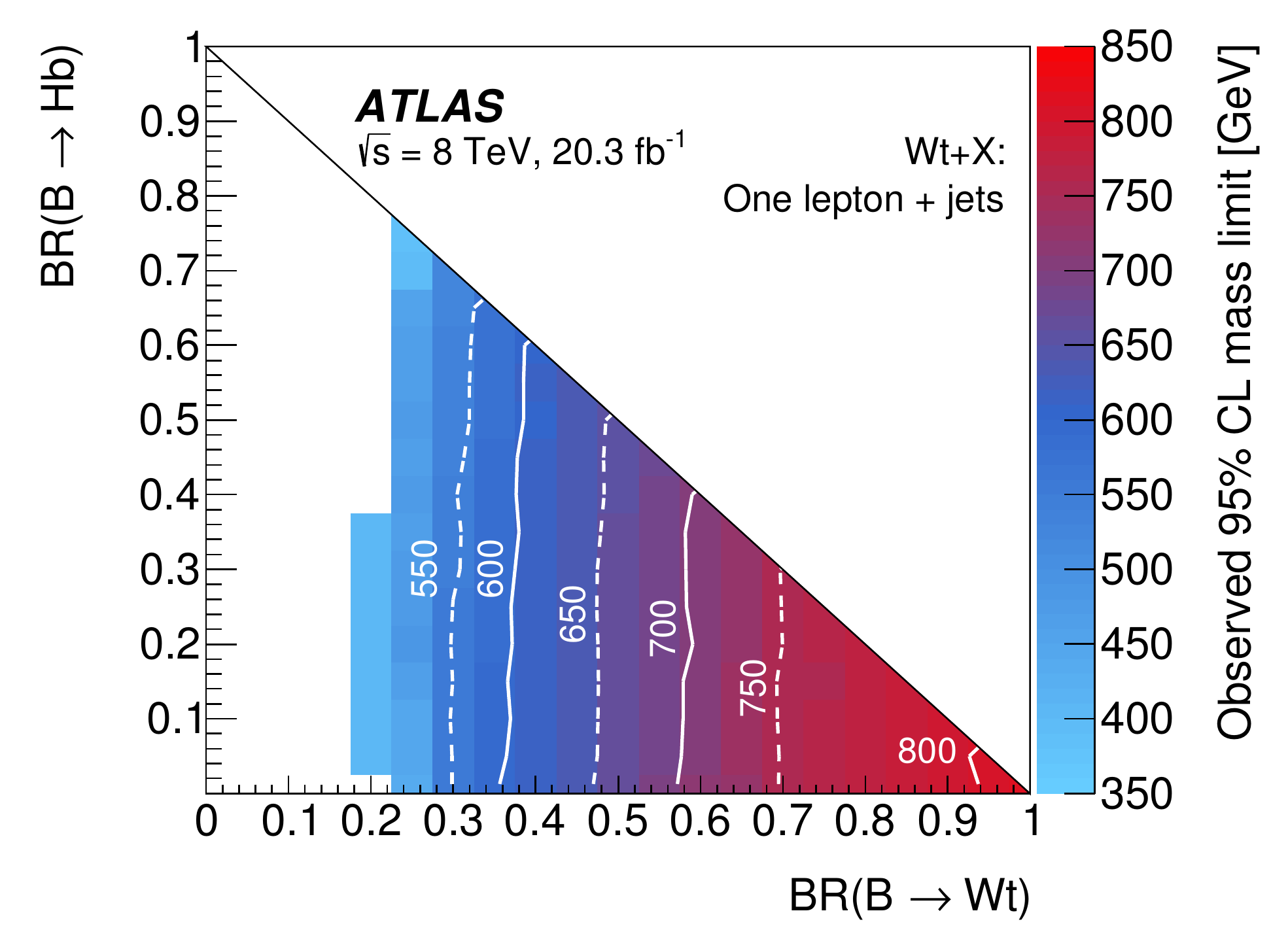}
\caption{Expected (left) and observed (right) 95\% CL lower limits on the mass of the VLQ $B$.
The results are shown 
in terms of the decay branching ratios, with BR($B \rightarrow Hb$) plotted on the vertical axis,
 BR($B \rightarrow Wt$) on the horizontal axis, and the branching ratios of the three decay
 modes considered, namely $B \rightarrow Wt/Hb/Zb$, assumed to sum to unity. } 
\label{fig:temp_plot_bdt}
\end{figure*}  

The limits extracted using the cuts-based analysis are qualitatively similar to those of the BDT analysis,
but less restrictive due to the lower sensitivity of the cuts-based approach. 
For example, for a VLQ $B$ mass of 600~GeV, the cuts-based analysis is 
expected to be sensitive down to a value of BR$(B \rightarrow Wt) \approx 0.8$,
while the improved sensitivity of the BDT analysis extends this coverage down to $\approx 0.55$.
A similar pattern is seen for the observed limits, which are slightly more restrictive for both analyses 
than the corresponding expected limits.

\subsection{Limits on $T_{5/3}$ production}
\label{sec:limitsT}

The analysis does not attempt to measure the charge of the hadronically decaying $W$ bosons and is 
therefore also sensitive to pair production of a colored, charge $5/3$ exotic fermion, $T_{5/3}$. Assuming the decay into
a same-sign $W$ boson and top quark, via $T_{5/3} \rightarrow Wt$, pair production
of the $T_{5/3}$ would be kinematically similar to that of the VLQ $B$ in the chiral limit 
where BR$(B \rightarrow Wt) = 100$\%. Therefore, the BDT analysis, which was trained using
chiral heavy-quark signal samples, can be simply applied to the investigation of pair production
of the $T_{5/3}$.

Figure~\ref{fig:limit_vlbT53_bdt} shows the results of this study, providing the observed and 
expected 95\% CL upper limits on the production cross section for $T_{5/3}$
pair production, as a function of $T_{5/3}$ mass. Compared with the theoretical prediction
of the cross section, the results correspond to an observed (expected) 95\% CL lower limit on the
$T_{5/3}$ mass of \obslimitT~GeV (\explimitT~GeV).

\begin{figure}[hbpt]
\centering
\includegraphics[width=0.48\textwidth]{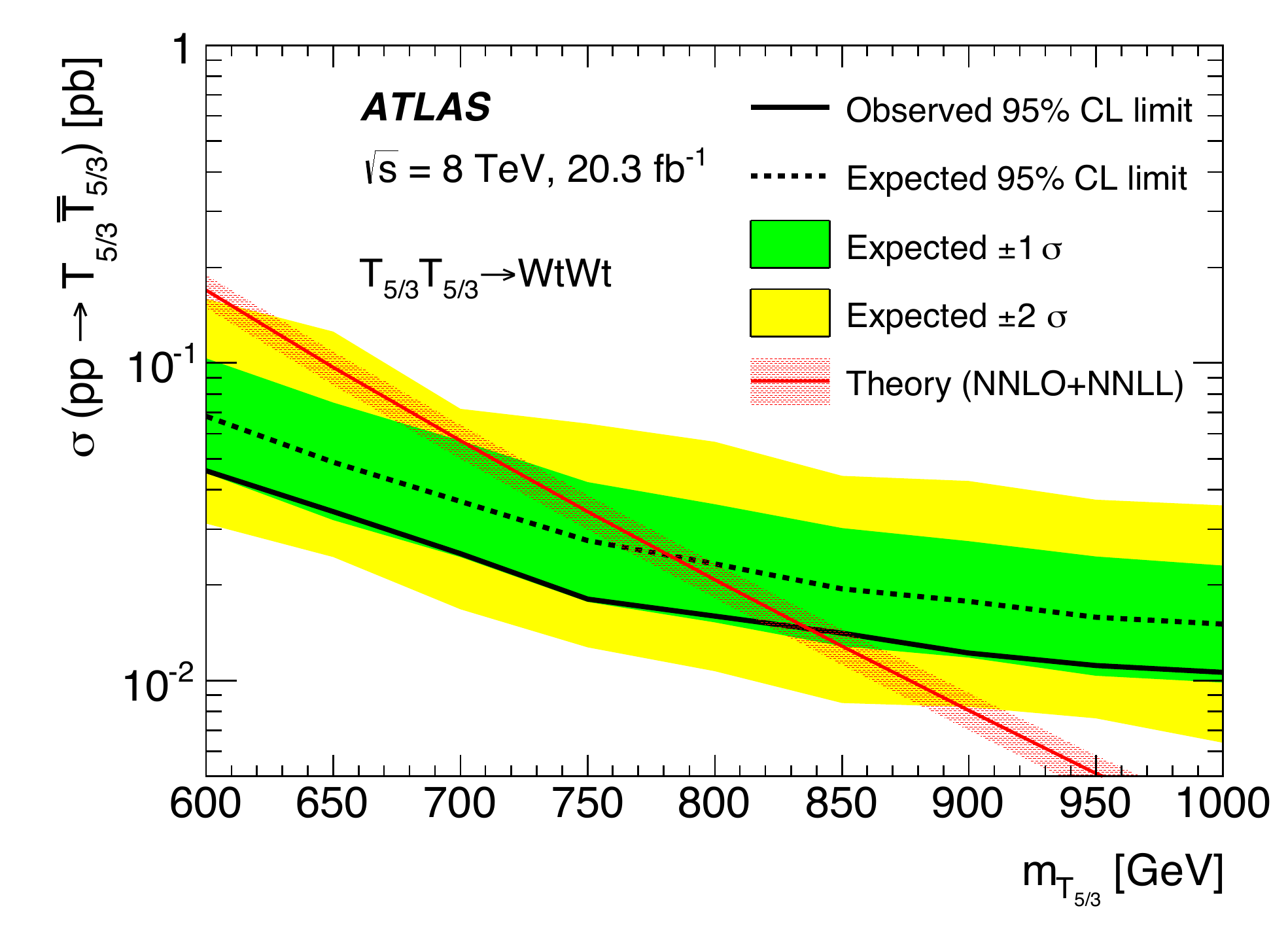}
\caption{Observed and expected 95\% CL upper limits, obtained with the BDT analysis,
 on the cross section for  $T_{5/3}$ pair production, as a function of $T_{5/3}$ mass. The colored,
 charge 5/3 exotic fermion is assumed to decay into a same-sign $W$ boson and top quark, 
 via $T_{5/3} \rightarrow Wt$. The uncertainty on the theoretical cross section is
indicated by the width of the theory band.}
\label{fig:limit_vlbT53_bdt}
\end{figure}  

\clearpage
\section{Conclusions}
\label{sec:conclusions}

Using the data sample of $20.3~\ifb$ of 8~TeV $pp$ collisions recorded in 2012 by the ATLAS detector at the LHC, 
a search has been performed for evidence of pair production of heavy vector-like quarks.
The analysis explores the lepton-plus-jets final state, characterized by events 
with one isolated charged lepton (electron or muon), significant 
missing transverse momentum and multiple jets.
One or more jets are required to be tagged as arising from $b$ quarks, and that at least one pair of jets is tagged as arising from the hadronic decay of an electroweak boson.
The analysis  finds no significant excess
above the expectations for Standard Model backgrounds. Limits are set on
VLQ $B$ production, as a function of its branching ratios, assuming the allowable
decay modes are $B \rightarrow Wt/Zb/Hb$. For a branching ratio of 100\% for
the decay $B \rightarrow Wt$, the observed (expected) 
95\% CL lower limit on the VLQ $B$ mass is \obslimitB~GeV (\explimitB~GeV). In the specific case where the VLQ $B$ has
branching ratios corresponding to those of an $SU(2)$ singlet state, the observed (expected)
95\% CL lower limit on the VLQ $B$ mass is \obslimitBsinglet~GeV (\explimitBsinglet~GeV). The same analysis also investigates
pair production  of a colored, charge 5/3 exotic fermion $T_{5/3}$, with subsequent decay $T_{5/3} \rightarrow Wt$, and
sets an observed (expected) 95\% CL lower limit on the $T_{5/3}$ mass of \obslimitT~GeV (\explimitT~GeV).

\section*{Acknowledgements}
We thank CERN for the very successful operation of the LHC, as well as the
support staff from our institutions without whom ATLAS could not be
operated efficiently.

We acknowledge the support of ANPCyT, Argentina; YerPhI, Armenia; ARC,
Australia; BMWFW and FWF, Austria; ANAS, Azerbaijan; SSTC, Belarus; CNPq and FAPESP,
Brazil; NSERC, NRC, and CFI, Canada; CERN; CONICYT, Chile; CAS, MOST, and NSFC,
China; COLCIENCIAS, Colombia; MSMT CR, MPO CR, and VSC CR, Czech Republic;
DNRF, DNSRC, and Lundbeck Foundation, Denmark; EPLANET, ERC, and NSRF, European Union;
IN2P3-CNRS, CEA-DSM/IRFU, France; GNSF, Georgia; BMBF, DFG, HGF, MPG, and AvH
Foundation, Germany; GSRT and NSRF, Greece; ISF, MINERVA, GIF, I-CORE, and Benoziyo Center,
Israel; INFN, Italy; MEXT and JSPS, Japan; CNRST, Morocco; FOM and NWO,
Netherlands; BRF and Research Council of Norway, Norway; MNiSW and NCN, Poland; GRICES and FCT,
Portugal; MNE/IFA, Romania; MES of Russia and ROSATOM, Russian Federation;
JINR, Russian Federation; MSTD, Serbia; MSSR, Slovakia; ARRS and MIZ\v{S}, Slovenia; DST/NRF, South
Africa;
MINECO, Spain; SRC and Wallenberg Foundation, Sweden; SER, SNSF, and Cantons of
Bern and Geneva, Switzerland; NSC, Taiwan; TAEK, Turkey; STFC, the Royal
Society and Leverhulme Trust, United Kingdom; and DOE and NSF, United States of
America.

The crucial computing support from all WLCG partners is acknowledged
gratefully, in particular from CERN and the ATLAS Tier-1 facilities at
TRIUMF (Canada), NDGF (Denmark, Norway, and Sweden), CC-IN2P3 (France),
KIT/GridKA (Germany), INFN-CNAF (Italy), NL-T1 (Netherlands), PIC (Spain),
ASGC (Taiwan), RAL (UK), and BNL (USA) and in the Tier-2 facilities
worldwide.

\bibliographystyle{atlasBibStyleWoTitle.bst}
\providecommand{\href}[2]{#2}\begingroup\raggedright\endgroup

\newpage
\begin{flushleft}
{\Large The ATLAS Collaboration}

\bigskip

G.~Aad$^{\rm 85}$,
B.~Abbott$^{\rm 113}$,
J.~Abdallah$^{\rm 152}$,
O.~Abdinov$^{\rm 11}$,
R.~Aben$^{\rm 107}$,
M.~Abolins$^{\rm 90}$,
O.S.~AbouZeid$^{\rm 159}$,
H.~Abramowicz$^{\rm 154}$,
H.~Abreu$^{\rm 153}$,
R.~Abreu$^{\rm 30}$
% ...
\bigskip
\\
% ...
$^{11}$ Institute of Physics, Azerbaijan Academy of Sciences, Baku, Azerbaijan\\
% ...
$^{30}$ CERN, Geneva, Switzerland\\
% ...
$^{85}$ CPPM, Aix-Marseille Universit{\'e} and CNRS/IN2P3, Marseille, France\\
% ...
$^{90}$ Department of Physics and Astronomy, Michigan State University, East Lansing MI, United States of America\\
% ...
$^{107}$ Nikhef National Institute for Subatomic Physics and University of Amsterdam, Amsterdam, Netherlands\\
% ...
$^{113}$ Homer L. Dodge Department of Physics and Astronomy, University of Oklahoma, Norman OK, United States of America\\
% ...
$^{152}$ Institute of Physics, Academia Sinica, Taipei, Taiwan\\
$^{153}$ Department of Physics, Technion: Israel Institute of Technology, Haifa, Israel\\
$^{154}$ Raymond and Beverly Sackler School of Physics and Astronomy, Tel Aviv University, Tel Aviv, Israel\\
% ...
$^{159}$ Department of Physics, University of Toronto, Toronto ON, Canada\\
% ...
\end{flushleft}


\end{document}